\documentclass[12pt]{article}
\begin{document}
\newcommand{\beq}{\begin{equation}}
\newcommand{\eeq}{\end{equation}}
\newcommand{\beqa}{\begin{eqnarray}}
\newcommand{\eeqa}{\end{eqnarray}}
\newcommand{\beqar}{\begin{eqnarray*}}
\newcommand{\eeqar}{\end{eqnarray*}}
\newcommand{\al}{\alpha}
\newcommand{\be}{\beta}
\newcommand{\del}{\delta}
\newcommand{\D}{\Delta}
\newcommand{\eps}{\epsilon}
\newcommand{\ga}{\gamma}
\newcommand{\Ga}{\Gamma}
\newcommand{\ka}{\kappa}
\newcommand{\nn}{\nonumber}
\newcommand{\inn}{\!\cdot\!}
\newcommand{\h}{\eta}
\newcommand{\ii}{\iota}
\newcommand{\kk}{\varphi}
\newcommand\F{{}_3F_2}
\newcommand{\la}{\lambda}
\newcommand{\La}{\Lambda}
\newcommand{\na}{\prt}
\newcommand{\Om}{\Omega}
\newcommand{\om}{\omega}
\newcommand{\p}{\phi}
\newcommand{\sig}{\sigma}
\renewcommand{\t}{\theta}
\newcommand{\z}{\zeta}
\newcommand{\ssc}{\scriptscriptstyle}
\newcommand{\eg}{{\it e.g.,}\ }
\newcommand{\ie}{{\it i.e.,}\ }
\newcommand{\labell}[1]{\label{#1}} %{\label{#1}} %
\newcommand{\reef}[1]{(\ref{#1})}
\newcommand\prt{\partial}
\newcommand\veps{\varepsilon}
\newcommand{\pol}{\varepsilon}
\newcommand\vp{\varphi}
\newcommand\ls{\ell_s}
\newcommand\cF{{\cal F}}
\newcommand\cA{{\cal A}}
\newcommand\cS{{\cal S}}
\newcommand\cT{{\cal T}}
\newcommand\cV{{\cal V}}
\newcommand\cL{{\cal L}}
\newcommand\cM{{\cal M}}
\newcommand\cN{{\cal N}}
\newcommand\cG{{\cal G}}
\newcommand\cH{{\cal H}}
\newcommand\cI{{\cal I}}
\newcommand\cJ{{\cal J}}
\newcommand\cl{{\iota}}
\newcommand\cP{{\cal P}}
\newcommand\cQ{{\cal Q}}
\newcommand\cg{{\it g}}
\newcommand\cR{{\cal R}}
\newcommand\cB{{\cal B}}
\newcommand\cO{{\cal O}}
\newcommand\tcO{{\tilde {{\cal O}}}}
\newcommand\bg{\bar{g}}
\newcommand\bb{\bar{b}}
\newcommand\bH{\bar{H}}
\newcommand\bX{\bar{X}}
\newcommand\bK{\bar{K}}
\newcommand\bA{\bar{A}}
\newcommand\bZ{\bar{Z}}
\newcommand\bxi{\bar{\xi}}
\newcommand\bphi{\bar{\phi}}
\newcommand\bpsi{\bar{\psi}}
\newcommand\bprt{\bar{\prt}}
\newcommand\bet{\bar{\eta}}
\newcommand\btau{\bar{\tau}}
\newcommand\hF{\hat{F}}
\newcommand\hA{\hat{A}}
\newcommand\hT{\hat{T}}
\newcommand\htau{\hat{\tau}}
\newcommand\hD{\hat{D}}
\newcommand\hf{\hat{f}}
\newcommand\hg{\hat{g}}
\newcommand\hp{\hat{\phi}}
\newcommand\hi{\hat{i}}
\newcommand\ha{\hat{a}}
\newcommand\hb{\hat{b}}
\newcommand\hQ{\hat{Q}}
\newcommand\hP{\hat{\Phi}}
\newcommand\hS{\hat{S}}
\newcommand\hX{\hat{X}}
\newcommand\tL{\tilde{\cal L}}
\newcommand\hL{\hat{\cal L}}
\newcommand\tG{{\widetilde G}}
\newcommand\tg{{\widetilde g}}
\newcommand\tphi{{\widetilde \phi}}
\newcommand\tPhi{{\widetilde \Phi}}
\newcommand\te{{\tilde e}}
\newcommand\tk{{\tilde k}}
\newcommand\tf{{\tilde f}}
\newcommand\ta{{\tilde a}}
\newcommand\tb{{\tilde b}}
\newcommand\tR{{\tilde R}}
\newcommand\teta{{\tilde \eta}}
\newcommand\tF{{\widetilde F}}
\newcommand\tK{{\widetilde K}}
\newcommand\tE{{\widetilde E}}
\newcommand\tpsi{{\tilde \psi}}
\newcommand\tX{{\widetilde X}}
\newcommand\tD{{\widetilde D}}
\newcommand\tO{{\widetilde O}}
\newcommand\tS{{\tilde S}}
\newcommand\tB{{\widetilde B}}
\newcommand\tA{{\widetilde A}}
\newcommand\tT{{\widetilde T}}
\newcommand\tC{{\widetilde C}}
\newcommand\tV{{\widetilde V}}
\newcommand\thF{{\widetilde {\hat {F}}}}
\newcommand\Tr{{\rm Tr}}
\newcommand\tr{{\rm tr}}
\newcommand\STr{{\rm STr}}
\newcommand\hR{\hat{R}}
\newcommand\M[2]{M^{#1}{}_{#2}}

\newcommand\bS{\textbf{ S}}
\newcommand\bI{\textbf{ I}}
\newcommand\bJ{\textbf{ J}}

%\begin{document}
\begin{titlepage}
\begin{center}

\vskip 2 cm
{\LARGE \bf   Minimal  gauge invariant couplings\\  \vskip 0.25 cm at  order $\alpha'^3$: NS-NS fields  
 }\\
\vskip 1.25 cm
 Mohammad R. Garousi\footnote{garousi@um.ac.ir}
 
\vskip 1 cm
{{\it Department of Physics, Faculty of Science, Ferdowsi University of Mashhad\\}{\it P.O. Box 1436, Mashhad, Iran}\\}
\vskip .1 cm
 \end{center}
\begin{abstract}
Removing   the field redefinitions, the  Bianchi identities and the total derivative  freedoms from the general form of  gauge invariant NS-NS couplings  at order $\alpha'^3$,  we have found that the minimum number of  independent couplings is  872. We find that there are  schemes in which there is no term with structures   $R,\,R_{\mu\nu},\,\nabla_\mu H^{\mu\alpha\beta}$, $ \nabla_\mu\nabla^\mu\Phi$. In these schemes,  there are sub-schemes in which, except one term, the couplings can have no term with more than two derivatives. In the sub-scheme that we have chosen, the 872 couplings appear in 55 different structures.  We fix some of the parameters in type II supersting theory by its corresponding  four-point functions.  The coupling which has term with more than two derivatives is constraint to be zero by the  four-point functions. %The parameters of the couplings which have no B-field, are fixed by the  T-duality constraint on diagonal metric and dilaton. 

\end{abstract}
\end{titlepage}
%\newpage
\tableofcontents

%\newpage
\section{Introduction} \label{intro}
String theory is a quantum theory of gravity   with  a finite number of massless fields and a  tower of infinite number of  massive fields reflecting the stringy nature of the gravity.
An efficient way to study different phenomena in this theory is to use an effective action  which includes     only  massless fields and their derivatives \cite{Howe:1983sra,Witten:1995ex}. %The terms in the effective action which have more than two derivatives, reflect   the  effects of the massive fields.  
The effective action  has a double expansions. The genus expansion which includes  the  classical and a tower of quantum corrections, and the   stringy expansion which is an expansion   in powers of the Regge slope parameter $\alpha'$. The latter expansion for metric yields  the Einstein gravity and the stringy corrections  are   a specific  tower of  higher orders of the  curvature tensors.
A challenge is to explore different techniques   to find the effective action that incorporates  all such corrections, including  non-perturbative effects \cite{Green:1997tv}.
In the bosonic and in the heterotic string theories, the higher derivative couplings  begin at order $\alpha'$, whereas, in type II superstring theory, they   begin at order $\alpha'^3$.

There are various techniques  in the string theory for finding these higher derivative couplings:
S-matrix element approach \cite{Scherk:1974mc,Yoneya:1974jg, Nepomechie:1985us,Metsaev:1986yb,Cai:1986sa,Gross:1986mw}, sigma-model approach \cite{Callan:1985ia,Fradkin:1984pq,Fradkin:1985fq,Metsaev:1987zx,Ketov:1990hw},  supersymmetry approach \cite{Gates:1986dm,Gates:1985wh,Bergshoeff:1986wc,Bergshoeff:1989de}, double field theory  approach  \cite{Siegel:1993xq,Hull:2009mi,Hohm:2010jy, Marques:2015vua,Baron:2017dvb},  and duality approach \cite{Ferrara:1989bc,Font:1990gx,Green:1997tv,Green:2016tfs,Garousi:2017fbe,Green:2019rhz}.
In the duality approach, the  consistency of the effective actions with duality transformations are imposed to find the higher derivative couplings \cite{Green:1997tv,Garousi:2017fbe}.
 In particular, it has been speculated in   \cite{Razaghian:2017okr} that the consistency of the effective actions at any order of $\alpha'$ with the T-duality transformations may fix both the effective actions and the corrections to the Buscher rules   \cite{Buscher:1987sk,Buscher:1987qj}. It has been shown explicitly  in \cite{Garousi:2019wgz} that the T-duality constraint fixes   the effective action and the corrections to the Buscher rules at order $\alpha'$, up to an overall factor. The T-duality constraint has been also used in \cite{Garousi:2019mca} to find the effective action of bosonic string theory at order $\alpha'^2$.

In using the T-duality technique for finding the effective actions at the higher-derivative orders in the string theory, one needs the  minimal   gauge invariant   couplings at each order of $\alpha'$. To find such couplings, one has to impose various Bianchi identities, use  field redefinitions freedom \cite{Gross:1986iv,Tseytlin:1986ti,Deser:1986xr} and remove total derivative terms from the most general gauge invariant couplings. In the literature,  the  Bianchi identities and total derivative terms are first  imposed to find the minimum number of couplings at  each order of $\alpha'$, up to  field redefinitions. The parameters in the resulting action are then either unambiguous which are not changed under field redefinition, or ambiguous which are changed under the field redefinitions. Some combinations of the latter parameters, however, remain invariant under the field redefinitions  \cite{Metsaev:1987zx}.  This allows one to separate the ambiguous parameters to essential parameters which are fixed by \eg S-matrix calculations  \cite{Metsaev:1987zx,Metsaev:1986yb}, and some remaining arbitrary parameters. Depending on which set of parameters are chosen as essential parameters and how to choose the arbitrary parameters, one has different schemes. To find the minimum number of independent couplings, one sets all the arbitrary parameters to zero.    This method has been used  to find the 8 independent couplings for gravity, $B$-field and dilaton at order $\alpha'$ in  \cite{Metsaev:1987zx}. 

One may impose  the Bianchi identities,  remove  the total derivative terms and  use the  field redefinition freedom at the same time. That is, one may first write  all  gauge invariant couplings at each order of $\alpha'$ and then impose the above freedoms to reduce the couplings to the minimal couplings. The parameters in the gauge invariant  action are then either unambiguous or ambiguous depending on whether or not they are changed under these freedoms.   Some combinations of the ambiguous  parameters, however, remain invariant.  This allows one to separate the ambiguous parameters to essential parameters which may be found  by  S-matrix calculations, and some arbitrary parameters. Again, depending on which set of parameters are choosing as essential parameters and how to choose the arbitrary parameters, one has different schemes. The minimum number of independent couplings are found in the schemes that all the arbitrary parameters are set to zero. This method has been used in \cite{Garousi:2019cdn} to find the 60  minimal gauge invariant couplings  for gravity, $B$-field and dilaton  at order $\alpha'^2$. In this paper, we are going to find such couplings at order $\alpha'^3$.

The outline of the paper is as follows: In section 2, we write the most general gauge invariant couplings involving metric, dilaton and B-field at order $\alpha'^3$. There are 23996 such  couplings. Then we add to them the most general total derivative terms and field redefinitions with arbitrary parameters. To impose various Bianchi identities, we rewrite  them in the local inertial frame, and rewrite the terms which have derivatives of B-field strength $H$,  in terms of potential, \ie $H=dB$.  We then use the arbitrary parameters in the total derivative terms and in the field redefinitions  to show that there are only 872 essential parameters and all other parameters are arbitrary which can be set to zero.  We show that there are minimal schemes in which there are 871 couplings  which have no term with  more than two derivatives and  no term  involving $R,R_{\mu\nu},\nabla_\mu H^{\mu\alpha\beta}$, $ \nabla_\mu\nabla^\mu\Phi$. There is also one essential coupling which has three derivatives on B-field. The 872 couplings in the scheme that we have chosen, appear in 55 different structures. We write the explicit form of these couplings in this section. In section 3, we impose the constraint that the couplings  in the type II superstring theory should reproduce the sphere-level S-matrix element of four NS-NS operators at order $\alpha'^3$, to fix some of the parameters in the superstring theory. In section 4, we impose the T-duality constraint when B-field is zero, to show that  the dilaton couplings are all zero. We postpone the evaluation of all 872 parameters by the T-duality constraint when B-field is non-zero, to the future works.

\section{Minimal couplings at order  $\alpha'^3$}\label{sec.2}

The effective action of string theory has a double expansions. One expansion is the genus expansion which includes  the  classical sphere-level and a tower of quantum effects. The other one is a stringy expansion which is an expansion in terms of higher-derivative couplings. The number of derivatives in each  coupling can be accounted by the order of $\alpha'$. The sphere-level effective action  of superstring theory has the following    power  series of $\alpha'$ in the string frame:
\beqa
S_{\rm eff}&=\sum^\infty_{n=0}\alpha'^nS_n=S_0+\alpha'^3 S_3+\alpha'^4 S_4+\cdots\ ; \quad S_n= \frac{\gamma_n}{\kappa^2}\int d^{10} x\sqrt{-g} e^{-2\Phi}\mathcal{L}_n\labell{seff}
\eeqa
where $\gamma_n$ is  normalization of the effective action at order $\alpha'^n$, \eg $\gamma_0=1/2$, and we have used the fact that the  superstring theory has no effective action at orders $\alpha',\alpha'^2$. The effective action must be invariant under the coordinate transformations and under the $B$-field and R-R  gauge transformations. So the NS-NS  fields    must  appear in the Lagrangian $ \mathcal{L}_n$  trough their field strengths and their covariant derivatives, \eg
the Lagrangian at the leading order of $\alpha'$ for NS-NS fields is
\beqa
\mathcal{L}_0&= R-\frac{1}{12}H_{\alpha\beta\gamma}H^{\alpha\beta\gamma}+4\nabla_\alpha\Phi\nabla^\alpha\Phi
\eeqa
Similarly for the R-R fields. There is also Chern-Simons Lagrangian involving  the R-R couplings. In this paper however  we are not interested in the R-R couplings. The  higher-derivative field redefinitions   and Bianchi identity can not change the form of this action. A systematic method has been used in \cite{Garousi:2019cdn} to find the minimum number of independent couplings at order $\alpha'^2$ in the bosonic string theory. It has been shown in \cite{Garousi:2019cdn} that there are 60 couplings at this order. The  coefficients of these  couplings have been found in \cite{Garousi:2019mca} by the T-duality constraint. In this paper we are going to find the independent couplings of NS-NS fields at order $\alpha'^3$ in the superstring theory.

Following \cite{Garousi:2019cdn}, one first should write all gauge invariant NS-NS couplings at eight-derivative order with even parity. Using the package   "xAct" \cite{Nutma:2013zea}, one finds there are 23996 such couplings in 202 different structures, \ie 
\beqa
 L'_3=& c'_1 H_{\alpha}{}^{\delta \epsilon} H^{\alpha \beta \gamma} \
H_{\beta}{}^{\varepsilon \mu} H_{\gamma}{}^{\zeta \eta} H_{\
\delta \varepsilon}{}^{\theta} H_{\epsilon \zeta}{}^{\iota} \
H_{\mu \iota}{}^{\kappa} H_{\eta \theta \kappa}+\cdots\labell{L3}
\eeqa
where $c'_1,\cdots, c'_{23996}$ are some parameters. The above couplings however are  not all independent. Some of them are related by total derivative terms, some of them are related by field redefinitions, and some others are related by various Bianchi identities.

To remove the total derivative terms from the above couplings, we consider the most general total derivative terms at order $\alpha'^3$ which has the following structure:
\beqa
\frac{\alpha'^3\gamma_n}{\kappa^2}\int d^{10}x \sqrt{-g}e^{-2\Phi} \mathcal{J}_3=\frac{\alpha'^3\gamma_n}{\kappa^2}\int d^{10}x\sqrt{-g} \nabla_\alpha (e^{-2\Phi}{\cal I}_3^\alpha) \labell{J3}
\eeqa
where the vector ${\cal I}_3^\alpha$ is   all possible  covariant and gauge invariant  terms at seven-derivative level with even parity, \ie, 
\beqa
{\cal I}_3^\alpha= &J_1 H^{\gamma\delta\epsilon}R^{\alpha\beta}R_{\beta\varepsilon\epsilon\theta}\nabla_{\delta}H_{\gamma}{}^{\varepsilon\theta}+\cdots
\eeqa
where the coefficients  $J_1,\cdots, J_{11941}$ are 11941 arbitrary parameters. Adding the total derivative terms with arbitrary coefficients  to $L'_3$, one finds the same Lagrangian     but with different parameters $c''_1, c''_2, \cdots$. We call the new Lagrangian  ${ L}''_3$. Hence 
\beqa
\Delta''_3-{\cal J}_3&=&0\labell{DL}
\eeqa
where $\Delta''_3={ L}''_3-L'_3$ is the same as $L'_3$ but with coefficients $\delta c''_1,\delta c''_2,\cdots$ where $\delta c''_i= c''_i-c'_i$. Solving the above equation, one finds some linear  relations between  only $\delta c''_1,\delta c''_2,\cdots$ which indicate how the couplings are related among themselves by the total derivative terms. The above equation also gives some relation between the coefficients of the total derivative terms and $\delta c''_1,\delta c''_2,\cdots$ in which we are not interested.

The couplings in \reef{DL}, however, are in a  fixed field variables. One is free to change the field variables as 
\begin{eqnarray}
g_{\mu\nu}&\rightarrow &g_{\mu\nu}+\alpha'^3 \delta g^{(3)}_{\mu\nu}\nn\\
B_{\mu\nu}&\rightarrow &B_{\mu\nu}+ \alpha'^3\delta B^{(3)}_{\mu\nu}\nn\\
\Phi &\rightarrow &\Phi+ \alpha'^3\delta\Phi^{(3)}\labell{gbp}
\end{eqnarray}
where the tensors $\delta g^{(3)}_{\mu\nu}$, $\delta B^{(3)}_{\mu\nu}$ and $\delta\Phi^{(3)}$ are all possible covariant and gauge invariant terms at 6-derivative level.  $\delta g^{(3)}_{\mu\nu}$, $\delta\Phi^{(3)}$ contain even-parity terms and $\delta B^{(3)}_{\mu\nu}$ contains odd-parity terms \ie,
\beqa
 \delta g^{(3)}_{\alpha\beta}&=& g_1 H_{\{\alpha}{}^{\gamma\delta}H_{\beta\}\gamma}{}^{\epsilon}H_{\delta}{}^{\varepsilon\theta}H_{\epsilon\varepsilon}{}^{\eta}H_{\theta}{}^{\mu\nu}H_{\eta\mu\nu}+\cdots \nn\\
 \delta B^{(3)}_{\alpha\beta}&=& e_1 R^{\gamma\delta}R_{\delta\epsilon\varepsilon [\alpha}\nabla_{\beta ]}H_{\gamma}{}^{\epsilon\varepsilon}+\cdots \nn\\
 \delta\Phi^{(3)}&=&f_1 H_{\alpha}{}^{\delta\epsilon}R^{\beta\gamma}\nabla^{\alpha}\Phi\nabla_{\gamma}H_{\beta\delta\epsilon}+\cdots\labell{eq.12}
\eeqa
The coefficients $g_1,\cdots, g_{3440}$, $e_1,\cdots, e_{2843}$ and $f_1,\cdots, f_{705}$ are arbitrary parameters. When the field variables in $S_3$  are changed according to the above  field redefinitions, they produce some couplings at orders $\alpha'^6$ and higher in which we are not interested in this paper. However, when the field variables in $S_0$  are changed,  up to some total derivative terms, the following   couplings  at order $\alpha'^3$ are produced:
\beqa
\delta S_0&\!\!\!\!\!=\!\!\!\!\!\!&\frac{\delta S_0}{\delta g_{\alpha\beta}}\delta g^{(3)}_{\alpha\beta}+\frac{\delta S_0}{\delta B_{\alpha\beta}}\delta B^{(3)}_{\alpha\beta}+\frac{\delta S_0}{\delta \Phi}\delta \Phi^{(3)}\equiv \frac{\alpha'^3\gamma_3}{\kappa^2}\int d^{10}x\sqrt{-g}e^{-2\Phi}\mathcal{K}_3\nn\\
&\!\!\!\!\!=\!\!\!\!\!\!& \frac{\alpha'^3\gamma_3}{\kappa^2}\int d^{10} x\sqrt{-g}e^{-2\Phi}\Big[(\frac{1}{2} \nabla_{\gamma}H^{\alpha \beta \gamma} -  H^{\alpha \beta}{}_{\gamma} \nabla^{\gamma}\Phi)\delta B^{(3)}_{\alpha\beta}\nn\\
&& -(  R^{\alpha \beta}-\frac{1}{4} H^{\alpha \gamma \delta} H^{\beta}{}_{\gamma \delta}+ 2 \nabla^{\beta}\nabla^{\alpha}\Phi)\delta g^{(3)}_{\alpha\beta}
\nn\\
&&-2( R -\frac{1}{12} H_{\alpha \beta \gamma} H^{\alpha \beta \gamma} + 4 \nabla_{\alpha}\nabla^{\alpha}\Phi -4 \nabla_{\alpha}\Phi \nabla^{\alpha}\Phi)(\delta\Phi^{(3)}-\frac{1}{4}\delta g^{(3)\mu}{}_\mu) \Big]\labell{eq.13}
\eeqa
where we have absorbed a factor of $\gamma_0/\gamma_3$ to the arbitrary parameters in \reef{eq.12}. Adding the total derivative terms and field redefinition terms  to $L'_3$, one finds the same Lagrangian     but with different parameters $c_1, c_2, \cdots$. We call the new Lagrangian  ${\cal L}_3$. Hence 
\beqa
\Delta_3-{\cal J}_3-{\cal K}_3&=&0\labell{DLK}
\eeqa
where $\Delta_3={ \cal L}_3-L'_3$ is the same as $L'_3$ but with coefficients $\delta c_1,\delta c_2,\cdots$ where $\delta c_i= c_i-c'_i$. Solving the above equation, one finds some linear  relations between  only $\delta c_1,\delta c_2,\cdots$ which indicate how the couplings are related among themselves by the total derivative and field redefinition terms. There are also many relations between $\delta c_1,\delta c_2,\cdots$ and the coefficients of total derivative terms and field redefinitions in which we are not interested,   

However, to solve the equation \reef{DLK} one should write it in terms of independent couplings, \ie    one has to impose the following Bianchi identities as well:
\beqa
 R_{\alpha[\beta\gamma\delta]}&=&0\nn\\
 \nabla_{[\mu}R_{\alpha\beta]\gamma\delta}&=&0\labell{bian}\\
\nabla_{[\mu}H_{\alpha\beta\gamma]}&=&0\nn\\
{[}\nabla,\nabla{]}\mathcal{O}-R\mathcal{O}&=&0\nn
\eeqa
 To impose these Bianchi identities in gauge invariant form, one may contract the  left-hand side of each Bianchi identity with the NS-NS field strengths and their derivatives  to produce terms at order $\alpha'^3$. The coefficients of these terms are arbitrary. Adding these terms to the equation \reef{DLK}, then one can solve the equation to find the linear relations between   only $\delta c_1,\delta c_2,\cdots$. Alternatively, to impose the  Bianchi identities in non-gauge invariant form, one may rewrite the terms in \reef{DLK} in  the local frame in which the first derivative of metric is zero, and  rewrite the terms in \reef{DLK} which have derivatives of $H$ in terms of B-field, \ie $H=dB$.  In this way,  the Bianchi identities satisfy automatically \cite{Garousi:2019cdn}. In fact, writing the couplings in terms of potential rather than field strength, there would be no Bianchi identity at all. We find that this latter approach is easier to impose the Bianchi identities by computer. Moreover, in this approach one does not need to introduce another  large number of arbitrary parameters to include the Bianchi identities to the equation \reef{DLK}.    

Using the above  steps, one can rewrite the different terms on the left-hand side of \reef{DLK} in terms of independent but non-gauge invariant couplings. The solution to the equation \reef{DLK} then has two parts. One part is 872 relations between only $\delta c_i$'s, and the other part is some relations between the coefficients of the total derivative terms, field redefinitions  and $\delta c_i$'s in which we are not interested. The number of relations in the first part gives the number of independent couplings in ${\cal L}_3$. In a particular scheme, one may set some of the coefficients in  $L_3'$ to zero, however, after replacing the non-zero terms in \reef{DLK}, the number of   relations between only $\delta c_i$'s should not be changed, \ie there must be always 872 relations.   We set the coefficients of the couplings in $L_3'$ in which each term has $R,\,R_{\mu\nu},\,\nabla_\mu H^{\mu\alpha\beta}$, $ \nabla_\mu\nabla^\mu\Phi$  zero.   After setting this coefficients to zero, there are still 872 relations between  $\delta c_i$'s.  This means we are  allowed  to  remove these terms.  

We then try to set zero couplings in  $L_3'$ which have term with more then two derivatives. Imposing this condition and  then solving \reef{DLK} again, one would find 871 relations between only  $\delta c_i$'s. It means that at least  one of the independent couplings has terms with more than two derivatives. We have found this independent coupling to be 
\beqa
{\cal L}_3^{R^2H\prt\prt H}&=&c_{520} H^{\alpha \beta \gamma} R_{\beta 
\mu \delta}{}^{\zeta} R_{\gamma \zeta 
\epsilon \varepsilon} \nabla^{\mu}\nabla^{\varepsilon}H_{\alpha}{}^{\delta \epsilon}\labell{T53}
\eeqa
The way we have found above coupling is that we divided the couplings involving more than two derivatives to two parts. We then set the coefficients of one part to zero. If the corresponding equations in \reef{DLK} gives 872 relations between the remaining  $\delta c_i$'s then that choice is allowed, otherwise the other part is allowed to set to zero. Again we divided the  non-zero part to two parts and set half of them to zero. If the  corresponding equations in \reef{DLK} gives 872 relations between the remaining  $\delta c_i$'s then that choice is allowed, otherwise the other part is allowed to set to zero. Repeating this strategy one finds the above couplings is one of the independent couplings. Apart from the above coupling, all other couplings which have terms with more than two derivatives are allowed to be zero. There are still 2882 couplings which have no term with more than two derivatives and have no terms with structures $R,\,R_{\mu\nu},\,\nabla_\mu H^{\mu\alpha\beta}$, $ \nabla_\mu\nabla^\mu\Phi$. Hence, there are still  many choices for choosing the  non-zero  coefficients such that they satisfy the 872 relations $\delta c_i=0$.    In the particular scheme that we have chosen, the 872 couplings  appear in 55  structures.

One structure is  \reef{T53} which has only one coupling. All other couplings appear in the following 54 structures:
\beqa
{\cal L}_3^{H^8}&\!\!\!\!\!=\!\!\!\!\! &c_1 H_{\alpha}{}^{\delta \epsilon} H^{\alpha \beta \
\gamma} H_{\beta}{}^{\varepsilon \mu} \
H_{\gamma}{}^{\zeta \eta} H_{\delta \varepsilon}{}^{\theta} H_{\
\epsilon \zeta}{}^{\iota} H_{\mu \iota}{}^{\kappa} 
H_{\eta \theta \kappa} + c_2 H_{\alpha}{}^{\delta 
\epsilon} H^{\alpha \beta \gamma} H_{\beta 
\delta}{}^{\varepsilon} H_{\gamma}{}^{\mu \zeta} 
H_{\epsilon \mu}{}^{\eta} H_{\varepsilon}{}^{\theta 
\iota} H_{\zeta \theta}{}^{\kappa} H_{\eta \iota \kappa}\nn\\&& + 
c_3 H_{\alpha \beta}{}^{\delta} H^{\alpha \beta 
\gamma} H_{\gamma}{}^{\epsilon \varepsilon} 
H_{\delta}{}^{\mu \zeta} H_{\epsilon\mu}{}^{\eta} H_{\varepsilon}{}^{\theta \iota} H_{\zeta 
\theta}{}^{\kappa} H_{\eta \iota \kappa} + c_4 
H_{\alpha \beta}{}^{\delta} H^{\alpha \beta \gamma} 
H_{\gamma}{}^{\epsilon \varepsilon} H_{\delta}{}^{\mu 
\zeta} H_{\epsilon \varepsilon}{}^{\eta} H_{\mu}{}^{\theta \iota} H_{\zeta \theta}{}^{\kappa} H_{\eta \iota 
\kappa}\nn\\&& + c_5 H_{\alpha}{}^{\delta \epsilon} 
H^{\alpha \beta \gamma} H_{\beta \delta}{}^{\varepsilon} 
H_{\gamma \epsilon}{}^{\mu} H_{\varepsilon}{}^{\zeta 
\eta} H_{\mu}{}^{\theta \iota} H_{\zeta 
\theta}{}^{\kappa} H_{\eta \iota \kappa} + c_6 
H_{\alpha \beta}{}^{\delta} H^{\alpha \beta \gamma} 
H_{\gamma}{}^{\epsilon \varepsilon} H_{\delta 
\epsilon}{}^{\mu} H_{\varepsilon}{}^{\zeta \eta} 
H_{\mu}{}^{\theta \iota} H_{\zeta \theta}{}^{\kappa} 
H_{\eta \iota \kappa}\nn\\&& + c_7 H_{\alpha}{}^{\delta 
\epsilon} H^{\alpha \beta \gamma} H_{\beta 
\delta}{}^{\varepsilon} H_{\gamma \epsilon \varepsilon} 
H_{\mu}{}^{\theta \iota} H^{\mu \zeta \eta} 
H_{\zeta \theta}{}^{\kappa} H_{\eta \iota \kappa} + 
c_8 H_{\alpha}{}^{\delta \epsilon} H^{\alpha \beta 
\gamma} H_{\beta}{}^{\varepsilon \mu} 
H_{\gamma}{}^{\zeta \eta} H_{\delta \varepsilon}{}^{\theta} H_{
\epsilon \zeta}{}^{\iota} H_{\mu \eta}{}^{\kappa} H_{
\theta \iota \kappa}\labell{T1}\nn
\eeqa
\beqa
{\cal L}_3^{H^6R}&\!\!\!\!\!=\!\!\!\!\! & c_9H_{\alpha \beta}{}^{\delta} H^{\alpha \beta \gamma} H_{\epsilon}{}^{\zeta \eta} H^{\epsilon \varepsilon \mu} H_{\varepsilon \zeta}{}^{\theta} H_{\mu \eta}{}^{\iota} R_{\gamma \theta \delta \iota} + c_{13} H_{\alpha \beta}{}^{\delta} H^{\alpha \beta \gamma} H_{\gamma}{}^{\epsilon \varepsilon} H_{\epsilon}{}^{\mu \zeta} H_{\mu}{}^{\eta \theta} H_{\eta \theta}{}^{\iota} R_{\delta \zeta \varepsilon \iota}\nn\\&& + c_{14} H_{\alpha \beta}{}^{\delta} H^{\alpha \beta \gamma} H_{\gamma}{}^{\epsilon \varepsilon} H_{\epsilon}{}^{\mu \zeta} H_{\varepsilon}{}^{\eta \theta} H_{\mu \eta}{}^{\iota} R_{\delta \zeta \theta \iota} + c_{15} H_{\alpha \beta}{}^{\delta} H^{\alpha \beta \gamma} H_{\gamma}{}^{\epsilon \varepsilon} H_{\epsilon}{}^{\mu \zeta} H_{\mu}{}^{\eta \theta} H_{\zeta \eta}{}^{\iota} R_{\delta \theta \varepsilon \iota}\nn\\&& + c_{16} H_{\alpha \beta}{}^{\delta} H^{\alpha \beta \gamma} H_{\gamma}{}^{\epsilon \varepsilon} H_{\epsilon \varepsilon}{}^{\mu} H_{\zeta \eta}{}^{\iota} H^{\zeta \eta \theta} R_{\delta \theta \mu \iota} + c_{17} H_{\alpha \beta}{}^{\delta} H^{\alpha \beta \gamma} H_{\gamma}{}^{\epsilon \varepsilon} H_{\epsilon}{}^{\mu \zeta} H_{\varepsilon \mu}{}^{\eta} H_{\zeta}{}^{\theta \iota} R_{\delta \theta \eta \iota} \nn\\&&+ c_{18} H_{\alpha \beta}{}^{\delta} H^{\alpha \beta \gamma} H_{\gamma}{}^{\epsilon \varepsilon} H_{\epsilon}{}^{\mu \zeta} H_{\mu}{}^{\eta \theta} H_{\eta \theta}{}^{\iota} R_{\delta \iota \varepsilon \zeta} + c_{19} H_{\alpha \beta}{}^{\delta} H^{\alpha \beta \gamma} H_{\gamma}{}^{\epsilon \varepsilon} H_{\epsilon}{}^{\mu \zeta} H_{\varepsilon}{}^{\eta \theta} H_{\mu \zeta}{}^{\iota} R_{\delta \iota \eta \theta}\nn\\&& + c_{31} H_{\alpha}{}^{\delta \epsilon} H^{\alpha \beta \gamma} H_{\beta}{}^{\varepsilon \mu} H_{\gamma}{}^{\zeta \eta} H_{\delta \varepsilon}{}^{\theta} H_{\zeta \theta}{}^{\iota} R_{\epsilon \eta \mu \iota} + c_{32} H_{\alpha}{}^{\delta \epsilon} H^{\alpha \beta \gamma} H_{\beta \delta}{}^{\varepsilon} H_{\gamma}{}^{\mu \zeta} H_{\mu}{}^{\eta \theta} H_{\zeta \eta}{}^{\iota} R_{\epsilon \theta \varepsilon \iota}\nn\\&& + c_{50} H_{\alpha}{}^{\delta \epsilon} H^{\alpha \beta \gamma} H_{\beta \delta}{}^{\varepsilon} H_{\gamma}{}^{\mu \zeta} H_{\epsilon}{}^{\eta \theta} H_{\mu \eta}{}^{\iota} R_{\varepsilon \zeta \theta \iota} + c_{51} H_{\alpha \beta}{}^{\delta} H^{\alpha \beta \gamma} H_{\gamma}{}^{\epsilon \varepsilon} H_{\delta \epsilon}{}^{\mu} H_{\zeta \eta}{}^{\iota} H^{\zeta \eta \theta} R_{\varepsilon \theta \mu \iota} \nn\\&&+ c_{52} H_{\alpha \beta}{}^{\delta} H^{\alpha \beta \gamma} H_{\gamma}{}^{\epsilon \varepsilon} H_{\delta}{}^{\mu \zeta} H_{\epsilon}{}^{\eta \theta} H_{\mu \eta}{}^{\iota} R_{\varepsilon \theta \zeta \iota} + c_{53} H_{\alpha \beta}{}^{\delta} H^{\alpha \beta \gamma} H_{\gamma}{}^{\epsilon \varepsilon} H_{\delta}{}^{\mu \zeta} H_{\epsilon \mu}{}^{\eta} H_{\eta}{}^{\theta \iota} R_{\varepsilon \theta \zeta \iota}\nn\\&& + c_{54} H_{\alpha}{}^{\delta \epsilon} H^{\alpha \beta \gamma} H_{\beta \delta}{}^{\varepsilon} H_{\gamma}{}^{\mu \zeta} H_{\epsilon \mu}{}^{\eta} H_{\zeta}{}^{\theta \iota} R_{\varepsilon \theta \eta \iota} + c_{55} H_{\alpha \beta}{}^{\delta} H^{\alpha \beta \gamma} H_{\gamma}{}^{\epsilon \varepsilon} H_{\delta}{}^{\mu \zeta} H_{\epsilon}{}^{\eta \theta} H_{\eta \theta}{}^{\iota} R_{\varepsilon \iota \mu \zeta}\nn\\&& + c_{56} H_{\alpha \beta}{}^{\delta} H^{\alpha \beta \gamma} H_{\gamma}{}^{\epsilon \varepsilon} H_{\delta}{}^{\mu \zeta} H_{\epsilon}{}^{\eta \theta} H_{\mu \eta}{}^{\iota} R_{\varepsilon \iota \zeta \theta} + c_{72} H_{\alpha \beta}{}^{\delta} H^{\alpha \beta \gamma} H_{\gamma}{}^{\epsilon \varepsilon} H_{\delta}{}^{\mu \zeta} H_{\epsilon}{}^{\eta \theta} H_{\varepsilon \eta}{}^{\iota} R_{\mu \theta \zeta \iota}\nn\\&& + c_{73} H_{\alpha \beta}{}^{\delta} H^{\alpha \beta \gamma} H_{\gamma}{}^{\epsilon \varepsilon} H_{\delta}{}^{\mu \zeta} H_{\epsilon \varepsilon}{}^{\eta} H_{\eta}{}^{\theta \iota} R_{\mu \theta \zeta \iota} + c_{74} H_{\alpha}{}^{\delta \epsilon} H^{\alpha \beta \gamma} H_{\beta}{}^{\varepsilon \mu} H_{\gamma}{}^{\zeta \eta} H_{\delta \varepsilon}{}^{\theta} H_{\epsilon \zeta}{}^{\iota} R_{\mu \theta \eta \iota}\nn\\&& + c_{75} H_{\alpha}{}^{\delta \epsilon} H^{\alpha \beta \gamma} H_{\beta \delta}{}^{\varepsilon} H_{\gamma \epsilon}{}^{\mu} H_{\varepsilon}{}^{\zeta \eta} H_{\zeta}{}^{\theta \iota} R_{\mu \theta \eta \iota} + c_{76} H_{\alpha \beta}{}^{\delta} H^{\alpha \beta \gamma} H_{\gamma}{}^{\epsilon \varepsilon} H_{\delta \epsilon}{}^{\mu} H_{\varepsilon}{}^{\zeta \eta} H_{\zeta}{}^{\theta \iota} R_{\mu \theta \eta \iota}\nn\\&& + c_{77} H_{\alpha}{}^{\delta \epsilon} H^{\alpha \beta \gamma} H_{\beta}{}^{\varepsilon \mu} H_{\gamma}{}^{\zeta \eta} H_{\delta \varepsilon}{}^{\theta} H_{\epsilon \zeta}{}^{\iota} R_{\mu \iota \eta \theta} + c_{80} H_{\alpha}{}^{\delta \epsilon} H^{\alpha \beta \gamma} H_{\beta \delta}{}^{\varepsilon} H_{\gamma}{}^{\mu \zeta} H_{\epsilon \mu}{}^{\eta} H_{\varepsilon}{}^{\theta \iota} R_{\zeta \theta \eta \iota} \nn\\&&+ c_{81} H_{\alpha \beta}{}^{\delta} H^{\alpha \beta \gamma} H_{\gamma}{}^{\epsilon \varepsilon} H_{\delta}{}^{\mu \zeta} H_{\epsilon \mu}{}^{\eta} H_{\varepsilon}{}^{\theta \iota} R_{\zeta \theta \eta \iota} + c_{82} H_{\alpha \beta}{}^{\delta} H^{\alpha \beta \gamma} H_{\gamma}{}^{\epsilon \varepsilon} H_{\delta}{}^{\mu \zeta} H_{\epsilon \varepsilon}{}^{\eta} H_{\mu}{}^{\theta \iota} R_{\zeta \theta \eta \iota}\nn\\&& + c_{83} H_{\alpha}{}^{\delta \epsilon} H^{\alpha \beta \gamma} H_{\beta \delta}{}^{\varepsilon} H_{\gamma \epsilon}{}^{\mu} H_{\varepsilon}{}^{\zeta \eta} H_{\mu}{}^{\theta \iota} R_{\zeta \theta \eta \iota} + c_{84} H_{\alpha \beta}{}^{\delta} H^{\alpha \beta \gamma} H_{\gamma}{}^{\epsilon \varepsilon} H_{\delta \epsilon}{}^{\mu} H_{\varepsilon}{}^{\zeta \eta} H_{\mu}{}^{\theta \iota} R_{\zeta \theta \eta \iota}\nn\\&& + c_{85} H_{\alpha}{}^{\delta \epsilon} H^{\alpha \beta \gamma} H_{\beta \delta}{}^{\varepsilon} H_{\gamma \epsilon \varepsilon} H_{\mu}{}^{\theta \iota} H^{\mu \zeta \eta} R_{\zeta \theta \eta \iota} + c_{86} H_{\alpha \beta}{}^{\delta} H^{\alpha \beta \gamma} H_{\gamma}{}^{\epsilon \varepsilon} H_{\delta \epsilon \varepsilon} H_{\mu}{}^{\theta \iota} H^{\mu \zeta \eta} R_{\zeta \theta \eta \iota}\labell{T2}\nn
\eeqa
\beqa
{\cal L}_3^{R^4}&\!\!\!\!\!=\!\!\!\!\! & c_{10} R_{\alpha \beta}{}^{\epsilon \varepsilon} R^{\alpha \beta \gamma \delta} R_{\gamma}{}^{\mu}{}_{\epsilon}{}^{\zeta} R_{\delta \mu \varepsilon \zeta} + c_{11} R_{\alpha}{}^{\epsilon}{}_{\gamma}{}^{\varepsilon} R^{\alpha \beta \gamma \delta} R_{\beta}{}^{\mu}{}_{\epsilon}{}^{\zeta} R_{\delta \zeta \varepsilon \mu} + c_{12} R_{\alpha \beta}{}^{\epsilon \varepsilon} R^{\alpha \beta \gamma \delta} R_{\gamma}{}^{\mu}{}_{\epsilon}{}^{\zeta} R_{\delta \zeta \varepsilon \mu}\nn\\&& + c_{20} R_{\alpha \beta}{}^{\epsilon \varepsilon} R^{\alpha \beta \gamma \delta} R_{\gamma}{}^{\mu}{}_{\delta}{}^{\zeta} R_{\epsilon \mu \varepsilon \zeta} + c_{21} R_{\alpha \gamma \beta}{}^{\epsilon} R^{\alpha \beta \gamma \delta} R_{\delta}{}^{\varepsilon \mu \zeta} R_{\epsilon \mu \varepsilon \zeta} + c_{22} R_{\alpha}{}^{\epsilon}{}_{\gamma}{}^{\varepsilon} R^{\alpha \beta \gamma \delta} R_{\beta}{}^{\mu}{}_{\delta}{}^{\zeta} R_{\epsilon \zeta \varepsilon \mu} \nn\\&&+ c_{33} R_{\alpha \gamma \beta \delta} R^{\alpha \beta \gamma \delta} R_{\epsilon \mu \varepsilon \zeta} R^{\epsilon \varepsilon \mu \zeta}\labell{T3}
\eeqa
\beqa
{\cal L}_3^{H^2R^3}&\!\!\!\!\!=\!\!\!\!\! & c_{23} H^{\alpha \beta \gamma} H^{\delta \epsilon \varepsilon} R_{\alpha}{}^{\mu}{}_{\beta}{}^{\zeta} R_{\gamma}{}^{\eta}{}_{\delta \mu} R_{\epsilon \zeta \varepsilon \eta} + c_{24} H_{\alpha}{}^{\delta \epsilon} H^{\alpha \beta \gamma} R_{\beta \delta}{}^{\varepsilon \mu} R_{\gamma}{}^{\zeta}{}_{\varepsilon}{}^{\eta} R_{\epsilon \zeta \mu \eta}\nn\\&& + c_{25} H_{\alpha}{}^{\delta \epsilon} H^{\alpha \beta \gamma} R_{\beta}{}^{\varepsilon}{}_{\delta}{}^{\mu} R_{\gamma}{}^{\zeta}{}_{\varepsilon}{}^{\eta} R_{\epsilon \zeta \mu \eta} + c_{26} H_{\alpha}{}^{\delta \epsilon} H^{\alpha \beta \gamma} R_{\beta}{}^{\varepsilon}{}_{\gamma}{}^{\mu} R_{\delta}{}^{\zeta}{}_{\varepsilon}{}^{\eta} R_{\epsilon \zeta \mu \eta}\nn\\&& + c_{27} H_{\alpha}{}^{\delta \epsilon} H^{\alpha \beta \gamma} R_{\beta}{}^{\varepsilon}{}_{\delta}{}^{\mu} R_{\gamma}{}^{\zeta}{}_{\varepsilon}{}^{\eta} R_{\epsilon \eta \mu \zeta} + c_{28} H_{\alpha}{}^{\delta \epsilon} H^{\alpha \beta \gamma} R_{\beta}{}^{\varepsilon}{}_{\gamma}{}^{\mu} R_{\delta}{}^{\zeta}{}_{\varepsilon}{}^{\eta} R_{\epsilon \eta \mu \zeta} \nn\\&&+ c_{29} H_{\alpha \beta}{}^{\delta} H^{\alpha \beta \gamma} R_{\gamma}{}^{\epsilon \varepsilon \mu} R_{\delta}{}^{\zeta}{}_{\varepsilon}{}^{\eta} R_{\epsilon \eta \mu \zeta} + c_{34} H^{\alpha \beta \gamma} H^{\delta \epsilon \varepsilon} R_{\alpha \delta \beta \epsilon} R_{\gamma}{}^{\mu \zeta \eta} R_{\varepsilon \zeta \mu \eta} \nn\\&&+ c_{35} H_{\alpha}{}^{\delta \epsilon} H^{\alpha \beta \gamma} R_{\beta}{}^{\varepsilon}{}_{\delta}{}^{\mu} R_{\gamma}{}^{\zeta}{}_{\epsilon}{}^{\eta} R_{\varepsilon \zeta \mu \eta} + c_{36} H_{\alpha \beta}{}^{\delta} H^{\alpha \beta \gamma} R_{\gamma}{}^{\epsilon \varepsilon \mu} R_{\delta \epsilon}{}^{\zeta \eta} R_{\varepsilon \zeta \mu \eta} \nn\\&&+ c_{37} H_{\alpha}{}^{\delta \epsilon} H^{\alpha \beta \gamma} R_{\beta}{}^{\varepsilon}{}_{\gamma}{}^{\mu} R_{\delta}{}^{\zeta}{}_{\epsilon}{}^{\eta} R_{\varepsilon \zeta \mu \eta} + c_{38} H_{\alpha \beta \gamma} H^{\alpha \beta \gamma} R_{\delta \epsilon}{}^{\zeta \eta} R^{\delta \epsilon \varepsilon \mu} R_{\varepsilon \zeta \mu \eta} \nn\\&&+ c_{39} H_{\alpha}{}^{\delta \epsilon} H^{\alpha \beta \gamma} R_{\beta \delta \gamma}{}^{\varepsilon} R_{\epsilon}{}^{\mu \zeta \eta} R_{\varepsilon \zeta \mu \eta} + c_{40} H_{\alpha \beta}{}^{\delta} H^{\alpha \beta \gamma} R_{\gamma}{}^{\epsilon}{}_{\delta}{}^{\varepsilon} R_{\epsilon}{}^{\mu \zeta \eta} R_{\varepsilon \zeta \mu \eta} \nn\\&&+ c_{41} H^{\alpha \beta \gamma} H^{\delta \epsilon \varepsilon} R_{\alpha \beta}{}^{\mu \zeta} R_{\gamma \delta \epsilon}{}^{\eta} R_{\varepsilon \eta \mu \zeta} + c_{42} H^{\alpha \beta \gamma} H^{\delta \epsilon \varepsilon} R_{\alpha \delta \beta}{}^{\mu} R_{\gamma}{}^{\zeta}{}_{\epsilon}{}^{\eta} R_{\varepsilon \eta \mu \zeta} \nn\\&&+ c_{43} H_{\alpha}{}^{\delta \epsilon} H^{\alpha \beta \gamma} R_{\beta}{}^{\varepsilon}{}_{\delta}{}^{\mu} R_{\gamma}{}^{\zeta}{}_{\epsilon}{}^{\eta} R_{\varepsilon \eta \mu \zeta} + c_{57} H_{\alpha}{}^{\delta \epsilon} H^{\alpha \beta \gamma} R_{\beta \delta \gamma \epsilon} R_{\varepsilon \zeta \mu \eta} R^{\varepsilon \mu \zeta \eta}\labell{T4}
\eeqa
\beqa
{\cal L}_3^{H^4R^2}&\!\!\!\!\!=\!\!\!\!\! & c_{30} H_{\alpha}{}^{\delta \epsilon} H^{\alpha \beta \gamma} H_{\beta}{}^{\varepsilon \mu} H_{\delta}{}^{\zeta \eta} R_{\gamma}{}^{\theta}{}_{\varepsilon \zeta} R_{\epsilon \theta \mu \eta} + c_{44} H_{\alpha \beta}{}^{\delta} H^{\alpha \beta \gamma} H^{\epsilon \varepsilon \mu} H^{\zeta \eta \theta} R_{\gamma \epsilon \delta \zeta} R_{\varepsilon \eta \mu \theta} \nn\\&&+ c_{45} H_{\alpha}{}^{\delta \epsilon} H^{\alpha \beta \gamma} H_{\beta}{}^{\varepsilon \mu} H^{\zeta \eta \theta} R_{\gamma \zeta \delta \epsilon} R_{\varepsilon \eta \mu \theta} + c_{46} H_{\alpha \beta}{}^{\delta} H^{\alpha \beta \gamma} H_{\gamma}{}^{\epsilon \varepsilon} H_{\epsilon}{}^{\mu \zeta} R_{\delta}{}^{\eta}{}_{\mu}{}^{\theta} R_{\varepsilon \theta \zeta \eta} \nn\\&&+ c_{47} H_{\alpha \beta}{}^{\delta} H^{\alpha \beta \gamma} H_{\gamma}{}^{\epsilon \varepsilon} H^{\mu \zeta \eta} R_{\delta}{}^{\theta}{}_{\epsilon \mu} R_{\varepsilon \theta \zeta \eta} + c_{48} H_{\alpha}{}^{\delta \epsilon} H^{\alpha \beta \gamma} H_{\beta \delta}{}^{\varepsilon} H_{\gamma}{}^{\mu \zeta} R_{\epsilon}{}^{\eta}{}_{\mu}{}^{\theta} R_{\varepsilon \theta \zeta \eta}\nn\\&& + c_{49} H_{\alpha \beta}{}^{\delta} H^{\alpha \beta \gamma} H_{\gamma}{}^{\epsilon \varepsilon} H_{\delta}{}^{\mu \zeta} R_{\epsilon}{}^{\eta}{}_{\mu}{}^{\theta} R_{\varepsilon \theta \zeta \eta} + c_{58} H_{\alpha}{}^{\delta \epsilon} H^{\alpha \beta \gamma} H_{\varepsilon}{}^{\eta \theta} H^{\varepsilon \mu \zeta} R_{\beta \delta \gamma \epsilon} R_{\mu \eta \zeta \theta}\nn\\&& + c_{59} H_{\alpha \beta}{}^{\delta} H^{\alpha \beta \gamma} H_{\epsilon \varepsilon}{}^{\zeta} H^{\epsilon \varepsilon \mu} R_{\gamma}{}^{\eta}{}_{\delta}{}^{\theta} R_{\mu \eta \zeta \theta} + c_{60} H_{\alpha}{}^{\delta \epsilon} H^{\alpha \beta \gamma} H_{\beta}{}^{\varepsilon \mu} H_{\delta \varepsilon}{}^{\zeta} R_{\gamma}{}^{\eta}{}_{\epsilon}{}^{\theta} R_{\mu \eta \zeta \theta}\nn\\&& + c_{61} H_{\alpha \beta}{}^{\delta} H^{\alpha \beta \gamma} H_{\gamma}{}^{\epsilon \varepsilon} H_{\epsilon \varepsilon}{}^{\mu} R_{\delta}{}^{\zeta \eta \theta} R_{\mu \eta \zeta \theta} + c_{62} H_{\alpha \beta}{}^{\delta} H^{\alpha \beta \gamma} H_{\gamma}{}^{\epsilon \varepsilon} H_{\epsilon}{}^{\mu \zeta} R_{\delta}{}^{\eta}{}_{\varepsilon}{}^{\theta} R_{\mu \eta \zeta \theta}\nn\\&& + c_{63} H_{\alpha}{}^{\delta \epsilon} H^{\alpha \beta \gamma} H_{\beta \delta}{}^{\varepsilon} H_{\gamma}{}^{\mu \zeta} R_{\epsilon}{}^{\eta}{}_{\varepsilon}{}^{\theta} R_{\mu \eta \zeta \theta} + c_{64} H_{\alpha \beta}{}^{\delta} H^{\alpha \beta \gamma} H_{\gamma}{}^{\epsilon \varepsilon} H_{\delta}{}^{\mu \zeta} R_{\epsilon}{}^{\eta}{}_{\varepsilon}{}^{\theta} R_{\mu \eta \zeta \theta} \nn\\&&+ c_{65} H_{\alpha}{}^{\delta \epsilon} H^{\alpha \beta \gamma} H_{\beta \delta}{}^{\varepsilon} H_{\gamma \epsilon}{}^{\mu} R_{\varepsilon}{}^{\zeta \eta \theta} R_{\mu \eta \zeta \theta} + c_{66} H_{\alpha \beta}{}^{\delta} H^{\alpha \beta \gamma} H_{\gamma}{}^{\epsilon \varepsilon} H_{\delta \epsilon}{}^{\mu} R_{\varepsilon}{}^{\zeta \eta \theta} R_{\mu \eta \zeta \theta} \nn\\&&+ c_{67} H_{\alpha}{}^{\delta \epsilon} H^{\alpha \beta \gamma} H_{\beta}{}^{\varepsilon \mu} H_{\delta}{}^{\zeta \eta} R_{\gamma \epsilon \varepsilon}{}^{\theta} R_{\mu \theta \zeta \eta} + c_{68} H_{\alpha \beta}{}^{\delta} H^{\alpha \beta \gamma} H_{\epsilon}{}^{\zeta \eta} H^{\epsilon \varepsilon \mu} R_{\gamma \varepsilon \delta}{}^{\theta} R_{\mu \theta \zeta \eta} \nn\\&&+ c_{69} H_{\alpha}{}^{\delta \epsilon} H^{\alpha \beta \gamma} H_{\beta}{}^{\varepsilon \mu} H_{\delta}{}^{\zeta \eta} R_{\gamma \varepsilon \epsilon}{}^{\theta} R_{\mu \theta \zeta \eta} + c_{70} H_{\alpha}{}^{\delta \epsilon} H^{\alpha \beta \gamma} H_{\beta}{}^{\varepsilon \mu} H_{\delta \varepsilon}{}^{\zeta} R_{\gamma}{}^{\eta}{}_{\epsilon}{}^{\theta} R_{\mu \theta \zeta \eta} \nn\\&&+ c_{71} H_{\alpha}{}^{\delta \epsilon} H^{\alpha \beta \gamma} H_{\beta}{}^{\varepsilon \mu} H_{\gamma}{}^{\zeta \eta} R_{\delta \epsilon \varepsilon}{}^{\theta} R_{\mu \theta \zeta \eta} + c_{78} H_{\alpha}{}^{\delta \epsilon} H^{\alpha \beta \gamma} H_{\beta \delta}{}^{\varepsilon} H_{\gamma \epsilon \varepsilon} R_{\mu \eta \zeta \theta} R^{\mu \zeta \eta \theta} \nn\\&&+ c_{79} H_{\alpha \beta}{}^{\delta} H^{\alpha \beta \gamma} H_{\gamma}{}^{\epsilon \varepsilon} H_{\delta \epsilon \varepsilon} R_{\mu \eta \zeta \theta} R^{\mu \zeta \eta \theta}\labell{T5}
\eeqa
\beqa
&&{\cal L}_3^{H^6(\prt\Phi)^2}=\nn\\&& c_{87} H_{\beta}{}^{\epsilon \varepsilon} H^{\beta \gamma \delta} H_{\gamma \epsilon}{}^{\mu} H_{\delta}{}^{\zeta \eta} H_{\varepsilon \zeta}{}^{\theta} H_{\mu \eta \theta} \nabla_{\alpha}\Phi \nabla^{\alpha}\Phi + c_{88} H_{\beta \gamma}{}^{\epsilon} H^{\beta \gamma \delta} H_{\delta}{}^{\varepsilon \mu} H_{\epsilon}{}^{\zeta \eta} H_{\varepsilon \zeta}{}^{\theta} H_{\mu \eta \theta} \nabla_{\alpha}\Phi \nabla^{\alpha}\Phi \nn\\&&+ c_{89} H_{\beta \gamma \delta} H^{\beta \gamma \delta} H_{\epsilon}{}^{\zeta \eta} H^{\epsilon \varepsilon \mu} H_{\varepsilon \zeta}{}^{\theta} H_{\mu \eta \theta} \nabla_{\alpha}\Phi \nabla^{\alpha}\Phi + c_{90} H_{\beta \gamma}{}^{\epsilon} H^{\beta \gamma \delta} H_{\delta}{}^{\varepsilon \mu} H_{\epsilon}{}^{\zeta \eta} H_{\varepsilon \mu}{}^{\theta} H_{\zeta \eta \theta} \nabla_{\alpha}\Phi \nabla^{\alpha}\Phi\nn\\&& + c_{91} H_{\beta \gamma}{}^{\epsilon} H^{\beta \gamma \delta} H_{\delta}{}^{\varepsilon \mu} H_{\epsilon \varepsilon}{}^{\zeta} H_{\mu}{}^{\eta \theta} H_{\zeta \eta \theta} \nabla_{\alpha}\Phi \nabla^{\alpha}\Phi + c_{92} H_{\beta \gamma \delta} H^{\beta \gamma \delta} H_{\epsilon \varepsilon}{}^{\zeta} H^{\epsilon \varepsilon \mu} H_{\mu}{}^{\eta \theta} H_{\zeta \eta \theta} \nabla_{\alpha}\Phi \nabla^{\alpha}\Phi \nn\\&&+ c_{93} H_{\beta \gamma \delta} H^{\beta \gamma \delta} H_{\epsilon \varepsilon \mu} H^{\epsilon \varepsilon \mu} H_{\zeta \eta \theta} H^{\zeta \eta \theta} \nabla_{\alpha}\Phi \nabla^{\alpha}\Phi + c_{108} H_{\alpha}{}^{\gamma \delta} H_{\beta}{}^{\epsilon \varepsilon} H_{\gamma}{}^{\mu \zeta} H_{\delta}{}^{\eta \theta} H_{\epsilon \mu \zeta} H_{\varepsilon \eta \theta} \nabla^{\alpha}\Phi \nabla^{\beta}\Phi\nn\\&& + c_{109} H_{\alpha}{}^{\gamma \delta} H_{\beta}{}^{\epsilon \varepsilon} H_{\gamma}{}^{\mu \zeta} H_{\delta \mu}{}^{\eta} H_{\epsilon \zeta}{}^{\theta} H_{\varepsilon \eta \theta} \nabla^{\alpha}\Phi \nabla^{\beta}\Phi + c_{110} H_{\alpha}{}^{\gamma \delta} H_{\beta}{}^{\epsilon \varepsilon} H_{\gamma \epsilon}{}^{\mu} H_{\delta}{}^{\zeta \eta} H_{\varepsilon \zeta}{}^{\theta} H_{\mu \eta \theta} \nabla^{\alpha}\Phi \nabla^{\beta}\Phi\nn\\&& + c_{111} H_{\alpha}{}^{\gamma \delta} H_{\beta}{}^{\epsilon \varepsilon} H_{\gamma \delta}{}^{\mu} H_{\epsilon}{}^{\zeta \eta} H_{\varepsilon \zeta}{}^{\theta} H_{\mu \eta \theta} \nabla^{\alpha}\Phi \nabla^{\beta}\Phi + c_{112} H_{\alpha}{}^{\gamma \delta} H_{\beta \gamma}{}^{\epsilon} H_{\delta}{}^{\varepsilon \mu} H_{\epsilon}{}^{\zeta \eta} H_{\varepsilon \zeta}{}^{\theta} H_{\mu \eta \theta} \nabla^{\alpha}\Phi \nabla^{\beta}\Phi\nn\\&& + c_{113} H_{\alpha}{}^{\gamma \delta} H_{\beta \gamma \delta} H_{\epsilon}{}^{\zeta \eta} H^{\epsilon \varepsilon \mu} H_{\varepsilon \zeta}{}^{\theta} H_{\mu \eta \theta} \nabla^{\alpha}\Phi \nabla^{\beta}\Phi + c_{114} H_{\alpha}{}^{\gamma \delta} H_{\beta \gamma}{}^{\epsilon} H_{\delta}{}^{\varepsilon \mu} H_{\epsilon}{}^{\zeta \eta} H_{\varepsilon \mu}{}^{\theta} H_{\zeta \eta \theta} \nabla^{\alpha}\Phi \nabla^{\beta}\Phi\nn\\&& + c_{115} H_{\alpha}{}^{\gamma \delta} H_{\beta}{}^{\epsilon \varepsilon} H_{\gamma \epsilon}{}^{\mu} H_{\delta \mu}{}^{\zeta} H_{\varepsilon}{}^{\eta \theta} H_{\zeta \eta \theta} \nabla^{\alpha}\Phi \nabla^{\beta}\Phi + c_{116} H_{\alpha}{}^{\gamma \delta} H_{\beta}{}^{\epsilon \varepsilon} H_{\gamma \delta}{}^{\mu} H_{\epsilon \mu}{}^{\zeta} H_{\varepsilon}{}^{\eta \theta} H_{\zeta \eta \theta} \nabla^{\alpha}\Phi \nabla^{\beta}\Phi \nn\\&&+ c_{117} H_{\alpha}{}^{\gamma \delta} H_{\beta}{}^{\epsilon \varepsilon} H_{\gamma \epsilon}{}^{\mu} H_{\delta \varepsilon}{}^{\zeta} H_{\mu}{}^{\eta \theta} H_{\zeta \eta \theta} \nabla^{\alpha}\Phi \nabla^{\beta}\Phi + c_{118} H_{\alpha}{}^{\gamma \delta} H_{\beta}{}^{\epsilon \varepsilon} H_{\gamma \delta}{}^{\mu} H_{\epsilon \varepsilon}{}^{\zeta} H_{\mu}{}^{\eta \theta} H_{\zeta \eta \theta} \nabla^{\alpha}\Phi \nabla^{\beta}\Phi \nn\\&&+ c_{119} H_{\alpha}{}^{\gamma \delta} H_{\beta \gamma}{}^{\epsilon} H_{\delta}{}^{\varepsilon \mu} H_{\epsilon \varepsilon}{}^{\zeta} H_{\mu}{}^{\eta \theta} H_{\zeta \eta \theta} \nabla^{\alpha}\Phi \nabla^{\beta}\Phi + c_{120} H_{\alpha}{}^{\gamma \delta} H_{\beta \gamma \delta} H_{\epsilon \varepsilon}{}^{\zeta} H^{\epsilon \varepsilon \mu} H_{\mu}{}^{\eta \theta} H_{\zeta \eta \theta} \nabla^{\alpha}\Phi \nabla^{\beta}\Phi\nn\\&& + c_{121} H_{\alpha}{}^{\gamma \delta} H_{\beta}{}^{\epsilon \varepsilon} H_{\gamma \epsilon}{}^{\mu} H_{\delta \varepsilon \mu} H_{\zeta \eta \theta} H^{\zeta \eta \theta} \nabla^{\alpha}\Phi \nabla^{\beta}\Phi + c_{122} H_{\alpha}{}^{\gamma \delta} H_{\beta}{}^{\epsilon \varepsilon} H_{\gamma \delta}{}^{\mu} H_{\epsilon \varepsilon \mu} H_{\zeta \eta \theta} H^{\zeta \eta \theta} \nabla^{\alpha}\Phi \nabla^{\beta}\Phi\nn\\&& + c_{123} H_{\alpha}{}^{\gamma \delta} H_{\beta \gamma}{}^{\epsilon} H_{\delta}{}^{\varepsilon \mu} H_{\epsilon \varepsilon \mu} H_{\zeta \eta \theta} H^{\zeta \eta \theta} \nabla^{\alpha}\Phi \nabla^{\beta}\Phi + c_{124} H_{\alpha}{}^{\gamma \delta} H_{\beta \gamma \delta} H_{\epsilon \varepsilon \mu} H^{\epsilon \varepsilon \mu} H_{\zeta \eta \theta} H^{\zeta \eta \theta} \nabla^{\alpha}\Phi \nabla^{\beta}\Phi\labell{T6}\nn
\eeqa
\beqa
{\cal L}_3^{R^3(\prt\Phi)^2}&\!\!\!\!\!=\!\!\!\!\! & c_{94} R_{\beta}{}^{\varepsilon}{}_{\delta}{}^{\mu} R^{\beta \gamma \delta \epsilon} R_{\gamma \mu \epsilon \varepsilon} \nabla_{\alpha}\Phi \nabla^{\alpha}\Phi + c_{95} R_{\beta \gamma}{}^{\varepsilon \mu} R^{\beta \gamma \delta \epsilon} R_{\delta \varepsilon \epsilon \mu} \nabla_{\alpha}\Phi \nabla^{\alpha}\Phi\nn\\&&  + c_{148} R_{\alpha}{}^{\gamma \delta \epsilon} R_{\beta}{}^{\varepsilon}{}_{\delta}{}^{\mu} R_{\gamma \mu \epsilon \varepsilon} \nabla^{\alpha}\Phi \nabla^{\beta}\Phi + c_{149} R_{\alpha}{}^{\gamma \delta \epsilon} R_{\beta \gamma}{}^{\varepsilon \mu} R_{\delta \varepsilon \epsilon \mu} \nabla^{\alpha}\Phi \nabla^{\beta}\Phi\nn\\&&  + c_{150} R_{\alpha}{}^{\gamma}{}_{\beta}{}^{\delta} R_{\gamma}{}^{\epsilon \varepsilon \mu} R_{\delta \varepsilon \epsilon \mu} \nabla^{\alpha}\Phi \nabla^{\beta}\Phi\labell{T7}
\eeqa
\beqa
{\cal L}_3^{H^4R(\prt\Phi)^2}&\!\!\!\!\!=\!\!\!\!\! & c_{96} H_{\beta}{}^{\epsilon \varepsilon} H^{\beta \gamma \delta} H_{\gamma}{}^{\mu \zeta} H_{\epsilon \mu}{}^{\eta} R_{\delta \varepsilon \zeta \eta} \nabla_{\alpha}\Phi \nabla^{\alpha}\Phi + c_{100} H_{\beta \gamma}{}^{\epsilon} H^{\beta \gamma \delta} H_{\varepsilon \mu}{}^{\eta} H^{\varepsilon \mu \zeta} R_{\delta \zeta \epsilon \eta} \nabla_{\alpha}\Phi \nabla^{\alpha}\Phi\nn\\&& + c_{103} H_{\beta \gamma}{}^{\epsilon} H^{\beta \gamma \delta} H_{\delta}{}^{\varepsilon \mu} H_{\varepsilon}{}^{\zeta \eta} R_{\epsilon \zeta \mu \eta} \nabla_{\alpha}\Phi \nabla^{\alpha}\Phi + c_{105} H_{\beta}{}^{\epsilon \varepsilon} H^{\beta \gamma \delta} H_{\gamma \epsilon}{}^{\mu} H_{\delta}{}^{\zeta \eta} R_{\varepsilon \zeta \mu \eta} \nabla_{\alpha}\Phi \nabla^{\alpha}\Phi \nn\\&&+ c_{106} H_{\beta \gamma}{}^{\epsilon} H^{\beta \gamma \delta} H_{\delta}{}^{\varepsilon \mu} H_{\epsilon}{}^{\zeta \eta} R_{\varepsilon \zeta \mu \eta} \nabla_{\alpha}\Phi \nabla^{\alpha}\Phi + c_{107} H_{\beta \gamma \delta} H^{\beta \gamma \delta} H_{\epsilon}{}^{\zeta \eta} H^{\epsilon \varepsilon \mu} R_{\varepsilon \zeta \mu \eta} \nabla_{\alpha}\Phi \nabla^{\alpha}\Phi\nn\\&& + c_{125} H_{\gamma \delta}{}^{\varepsilon} H^{\gamma \delta \epsilon} H_{\epsilon}{}^{\mu \zeta} H_{\mu \zeta}{}^{\eta} R_{\alpha \varepsilon \beta \eta} \nabla^{\alpha}\Phi \nabla^{\beta}\Phi + c_{126} H_{\gamma}{}^{\varepsilon \mu} H^{\gamma \delta \epsilon} H_{\delta \varepsilon}{}^{\zeta} H_{\epsilon \mu}{}^{\eta} R_{\alpha \zeta \beta \eta} \nabla^{\alpha}\Phi \nabla^{\beta}\Phi \nn\\&&+ c_{127} H_{\gamma \delta}{}^{\varepsilon} H^{\gamma \delta \epsilon} H_{\epsilon}{}^{\mu \zeta} H_{\varepsilon \mu}{}^{\eta} R_{\alpha \zeta \beta \eta} \nabla^{\alpha}\Phi \nabla^{\beta}\Phi + c_{128} H_{\gamma \delta \epsilon} H^{\gamma \delta \epsilon} H_{\varepsilon \mu}{}^{\eta} H^{\varepsilon \mu \zeta} R_{\alpha \zeta \beta \eta} \nabla^{\alpha}\Phi \nabla^{\beta}\Phi \nn\\&&+ c_{129} H_{\alpha}{}^{\gamma \delta} H_{\gamma}{}^{\epsilon \varepsilon} H_{\mu \zeta \eta} H^{\mu \zeta \eta} R_{\beta \epsilon \delta \varepsilon} \nabla^{\alpha}\Phi \nabla^{\beta}\Phi + c_{130} H_{\alpha}{}^{\gamma \delta} H_{\gamma}{}^{\epsilon \varepsilon} H_{\epsilon}{}^{\mu \zeta} H_{\mu \zeta}{}^{\eta} R_{\beta \varepsilon \delta \eta} \nabla^{\alpha}\Phi \nabla^{\beta}\Phi\nn\\&& + c_{131} H_{\alpha}{}^{\gamma \delta} H_{\gamma}{}^{\epsilon \varepsilon} H_{\delta}{}^{\mu \zeta} H_{\epsilon \mu}{}^{\eta} R_{\beta \varepsilon \zeta \eta} \nabla^{\alpha}\Phi \nabla^{\beta}\Phi + c_{139} H_{\alpha}{}^{\gamma \delta} H_{\gamma}{}^{\epsilon \varepsilon} H_{\epsilon}{}^{\mu \zeta} H_{\varepsilon \mu}{}^{\eta} R_{\beta \zeta \delta \eta} \nabla^{\alpha}\Phi \nabla^{\beta}\Phi\nn\\&& + c_{140} H_{\alpha}{}^{\gamma \delta} H_{\gamma}{}^{\epsilon \varepsilon} H_{\epsilon \varepsilon}{}^{\mu} H_{\mu}{}^{\zeta \eta} R_{\beta \zeta \delta \eta} \nabla^{\alpha}\Phi \nabla^{\beta}\Phi + c_{141} H_{\alpha}{}^{\gamma \delta} H_{\gamma \delta}{}^{\epsilon} H_{\varepsilon \mu}{}^{\eta} H^{\varepsilon \mu \zeta} R_{\beta \zeta \epsilon \eta} \nabla^{\alpha}\Phi \nabla^{\beta}\Phi \nn\\&&+ c_{142} H_{\alpha}{}^{\gamma \delta} H_{\gamma}{}^{\epsilon \varepsilon} H_{\delta \epsilon}{}^{\mu} H_{\varepsilon}{}^{\zeta \eta} R_{\beta \zeta \mu \eta} \nabla^{\alpha}\Phi \nabla^{\beta}\Phi + c_{143} H_{\alpha}{}^{\gamma \delta} H_{\gamma \delta}{}^{\epsilon} H_{\epsilon}{}^{\varepsilon \mu} H_{\varepsilon}{}^{\zeta \eta} R_{\beta \zeta \mu \eta} \nabla^{\alpha}\Phi \nabla^{\beta}\Phi \nn\\&&+ c_{144} H_{\alpha}{}^{\gamma \delta} H_{\gamma}{}^{\epsilon \varepsilon} H_{\epsilon}{}^{\mu \zeta} H_{\mu \zeta}{}^{\eta} R_{\beta \eta \delta \varepsilon} \nabla^{\alpha}\Phi \nabla^{\beta}\Phi + c_{145} H_{\alpha}{}^{\gamma \delta} H_{\gamma}{}^{\epsilon \varepsilon} H_{\delta}{}^{\mu \zeta} H_{\epsilon \varepsilon}{}^{\eta} R_{\beta \eta \mu \zeta} \nabla^{\alpha}\Phi \nabla^{\beta}\Phi\nn\\&& + c_{146} H_{\alpha}{}^{\gamma \delta} H_{\beta}{}^{\epsilon \varepsilon} H_{\mu \zeta \eta} H^{\mu \zeta \eta} R_{\gamma \epsilon \delta \varepsilon} \nabla^{\alpha}\Phi \nabla^{\beta}\Phi + c_{158} H_{\alpha}{}^{\gamma \delta} H_{\beta \gamma}{}^{\epsilon} H_{\varepsilon \mu}{}^{\eta} H^{\varepsilon \mu \zeta} R_{\delta \zeta \epsilon \eta} \nabla^{\alpha}\Phi \nabla^{\beta}\Phi \nn\\&&+ c_{159} H_{\alpha}{}^{\gamma \delta} H_{\beta}{}^{\epsilon \varepsilon} H_{\gamma}{}^{\mu \zeta} H_{\epsilon \mu}{}^{\eta} R_{\delta \zeta \varepsilon \eta} \nabla^{\alpha}\Phi \nabla^{\beta}\Phi + c_{160} H_{\alpha}{}^{\gamma \delta} H_{\beta}{}^{\epsilon \varepsilon} H_{\gamma \epsilon}{}^{\mu} H_{\mu}{}^{\zeta \eta} R_{\delta \zeta \varepsilon \eta} \nabla^{\alpha}\Phi \nabla^{\beta}\Phi \nn\\&&+ c_{161} H_{\alpha}{}^{\gamma \delta} H_{\beta}{}^{\epsilon \varepsilon} H_{\gamma}{}^{\mu \zeta} H_{\mu \zeta}{}^{\eta} R_{\delta \eta \epsilon \varepsilon} \nabla^{\alpha}\Phi \nabla^{\beta}\Phi + c_{162} H_{\alpha}{}^{\gamma \delta} H_{\beta}{}^{\epsilon \varepsilon} H_{\gamma}{}^{\mu \zeta} H_{\epsilon \mu}{}^{\eta} R_{\delta \eta \varepsilon \zeta} \nabla^{\alpha}\Phi \nabla^{\beta}\Phi\nn\\&& + c_{169} H_{\alpha}{}^{\gamma \delta} H_{\beta}{}^{\epsilon \varepsilon} H_{\gamma}{}^{\mu \zeta} H_{\delta \mu}{}^{\eta} R_{\epsilon \zeta \varepsilon \eta} \nabla^{\alpha}\Phi \nabla^{\beta}\Phi + c_{170} H_{\alpha}{}^{\gamma \delta} H_{\beta}{}^{\epsilon \varepsilon} H_{\gamma \delta}{}^{\mu} H_{\mu}{}^{\zeta \eta} R_{\epsilon \zeta \varepsilon \eta} \nabla^{\alpha}\Phi \nabla^{\beta}\Phi \nn\\&&+ c_{171} H_{\alpha}{}^{\gamma \delta} H_{\beta \gamma}{}^{\epsilon} H_{\delta}{}^{\varepsilon \mu} H_{\varepsilon}{}^{\zeta \eta} R_{\epsilon \zeta \mu \eta} \nabla^{\alpha}\Phi \nabla^{\beta}\Phi + c_{173} H_{\alpha}{}^{\gamma \delta} H_{\beta}{}^{\epsilon \varepsilon} H_{\gamma \epsilon}{}^{\mu} H_{\delta}{}^{\zeta \eta} R_{\varepsilon \zeta \mu \eta} \nabla^{\alpha}\Phi \nabla^{\beta}\Phi \nn\\&&+ c_{174} H_{\alpha}{}^{\gamma \delta} H_{\beta}{}^{\epsilon \varepsilon} H_{\gamma \delta}{}^{\mu} H_{\epsilon}{}^{\zeta \eta} R_{\varepsilon \zeta \mu \eta} \nabla^{\alpha}\Phi \nabla^{\beta}\Phi + c_{175} H_{\alpha}{}^{\gamma \delta} H_{\beta \gamma}{}^{\epsilon} H_{\delta}{}^{\varepsilon \mu} H_{\epsilon}{}^{\zeta \eta} R_{\varepsilon \zeta \mu \eta} \nabla^{\alpha}\Phi \nabla^{\beta}\Phi \nn\\&&+ c_{176} H_{\alpha}{}^{\gamma \delta} H_{\beta \gamma \delta} H_{\epsilon}{}^{\zeta \eta} H^{\epsilon \varepsilon \mu} R_{\varepsilon \zeta \mu \eta} \nabla^{\alpha}\Phi \nabla^{\beta}\Phi\labell{T8}
\eeqa
\beqa
{\cal L}_3^{H^2R^2(\prt\Phi)^2}&\!\!\!\!\!=\!\!\!\!\! & c_{97} H_{\beta}{}^{\epsilon \varepsilon} H^{\beta \gamma \delta} R_{\gamma}{}^{\mu}{}_{\epsilon}{}^{\zeta} R_{\delta\mu \varepsilon \zeta} \nabla_{\alpha}\Phi \nabla^{\alpha}\Phi + c_{98} H^{\beta \gamma \delta} H^{\epsilon \varepsilon\mu} R_{\beta \gamma \epsilon}{}^{\zeta} R_{\delta \zeta \varepsilon\mu} \nabla_{\alpha}\Phi \nabla^{\alpha}\Phi\nn\\&& + c_{99} H_{\beta}{}^{\epsilon \varepsilon} H^{\beta \gamma \delta} R_{\gamma}{}^{\mu}{}_{\epsilon}{}^{\zeta} R_{\delta \zeta \varepsilon\mu} \nabla_{\alpha}\Phi \nabla^{\alpha}\Phi + c_{101} H_{\beta}{}^{\epsilon \varepsilon} H^{\beta \gamma \delta} R_{\gamma}{}^{\mu}{}_{\delta}{}^{\zeta} R_{\epsilon\mu \varepsilon \zeta} \nabla_{\alpha}\Phi \nabla^{\alpha}\Phi\nn\\&& + c_{102} H_{\beta \gamma}{}^{\epsilon} H^{\beta \gamma \delta} R_{\delta}{}^{\varepsilon\mu \zeta} R_{\epsilon\mu \varepsilon \zeta} \nabla_{\alpha}\Phi \nabla^{\alpha}\Phi + c_{104} H_{\beta \gamma \delta} H^{\beta \gamma \delta} R_{\epsilon\mu \varepsilon \zeta} R^{\epsilon \varepsilon\mu \zeta} \nabla_{\alpha}\Phi \nabla^{\alpha}\Phi \nn\\&&+ c_{132} H_{\gamma}{}^{\varepsilon\mu} H^{\gamma \delta \epsilon} R_{\alpha \delta \varepsilon}{}^{\zeta} R_{\beta\mu \epsilon \zeta} \nabla^{\alpha}\Phi \nabla^{\beta}\Phi + c_{133} H_{\gamma \delta \epsilon} H^{\gamma \delta \epsilon} R_{\alpha}{}^{\varepsilon\mu \zeta} R_{\beta\mu \varepsilon \zeta} \nabla^{\alpha}\Phi \nabla^{\beta}\Phi\nn\\&& + c_{134} H_{\gamma \delta}{}^{\varepsilon} H^{\gamma \delta \epsilon} R_{\alpha}{}^{\mu}{}_{\epsilon}{}^{\zeta} R_{\beta\mu \varepsilon \zeta} \nabla^{\alpha}\Phi \nabla^{\beta}\Phi + c_{135} H^{\gamma \delta \epsilon} H^{\varepsilon\mu \zeta} R_{\alpha \gamma \varepsilon\mu} R_{\beta \zeta \delta \epsilon} \nabla^{\alpha}\Phi \nabla^{\beta}\Phi \nn\\&&+ c_{136} H_{\gamma}{}^{\varepsilon\mu} H^{\gamma \delta \epsilon} R_{\alpha}{}^{\zeta}{}_{\delta \varepsilon} R_{\beta \zeta \epsilon\mu} \nabla^{\alpha}\Phi \nabla^{\beta}\Phi + c_{137} H_{\gamma \delta}{}^{\varepsilon} H^{\gamma \delta \epsilon} R_{\alpha}{}^{\mu}{}_{\epsilon}{}^{\zeta} R_{\beta \zeta \varepsilon\mu} \nabla^{\alpha}\Phi \nabla^{\beta}\Phi \nn\\&&+ c_{138} H_{\gamma}{}^{\varepsilon\mu} H^{\gamma \delta \epsilon} R_{\alpha}{}^{\zeta}{}_{\delta \epsilon} R_{\beta \zeta \varepsilon\mu} \nabla^{\alpha}\Phi \nabla^{\beta}\Phi + c_{147} H_{\alpha}{}^{\gamma \delta} H^{\epsilon \varepsilon\mu} R_{\beta}{}^{\zeta}{}_{\epsilon \varepsilon} R_{\gamma\mu \delta \zeta} \nabla^{\alpha}\Phi \nabla^{\beta}\Phi \nn\\&&+ c_{151} H^{\gamma \delta \epsilon} H^{\varepsilon\mu \zeta} R_{\alpha \gamma \beta \varepsilon} R_{\delta\mu \epsilon \zeta} \nabla^{\alpha}\Phi \nabla^{\beta}\Phi + c_{152} H_{\alpha}{}^{\gamma \delta} H_{\gamma}{}^{\epsilon \varepsilon} R_{\beta}{}^{\mu}{}_{\epsilon}{}^{\zeta} R_{\delta\mu \varepsilon \zeta} \nabla^{\alpha}\Phi \nabla^{\beta}\Phi \nn\\&&+ c_{153} H_{\alpha}{}^{\gamma \delta} H_{\beta}{}^{\epsilon \varepsilon} R_{\gamma}{}^{\mu}{}_{\epsilon}{}^{\zeta} R_{\delta\mu \varepsilon \zeta} \nabla^{\alpha}\Phi \nabla^{\beta}\Phi + c_{154} H_{\alpha}{}^{\gamma \delta} H^{\epsilon \varepsilon\mu} R_{\beta \epsilon \gamma}{}^{\zeta} R_{\delta \zeta \varepsilon\mu} \nabla^{\alpha}\Phi \nabla^{\beta}\Phi \nn\\&&+ c_{155} H_{\alpha}{}^{\gamma \delta} H_{\gamma}{}^{\epsilon \varepsilon} R_{\beta}{}^{\mu}{}_{\epsilon}{}^{\zeta} R_{\delta \zeta \varepsilon\mu} \nabla^{\alpha}\Phi \nabla^{\beta}\Phi + c_{156} H_{\alpha}{}^{\gamma \delta} H^{\epsilon \varepsilon\mu} R_{\beta}{}^{\zeta}{}_{\gamma \epsilon} R_{\delta \zeta \varepsilon\mu} \nabla^{\alpha}\Phi \nabla^{\beta}\Phi \nn\\&&+ c_{157} H_{\alpha}{}^{\gamma \delta} H_{\beta}{}^{\epsilon \varepsilon} R_{\gamma}{}^{\mu}{}_{\epsilon}{}^{\zeta} R_{\delta \zeta \varepsilon\mu} \nabla^{\alpha}\Phi \nabla^{\beta}\Phi + c_{163} H_{\gamma \delta}{}^{\varepsilon} H^{\gamma \delta \epsilon} R_{\alpha}{}^{\mu}{}_{\beta}{}^{\zeta} R_{\epsilon\mu \varepsilon \zeta} \nabla^{\alpha}\Phi \nabla^{\beta}\Phi\nn\\&& + c_{164} H_{\alpha}{}^{\gamma \delta} H_{\gamma \delta}{}^{\epsilon} R_{\beta}{}^{\varepsilon\mu \zeta} R_{\epsilon\mu \varepsilon \zeta} \nabla^{\alpha}\Phi \nabla^{\beta}\Phi + c_{165} H_{\alpha}{}^{\gamma \delta} H_{\gamma}{}^{\epsilon \varepsilon} R_{\beta}{}^{\mu}{}_{\delta}{}^{\zeta} R_{\epsilon\mu \varepsilon \zeta} \nabla^{\alpha}\Phi \nabla^{\beta}\Phi \nn\\&&+ c_{166} H_{\alpha}{}^{\gamma \delta} H_{\beta}{}^{\epsilon \varepsilon} R_{\gamma}{}^{\mu}{}_{\delta}{}^{\zeta} R_{\epsilon\mu \varepsilon \zeta} \nabla^{\alpha}\Phi \nabla^{\beta}\Phi + c_{167} H_{\alpha}{}^{\gamma \delta} H_{\beta \gamma}{}^{\epsilon} R_{\delta}{}^{\varepsilon\mu \zeta} R_{\epsilon\mu \varepsilon \zeta} \nabla^{\alpha}\Phi \nabla^{\beta}\Phi \nn\\&&+ c_{168} H_{\gamma}{}^{\varepsilon\mu} H^{\gamma \delta \epsilon} R_{\alpha \delta \beta}{}^{\zeta} R_{\epsilon \zeta \varepsilon\mu} \nabla^{\alpha}\Phi \nabla^{\beta}\Phi + c_{172} H_{\alpha}{}^{\gamma \delta} H_{\beta \gamma \delta} R_{\epsilon\mu \varepsilon \zeta} R^{\epsilon \varepsilon\mu \zeta} \nabla^{\alpha}\Phi \nabla^{\beta}\Phi\labell{T9}
\eeqa
\beqa
&&{\cal L}_3^{H^4(\prt\Phi)^4}=\nn\\&& c_{177} H_{\gamma}{}^{\varepsilon \mu} H^{\gamma \delta \epsilon} H_{\delta \varepsilon}{}^{\zeta} H_{\epsilon \mu \zeta} \nabla_{\alpha}\Phi \nabla^{\alpha}\Phi \nabla_{\beta}\Phi \nabla^{\beta}\Phi + c_{256} H_{\beta}{}^{\delta \epsilon} H_{\gamma}{}^{\varepsilon \mu} H_{\delta \varepsilon}{}^{\zeta} H_{\epsilon \mu \zeta} \nabla_{\alpha}\Phi \nabla^{\alpha}\Phi \nabla^{\beta}\Phi \nabla^{\gamma}\Phi \nn\\&& + c_{287} H_{\alpha}{}^{\epsilon \varepsilon} H_{\beta \epsilon}{}^{\mu} H_{\gamma \varepsilon}{}^{\zeta} H_{\delta \mu \zeta} \nabla^{\alpha}\Phi \nabla^{\beta}\Phi \nabla^{\gamma}\Phi \nabla^{\delta}\Phi + c_{288} H_{\alpha}{}^{\epsilon \varepsilon} H_{\beta \epsilon \varepsilon} H_{\gamma}{}^{\mu \zeta} H_{\delta \mu \zeta} \nabla^{\alpha}\Phi \nabla^{\beta}\Phi \nabla^{\gamma}\Phi \nabla^{\delta}\Phi\labell{T10}\nn
\eeqa
\beqa
{\cal L}_3^{R^2(\prt\Phi)^4}&\!\!\!\!\!=\!\!\!\!\! & c_{178} R_{\gamma \epsilon \delta \varepsilon} R^{\gamma \delta \epsilon \varepsilon} \nabla_{\alpha}\Phi \nabla^{\alpha}\Phi \nabla_{\beta}\Phi \nabla^{\beta}\Phi + c_{289} R_{\alpha}{}^{\epsilon}{}_{\beta}{}^{\varepsilon} R_{\gamma \epsilon \delta \varepsilon} \nabla^{\alpha}\Phi \nabla^{\beta}\Phi \nabla^{\gamma}\Phi \nabla^{\delta}\Phi\labell{T11}
\eeqa
\beqa
{\cal L}_3^{H^2R(\prt\Phi)^4}&\!\!\!\!\!=\!\!\!\!\!&  c_{179} H_{\gamma}{}^{\varepsilon\mu} H^{\gamma \delta \epsilon} R_{\delta \varepsilon \epsilon\mu} \nabla_{\alpha}\Phi \nabla^{\alpha}\Phi \nabla_{\beta}\Phi \nabla^{\beta}\Phi + c_{257} H_{\beta}{}^{\delta \epsilon} H_{\gamma}{}^{\varepsilon\mu} R_{\delta \varepsilon \epsilon\mu} \nabla_{\alpha}\Phi \nabla^{\alpha}\Phi \nabla^{\beta}\Phi \nabla^{\gamma}\Phi\labell{T12}\nn
\eeqa
\beqa
&&{\cal L}_3^{H^4(\prt\Phi)^2\prt\prt\Phi}= \nn\\&& c_{180} H_{\gamma}{}^{\varepsilon\mu} H^{\gamma \delta \epsilon} H_{\delta \varepsilon}{}^{\zeta} H_{\epsilon\mu \zeta} \nabla^{\alpha}\Phi \nabla_{\beta}\nabla_{\alpha}\Phi \nabla^{\beta}\Phi + c_{259} H_{\beta}{}^{\delta \epsilon} H_{\gamma}{}^{\varepsilon\mu} H_{\delta \varepsilon}{}^{\zeta} H_{\epsilon\mu \zeta} \nabla^{\alpha}\Phi \nabla^{\beta}\Phi \nabla^{\gamma}\nabla_{\alpha}\Phi\nn\\&& + c_{274} H_{\beta}{}^{\delta \epsilon} H_{\gamma}{}^{\varepsilon\mu} H_{\delta \varepsilon}{}^{\zeta} H_{\epsilon\mu \zeta} \nabla_{\alpha}\Phi \nabla^{\alpha}\Phi \nabla^{\gamma}\nabla^{\beta}\Phi + c_{275} H_{\beta}{}^{\delta \epsilon} H_{\gamma \delta}{}^{\varepsilon} H_{\epsilon}{}^{\mu \zeta} H_{\varepsilon\mu \zeta} \nabla_{\alpha}\Phi \nabla^{\alpha}\Phi \nabla^{\gamma}\nabla^{\beta}\Phi\nn\\&& + c_{309} H_{\alpha}{}^{\epsilon \varepsilon} H_{\beta}{}^{\mu \zeta} H_{\gamma \epsilon\mu} H_{\delta \varepsilon \zeta} \nabla^{\alpha}\Phi \nabla^{\beta}\Phi \nabla^{\delta}\nabla^{\gamma}\Phi + c_{310} H_{\alpha}{}^{\epsilon \varepsilon} H_{\beta}{}^{\mu \zeta} H_{\gamma \epsilon \varepsilon} H_{\delta\mu \zeta} \nabla^{\alpha}\Phi \nabla^{\beta}\Phi \nabla^{\delta}\nabla^{\gamma}\Phi\nn\\&& + c_{311} H_{\alpha}{}^{\epsilon \varepsilon} H_{\beta \epsilon}{}^{\mu} H_{\gamma \varepsilon}{}^{\zeta} H_{\delta\mu \zeta} \nabla^{\alpha}\Phi \nabla^{\beta}\Phi \nabla^{\delta}\nabla^{\gamma}\Phi + c_{312} H_{\alpha}{}^{\epsilon \varepsilon} H_{\beta \epsilon \varepsilon} H_{\gamma}{}^{\mu \zeta} H_{\delta\mu \zeta} \nabla^{\alpha}\Phi \nabla^{\beta}\Phi \nabla^{\delta}\nabla^{\gamma}\Phi \nn\\&&+ c_{313} H_{\alpha \gamma}{}^{\epsilon} H_{\beta}{}^{\varepsilon\mu} H_{\delta \varepsilon}{}^{\zeta} H_{\epsilon\mu \zeta} \nabla^{\alpha}\Phi \nabla^{\beta}\Phi \nabla^{\delta}\nabla^{\gamma}\Phi + c_{314} H_{\alpha \gamma}{}^{\epsilon} H_{\beta}{}^{\varepsilon\mu} H_{\delta \epsilon}{}^{\zeta} H_{\varepsilon\mu \zeta} \nabla^{\alpha}\Phi \nabla^{\beta}\Phi \nabla^{\delta}\nabla^{\gamma}\Phi \nn\\&&+ c_{315} H_{\alpha \gamma}{}^{\epsilon} H_{\beta \epsilon}{}^{\varepsilon} H_{\delta}{}^{\mu \zeta} H_{\varepsilon\mu \zeta} \nabla^{\alpha}\Phi \nabla^{\beta}\Phi \nabla^{\delta}\nabla^{\gamma}\Phi + c_{316} H_{\alpha \gamma}{}^{\epsilon} H_{\beta \delta}{}^{\varepsilon} H_{\epsilon}{}^{\mu \zeta} H_{\varepsilon\mu \zeta} \nabla^{\alpha}\Phi \nabla^{\beta}\Phi \nabla^{\delta}\nabla^{\gamma}\Phi \nn\\&&+ c_{317} H_{\alpha \gamma}{}^{\epsilon} H_{\beta \delta \epsilon} H_{\varepsilon\mu \zeta} H^{\varepsilon\mu \zeta} \nabla^{\alpha}\Phi \nabla^{\beta}\Phi \nabla^{\delta}\nabla^{\gamma}\Phi\labell{T13}
\eeqa
\beqa
{\cal L}_3^{R^2(\prt\Phi)^2\prt\prt\Phi}&\!\!\!\!\!=\!\!\!\!\! & c_{181} R_{\gamma \epsilon \delta \varepsilon} R^{\gamma \delta \epsilon \varepsilon} \nabla^{\alpha}\Phi \nabla_{\beta}\nabla_{\alpha}\Phi \nabla^{\beta}\Phi + c_{261} R_{\beta}{}^{\delta \epsilon \varepsilon} R_{\gamma \epsilon \delta \varepsilon} \nabla^{\alpha}\Phi \nabla^{\beta}\Phi \nabla^{\gamma}\nabla_{\alpha}\Phi \nn\\&&+ c_{277} R_{\beta}{}^{\delta \epsilon \varepsilon} R_{\gamma \epsilon \delta \varepsilon} \nabla_{\alpha}\Phi \nabla^{\alpha}\Phi \nabla^{\gamma}\nabla^{\beta}\Phi + c_{321} R_{\alpha}{}^{\epsilon}{}_{\gamma}{}^{\varepsilon} R_{\beta \epsilon \delta \varepsilon} \nabla^{\alpha}\Phi \nabla^{\beta}\Phi \nabla^{\delta}\nabla^{\gamma}\Phi \nn\\&&+ c_{322} R_{\alpha}{}^{\epsilon}{}_{\gamma}{}^{\varepsilon} R_{\beta \varepsilon \delta \epsilon} \nabla^{\alpha}\Phi \nabla^{\beta}\Phi \nabla^{\delta}\nabla^{\gamma}\Phi + c_{326} R_{\alpha}{}^{\epsilon}{}_{\beta}{}^{\varepsilon} R_{\gamma \epsilon \delta \varepsilon} \nabla^{\alpha}\Phi \nabla^{\beta}\Phi \nabla^{\delta}\nabla^{\gamma}\Phi\labell{T14}
\eeqa
\beqa
{\cal L}_3^{H^2R(\prt\Phi)^2\prt\prt\Phi}&\!\!\!\!\!=\!\!\!\!\! & c_{182} H_{\gamma}{}^{\varepsilon\mu} H^{\gamma \delta \epsilon} R_{\delta \varepsilon \epsilon\mu} \nabla^{\alpha}\Phi \nabla_{\beta}\nabla_{\alpha}\Phi \nabla^{\beta}\Phi + c_{260} H_{\delta \epsilon}{}^{\mu} H^{\delta \epsilon \varepsilon} R_{\beta \varepsilon \gamma\mu} \nabla^{\alpha}\Phi \nabla^{\beta}\Phi \nabla^{\gamma}\nabla_{\alpha}\Phi \nn\\&&+ c_{262} H_{\beta}{}^{\delta \epsilon} H_{\delta}{}^{\varepsilon\mu} R_{\gamma \varepsilon \epsilon\mu} \nabla^{\alpha}\Phi \nabla^{\beta}\Phi \nabla^{\gamma}\nabla_{\alpha}\Phi + c_{263} H_{\beta}{}^{\delta \epsilon} H_{\gamma}{}^{\varepsilon\mu} R_{\delta \varepsilon \epsilon\mu} \nabla^{\alpha}\Phi \nabla^{\beta}\Phi \nabla^{\gamma}\nabla_{\alpha}\Phi \nn\\&&+ c_{276} H_{\delta \epsilon}{}^{\mu} H^{\delta \epsilon \varepsilon} R_{\beta \varepsilon \gamma\mu} \nabla_{\alpha}\Phi \nabla^{\alpha}\Phi \nabla^{\gamma}\nabla^{\beta}\Phi + c_{278} H_{\beta}{}^{\delta \epsilon} H_{\delta}{}^{\varepsilon\mu} R_{\gamma \varepsilon \epsilon\mu} \nabla_{\alpha}\Phi \nabla^{\alpha}\Phi \nabla^{\gamma}\nabla^{\beta}\Phi\nn\\&& + c_{279} H_{\beta}{}^{\delta \epsilon} H_{\gamma}{}^{\varepsilon\mu} R_{\delta \varepsilon \epsilon\mu} \nabla_{\alpha}\Phi \nabla^{\alpha}\Phi \nabla^{\gamma}\nabla^{\beta}\Phi + c_{318} H_{\gamma}{}^{\epsilon \varepsilon} H_{\epsilon \varepsilon}{}^{\mu} R_{\alpha \delta \beta\mu} \nabla^{\alpha}\Phi \nabla^{\beta}\Phi \nabla^{\delta}\nabla^{\gamma}\Phi \nn\\&&+ c_{319} H_{\gamma}{}^{\epsilon \varepsilon} H_{\delta \epsilon}{}^{\mu} R_{\alpha \varepsilon \beta\mu} \nabla^{\alpha}\Phi \nabla^{\beta}\Phi \nabla^{\delta}\nabla^{\gamma}\Phi + c_{320} H_{\alpha}{}^{\epsilon \varepsilon} H_{\epsilon \varepsilon}{}^{\mu} R_{\beta \gamma \delta\mu} \nabla^{\alpha}\Phi \nabla^{\beta}\Phi \nabla^{\delta}\nabla^{\gamma}\Phi \nn\\&&+ c_{323} H_{\alpha}{}^{\epsilon \varepsilon} H_{\gamma \epsilon}{}^{\mu} R_{\beta \varepsilon \delta\mu} \nabla^{\alpha}\Phi \nabla^{\beta}\Phi \nabla^{\delta}\nabla^{\gamma}\Phi + c_{324} H_{\alpha \gamma}{}^{\epsilon} H_{\epsilon}{}^{\varepsilon\mu} R_{\beta \varepsilon \delta\mu} \nabla^{\alpha}\Phi \nabla^{\beta}\Phi \nabla^{\delta}\nabla^{\gamma}\Phi \nn\\&&+ c_{325} H_{\alpha}{}^{\epsilon \varepsilon} H_{\gamma \epsilon}{}^{\mu} R_{\beta\mu \delta \varepsilon} \nabla^{\alpha}\Phi \nabla^{\beta}\Phi \nabla^{\delta}\nabla^{\gamma}\Phi + c_{327} H_{\alpha}{}^{\epsilon \varepsilon} H_{\beta \epsilon}{}^{\mu} R_{\gamma \varepsilon \delta\mu} \nabla^{\alpha}\Phi \nabla^{\beta}\Phi \nabla^{\delta}\nabla^{\gamma}\Phi\nn\\&& + c_{328} H_{\alpha \gamma}{}^{\epsilon} H_{\beta}{}^{\varepsilon\mu} R_{\delta \varepsilon \epsilon\mu} \nabla^{\alpha}\Phi \nabla^{\beta}\Phi \nabla^{\delta}\nabla^{\gamma}\Phi\labell{T15}
\eeqa
\beqa
{\cal L}_3^{H^6\prt\prt\Phi}&\!\!\!\!\!=\!\!\!\!\! & c_{183} H_{\alpha}{}^{\gamma \delta} H_{\beta}{}^{\epsilon \varepsilon} H_{\gamma}{}^{\mu \zeta} H_{\delta}{}^{\eta \theta} H_{\epsilon \mu \zeta} H_{\varepsilon \eta \theta} \nabla^{\beta}\nabla^{\alpha}\Phi + c_{184} H_{\alpha}{}^{\gamma \delta} H_{\beta}{}^{\epsilon \varepsilon} H_{\gamma}{}^{\mu \zeta} H_{\delta \mu}{}^{\eta} H_{\epsilon \zeta}{}^{\theta} H_{\varepsilon \eta \theta} \nabla^{\beta}\nabla^{\alpha}\Phi \nn\\&&+ c_{185} H_{\alpha}{}^{\gamma \delta} H_{\beta}{}^{\epsilon \varepsilon} H_{\gamma \epsilon}{}^{\mu} H_{\delta}{}^{\zeta \eta} H_{\varepsilon \zeta}{}^{\theta} H_{\mu \eta \theta} \nabla^{\beta}\nabla^{\alpha}\Phi + c_{186} H_{\alpha}{}^{\gamma \delta} H_{\beta}{}^{\epsilon \varepsilon} H_{\gamma \delta}{}^{\mu} H_{\epsilon}{}^{\zeta \eta} H_{\varepsilon \zeta}{}^{\theta} H_{\mu \eta \theta} \nabla^{\beta}\nabla^{\alpha}\Phi \nn\\&&+ c_{187} H_{\alpha}{}^{\gamma \delta} H_{\beta \gamma}{}^{\epsilon} H_{\delta}{}^{\varepsilon \mu} H_{\epsilon}{}^{\zeta \eta} H_{\varepsilon \zeta}{}^{\theta} H_{\mu \eta \theta} \nabla^{\beta}\nabla^{\alpha}\Phi + c_{188} H_{\alpha}{}^{\gamma \delta} H_{\beta \gamma \delta} H_{\epsilon}{}^{\zeta \eta} H^{\epsilon \varepsilon \mu} H_{\varepsilon \zeta}{}^{\theta} H_{\mu \eta \theta} \nabla^{\beta}\nabla^{\alpha}\Phi \nn\\&&+ c_{189} H_{\alpha}{}^{\gamma \delta} H_{\beta \gamma}{}^{\epsilon} H_{\delta}{}^{\varepsilon \mu} H_{\epsilon}{}^{\zeta \eta} H_{\varepsilon \mu}{}^{\theta} H_{\zeta \eta \theta} \nabla^{\beta}\nabla^{\alpha}\Phi + c_{190} H_{\alpha}{}^{\gamma \delta} H_{\beta}{}^{\epsilon \varepsilon} H_{\gamma \epsilon}{}^{\mu} H_{\delta \mu}{}^{\zeta} H_{\varepsilon}{}^{\eta \theta} H_{\zeta \eta \theta} \nabla^{\beta}\nabla^{\alpha}\Phi\nn\\&&+ c_{191} H_{\alpha}{}^{\gamma \delta} H_{\beta}{}^{\epsilon \varepsilon} H_{\gamma \delta}{}^{\mu} H_{\epsilon \mu}{}^{\zeta} H_{\varepsilon}{}^{\eta \theta} H_{\zeta \eta \theta} \nabla^{\beta}\nabla^{\alpha}\Phi + c_{192} H_{\alpha}{}^{\gamma \delta} H_{\beta}{}^{\epsilon \varepsilon} H_{\gamma \epsilon}{}^{\mu} H_{\delta \varepsilon}{}^{\zeta} H_{\mu}{}^{\eta \theta} H_{\zeta \eta \theta} \nabla^{\beta}\nabla^{\alpha}\Phi \nn\\&&+ c_{193} H_{\alpha}{}^{\gamma \delta} H_{\beta}{}^{\epsilon \varepsilon} H_{\gamma \delta}{}^{\mu} H_{\epsilon \varepsilon}{}^{\zeta} H_{\mu}{}^{\eta \theta} H_{\zeta \eta \theta} \nabla^{\beta}\nabla^{\alpha}\Phi + c_{194} H_{\alpha}{}^{\gamma \delta} H_{\beta \gamma}{}^{\epsilon} H_{\delta}{}^{\varepsilon \mu} H_{\epsilon \varepsilon}{}^{\zeta} H_{\mu}{}^{\eta \theta} H_{\zeta \eta \theta} \nabla^{\beta}\nabla^{\alpha}\Phi \nn\\&&+ c_{195} H_{\alpha}{}^{\gamma \delta} H_{\beta \gamma \delta} H_{\epsilon \varepsilon}{}^{\zeta} H^{\epsilon \varepsilon \mu} H_{\mu}{}^{\eta \theta} H_{\zeta \eta \theta} \nabla^{\beta}\nabla^{\alpha}\Phi + c_{196} H_{\alpha}{}^{\gamma \delta} H_{\beta}{}^{\epsilon \varepsilon} H_{\gamma \epsilon}{}^{\mu} H_{\delta \varepsilon \mu} H_{\zeta \eta \theta} H^{\zeta \eta \theta} \nabla^{\beta}\nabla^{\alpha}\Phi \nn\\&&+ c_{197} H_{\alpha}{}^{\gamma \delta} H_{\beta}{}^{\epsilon \varepsilon} H_{\gamma \delta}{}^{\mu} H_{\epsilon \varepsilon \mu} H_{\zeta \eta \theta} H^{\zeta \eta \theta} \nabla^{\beta}\nabla^{\alpha}\Phi + c_{198} H_{\alpha}{}^{\gamma \delta} H_{\beta \gamma}{}^{\epsilon} H_{\delta}{}^{\varepsilon \mu} H_{\epsilon \varepsilon \mu} H_{\zeta \eta \theta} H^{\zeta \eta \theta} \nabla^{\beta}\nabla^{\alpha}\Phi \nn\\&&+ c_{199} H_{\alpha}{}^{\gamma \delta} H_{\beta \gamma \delta} H_{\epsilon \varepsilon \mu} H^{\epsilon \varepsilon \mu} H_{\zeta \eta \theta} H^{\zeta \eta \theta} \nabla^{\beta}\nabla^{\alpha}\Phi\labell{T16}
\eeqa
\beqa
{\cal L}_3^{H^4R\prt\prt\Phi}&\!\!\!\!\!=\!\!\!\!\! & c_{200} H_{\gamma \delta}{}^{\varepsilon} H^{\gamma \delta \epsilon} H_{\epsilon}{}^{\mu \zeta} H_{\mu \zeta}{}^{\eta} R_{\alpha \varepsilon \beta \eta} \nabla^{\beta}\nabla^{\alpha}\Phi + c_{201} H_{\gamma}{}^{\varepsilon \mu} H^{\gamma \delta \epsilon} H_{\delta \varepsilon}{}^{\zeta} H_{\epsilon \mu}{}^{\eta} R_{\alpha \zeta \beta \eta} \nabla^{\beta}\nabla^{\alpha}\Phi  \nn\\&&+ c_{202} H_{\gamma \delta}{}^{\varepsilon} H^{\gamma \delta \epsilon} H_{\epsilon}{}^{\mu \zeta} H_{\varepsilon \mu}{}^{\eta} R_{\alpha \zeta \beta \eta} \nabla^{\beta}\nabla^{\alpha}\Phi + c_{203} H_{\gamma \delta \epsilon} H^{\gamma \delta \epsilon} H_{\varepsilon \mu}{}^{\eta} H^{\varepsilon \mu \zeta} R_{\alpha \zeta \beta \eta} \nabla^{\beta}\nabla^{\alpha}\Phi \nn\\&& + c_{204} H_{\alpha}{}^{\gamma \delta} H_{\gamma}{}^{\epsilon \varepsilon} H_{\mu \zeta \eta} H^{\mu \zeta \eta} R_{\beta \epsilon \delta \varepsilon} \nabla^{\beta}\nabla^{\alpha}\Phi + c_{205} H_{\alpha}{}^{\gamma \delta} H_{\gamma}{}^{\epsilon \varepsilon} H_{\epsilon}{}^{\mu \zeta} H_{\mu \zeta}{}^{\eta} R_{\beta \varepsilon \delta \eta} \nabla^{\beta}\nabla^{\alpha}\Phi  \nn\\&&+ c_{206} H_{\alpha}{}^{\gamma \delta} H_{\gamma}{}^{\epsilon \varepsilon} H_{\delta}{}^{\mu \zeta} H_{\epsilon \mu}{}^{\eta} R_{\beta \varepsilon \zeta \eta} \nabla^{\beta}\nabla^{\alpha}\Phi + c_{214} H_{\alpha}{}^{\gamma \delta} H_{\gamma}{}^{\epsilon \varepsilon} H_{\epsilon}{}^{\mu \zeta} H_{\varepsilon \mu}{}^{\eta} R_{\beta \zeta \delta \eta} \nabla^{\beta}\nabla^{\alpha}\Phi \nn\\&& + c_{215} H_{\alpha}{}^{\gamma \delta} H_{\gamma}{}^{\epsilon \varepsilon} H_{\epsilon \varepsilon}{}^{\mu} H_{\mu}{}^{\zeta \eta} R_{\beta \zeta \delta \eta} \nabla^{\beta}\nabla^{\alpha}\Phi + c_{216} H_{\alpha}{}^{\gamma \delta} H_{\gamma \delta}{}^{\epsilon} H_{\varepsilon \mu}{}^{\eta} H^{\varepsilon \mu \zeta} R_{\beta \zeta \epsilon \eta} \nabla^{\beta}\nabla^{\alpha}\Phi  \nn\\&&+ c_{217} H_{\alpha}{}^{\gamma \delta} H_{\gamma}{}^{\epsilon \varepsilon} H_{\delta \epsilon}{}^{\mu} H_{\varepsilon}{}^{\zeta \eta} R_{\beta \zeta \mu \eta} \nabla^{\beta}\nabla^{\alpha}\Phi + c_{218} H_{\alpha}{}^{\gamma \delta} H_{\gamma \delta}{}^{\epsilon} H_{\epsilon}{}^{\varepsilon \mu} H_{\varepsilon}{}^{\zeta \eta} R_{\beta \zeta \mu \eta} \nabla^{\beta}\nabla^{\alpha}\Phi  \nn\\&&+ c_{219} H_{\alpha}{}^{\gamma \delta} H_{\gamma}{}^{\epsilon \varepsilon} H_{\epsilon}{}^{\mu \zeta} H_{\mu \zeta}{}^{\eta} R_{\beta \eta \delta \varepsilon} \nabla^{\beta}\nabla^{\alpha}\Phi + c_{220} H_{\alpha}{}^{\gamma \delta} H_{\gamma}{}^{\epsilon \varepsilon} H_{\delta}{}^{\mu \zeta} H_{\epsilon \varepsilon}{}^{\eta} R_{\beta \eta \mu \zeta} \nabla^{\beta}\nabla^{\alpha}\Phi  \nn\\&&+ c_{221} H_{\alpha}{}^{\gamma \delta} H_{\beta}{}^{\epsilon \varepsilon} H_{\mu \zeta \eta} H^{\mu \zeta \eta} R_{\gamma \epsilon \delta \varepsilon} \nabla^{\beta}\nabla^{\alpha}\Phi + c_{233} H_{\alpha}{}^{\gamma \delta} H_{\beta \gamma}{}^{\epsilon} H_{\varepsilon \mu}{}^{\eta} H^{\varepsilon \mu \zeta} R_{\delta \zeta \epsilon \eta} \nabla^{\beta}\nabla^{\alpha}\Phi  \nn\\&&+ c_{234} H_{\alpha}{}^{\gamma \delta} H_{\beta}{}^{\epsilon \varepsilon} H_{\gamma}{}^{\mu \zeta} H_{\epsilon \mu}{}^{\eta} R_{\delta \zeta \varepsilon \eta} \nabla^{\beta}\nabla^{\alpha}\Phi + c_{235} H_{\alpha}{}^{\gamma \delta} H_{\beta}{}^{\epsilon \varepsilon} H_{\gamma \epsilon}{}^{\mu} H_{\mu}{}^{\zeta \eta} R_{\delta \zeta \varepsilon \eta} \nabla^{\beta}\nabla^{\alpha}\Phi \nn\\&& + c_{236} H_{\alpha}{}^{\gamma \delta} H_{\beta}{}^{\epsilon \varepsilon} H_{\gamma}{}^{\mu \zeta} H_{\mu \zeta}{}^{\eta} R_{\delta \eta \epsilon \varepsilon} \nabla^{\beta}\nabla^{\alpha}\Phi + c_{237} H_{\alpha}{}^{\gamma \delta} H_{\beta}{}^{\epsilon \varepsilon} H_{\gamma}{}^{\mu \zeta} H_{\epsilon \mu}{}^{\eta} R_{\delta \eta \varepsilon \zeta} \nabla^{\beta}\nabla^{\alpha}\Phi  \nn\\&&+ c_{244} H_{\alpha}{}^{\gamma \delta} H_{\beta}{}^{\epsilon \varepsilon} H_{\gamma}{}^{\mu \zeta} H_{\delta \mu}{}^{\eta} R_{\epsilon \zeta \varepsilon \eta} \nabla^{\beta}\nabla^{\alpha}\Phi + c_{245} H_{\alpha}{}^{\gamma \delta} H_{\beta}{}^{\epsilon \varepsilon} H_{\gamma \delta}{}^{\mu} H_{\mu}{}^{\zeta \eta} R_{\epsilon \zeta \varepsilon \eta} \nabla^{\beta}\nabla^{\alpha}\Phi  \nn\\&&+ c_{246} H_{\alpha}{}^{\gamma \delta} H_{\beta \gamma}{}^{\epsilon} H_{\delta}{}^{\varepsilon \mu} H_{\varepsilon}{}^{\zeta \eta} R_{\epsilon \zeta \mu \eta} \nabla^{\beta}\nabla^{\alpha}\Phi + c_{248} H_{\alpha}{}^{\gamma \delta} H_{\beta}{}^{\epsilon \varepsilon} H_{\gamma \epsilon}{}^{\mu} H_{\delta}{}^{\zeta \eta} R_{\varepsilon \zeta \mu \eta} \nabla^{\beta}\nabla^{\alpha}\Phi \nn\\&& + c_{249} H_{\alpha}{}^{\gamma \delta} H_{\beta}{}^{\epsilon \varepsilon} H_{\gamma \delta}{}^{\mu} H_{\epsilon}{}^{\zeta \eta} R_{\varepsilon \zeta \mu \eta} \nabla^{\beta}\nabla^{\alpha}\Phi + c_{250} H_{\alpha}{}^{\gamma \delta} H_{\beta \gamma}{}^{\epsilon} H_{\delta}{}^{\varepsilon \mu} H_{\epsilon}{}^{\zeta \eta} R_{\varepsilon \zeta \mu \eta} \nabla^{\beta}\nabla^{\alpha}\Phi  \nn\\&&+ c_{251} H_{\alpha}{}^{\gamma \delta} H_{\beta \gamma \delta} H_{\epsilon}{}^{\zeta \eta} H^{\epsilon \varepsilon \mu} R_{\varepsilon \zeta \mu \eta} \nabla^{\beta}\nabla^{\alpha}\Phi\labell{T17}
\eeqa
\beqa
{\cal L}_3^{H^2R^2\prt\prt\Phi}&\!\!\!\!\!=\!\!\!\!\! &  c_{207} H_{\gamma}{}^{\varepsilon \mu} H^{\gamma \delta \epsilon} R_{\alpha \delta \varepsilon}{}^{\zeta} R_{\beta \mu \epsilon \zeta} \nabla^{\beta}\nabla^{\alpha}\Phi + c_{208} H_{\gamma \delta \epsilon} H^{\gamma \delta \epsilon} R_{\alpha}{}^{\varepsilon \mu \zeta} R_{\beta \mu \varepsilon \zeta} \nabla^{\beta}\nabla^{\alpha}\Phi \nn\\&& + c_{209} H_{\gamma \delta}{}^{\varepsilon} H^{\gamma \delta \epsilon} R_{\alpha}{}^{\mu}{}_{\epsilon}{}^{\zeta} R_{\beta \mu \varepsilon \zeta} \nabla^{\beta}\nabla^{\alpha}\Phi + c_{210} H^{\gamma \delta \epsilon} H^{\varepsilon \mu \zeta} R_{\alpha \gamma \varepsilon \mu} R_{\beta \zeta \delta \epsilon} \nabla^{\beta}\nabla^{\alpha}\Phi  \nn\\&&+ c_{211} H_{\gamma}{}^{\varepsilon \mu} H^{\gamma \delta \epsilon} R_{\alpha}{}^{\zeta}{}_{\delta \varepsilon} R_{\beta \zeta \epsilon \mu} \nabla^{\beta}\nabla^{\alpha}\Phi + c_{212} H_{\gamma \delta}{}^{\varepsilon} H^{\gamma \delta \epsilon} R_{\alpha}{}^{\mu}{}_{\epsilon}{}^{\zeta} R_{\beta \zeta \varepsilon \mu} \nabla^{\beta}\nabla^{\alpha}\Phi \nn\\&& + c_{213} H_{\gamma}{}^{\varepsilon \mu} H^{\gamma \delta \epsilon} R_{\alpha}{}^{\zeta}{}_{\delta \epsilon} R_{\beta \zeta \varepsilon \mu} \nabla^{\beta}\nabla^{\alpha}\Phi + c_{222} H_{\alpha}{}^{\gamma \delta} H^{\epsilon \varepsilon \mu} R_{\beta}{}^{\zeta}{}_{\epsilon \varepsilon} R_{\gamma \mu \delta \zeta} \nabla^{\beta}\nabla^{\alpha}\Phi  \nn\\&&+ c_{226} H^{\gamma \delta \epsilon} H^{\varepsilon \mu \zeta} R_{\alpha \gamma \beta \varepsilon} R_{\delta \mu \epsilon \zeta} \nabla^{\beta}\nabla^{\alpha}\Phi + c_{227} H_{\alpha}{}^{\gamma \delta} H_{\gamma}{}^{\epsilon \varepsilon} R_{\beta}{}^{\mu}{}_{\epsilon}{}^{\zeta} R_{\delta \mu \varepsilon \zeta} \nabla^{\beta}\nabla^{\alpha}\Phi \nn\\&& + c_{228} H_{\alpha}{}^{\gamma \delta} H_{\beta}{}^{\epsilon \varepsilon} R_{\gamma}{}^{\mu}{}_{\epsilon}{}^{\zeta} R_{\delta \mu \varepsilon \zeta} \nabla^{\beta}\nabla^{\alpha}\Phi + c_{229} H_{\alpha}{}^{\gamma \delta} H^{\epsilon \varepsilon \mu} R_{\beta \epsilon \gamma}{}^{\zeta} R_{\delta \zeta \varepsilon \mu} \nabla^{\beta}\nabla^{\alpha}\Phi  \nn\\&&+ c_{230} H_{\alpha}{}^{\gamma \delta} H_{\gamma}{}^{\epsilon \varepsilon} R_{\beta}{}^{\mu}{}_{\epsilon}{}^{\zeta} R_{\delta \zeta \varepsilon \mu} \nabla^{\beta}\nabla^{\alpha}\Phi + c_{231} H_{\alpha}{}^{\gamma \delta} H^{\epsilon \varepsilon \mu} R_{\beta}{}^{\zeta}{}_{\gamma \epsilon} R_{\delta \zeta \varepsilon \mu} \nabla^{\beta}\nabla^{\alpha}\Phi  \nn\\&&+ c_{232} H_{\alpha}{}^{\gamma \delta} H_{\beta}{}^{\epsilon \varepsilon} R_{\gamma}{}^{\mu}{}_{\epsilon}{}^{\zeta} R_{\delta \zeta \varepsilon \mu} \nabla^{\beta}\nabla^{\alpha}\Phi + c_{238} H_{\gamma \delta}{}^{\varepsilon} H^{\gamma \delta \epsilon} R_{\alpha}{}^{\mu}{}_{\beta}{}^{\zeta} R_{\epsilon \mu \varepsilon \zeta} \nabla^{\beta}\nabla^{\alpha}\Phi  \nn\\&&+ c_{239} H_{\alpha}{}^{\gamma \delta} H_{\gamma \delta}{}^{\epsilon} R_{\beta}{}^{\varepsilon \mu \zeta} R_{\epsilon \mu \varepsilon \zeta} \nabla^{\beta}\nabla^{\alpha}\Phi + c_{240} H_{\alpha}{}^{\gamma \delta} H_{\gamma}{}^{\epsilon \varepsilon} R_{\beta}{}^{\mu}{}_{\delta}{}^{\zeta} R_{\epsilon \mu \varepsilon \zeta} \nabla^{\beta}\nabla^{\alpha}\Phi \nn\\&& + c_{241} H_{\alpha}{}^{\gamma \delta} H_{\beta}{}^{\epsilon \varepsilon} R_{\gamma}{}^{\mu}{}_{\delta}{}^{\zeta} R_{\epsilon \mu \varepsilon \zeta} \nabla^{\beta}\nabla^{\alpha}\Phi + c_{242} H_{\alpha}{}^{\gamma \delta} H_{\beta \gamma}{}^{\epsilon} R_{\delta}{}^{\varepsilon \mu \zeta} R_{\epsilon \mu \varepsilon \zeta} \nabla^{\beta}\nabla^{\alpha}\Phi  \nn\\&&+ c_{243} H_{\gamma}{}^{\varepsilon \mu} H^{\gamma \delta \epsilon} R_{\alpha \delta \beta}{}^{\zeta} R_{\epsilon \zeta \varepsilon \mu} \nabla^{\beta}\nabla^{\alpha}\Phi + c_{247} H_{\alpha}{}^{\gamma \delta} H_{\beta \gamma \delta} R_{\epsilon \mu \varepsilon \zeta} R^{\epsilon \varepsilon \mu \zeta} \nabla^{\beta}\nabla^{\alpha}\Phi\labell{T18}
\eeqa
\beqa
{\cal L}_3^{R^3\prt\prt\Phi}&\!\!\!\!\!=\!\!\!\!\! & c_{223} R_{\alpha}{}^{\gamma \delta \epsilon} R_{\beta}{}^{\varepsilon}{}_{\delta}{}^{\mu} R_{\gamma \mu \epsilon \varepsilon} \nabla^{\beta}\nabla^{\alpha}\Phi + c_{224} R_{\alpha}{}^{\gamma \delta \epsilon} R_{\beta \gamma}{}^{\varepsilon \mu} R_{\delta \varepsilon \epsilon \mu} \nabla^{\beta}\nabla^{\alpha}\Phi \nn\\&& + c_{225} R_{\alpha}{}^{\gamma}{}_{\beta}{}^{\delta} R_{\gamma}{}^{\epsilon \varepsilon \mu} R_{\delta \varepsilon \epsilon \mu} \nabla^{\beta}\nabla^{\alpha}\Phi\labell{T19}
\eeqa
\beqa
{\cal L}_3^{H^4(\prt\prt\Phi)^2}&\!\!\!\!\!=\!\!\!\!\! & c_{252} H_{\gamma}{}^{\varepsilon \mu} H^{\gamma \delta \epsilon} H_{\delta \varepsilon}{}^{\zeta} H_{\epsilon \mu \zeta} \nabla_{\beta}\nabla_{\alpha}\Phi \nabla^{\beta}\nabla^{\alpha}\Phi + c_{253} H_{\gamma \delta \epsilon} H^{\gamma \delta \epsilon} H_{\varepsilon \mu \zeta} H^{\varepsilon \mu \zeta} \nabla_{\beta}\nabla_{\alpha}\Phi \nabla^{\beta}\nabla^{\alpha}\Phi\nn\\&& + c_{264} H_{\beta}{}^{\delta \epsilon} H_{\gamma}{}^{\varepsilon \mu} H_{\delta \varepsilon}{}^{\zeta} H_{\epsilon \mu \zeta} \nabla^{\beta}\nabla^{\alpha}\Phi \nabla^{\gamma}\nabla_{\alpha}\Phi + c_{265} H_{\beta}{}^{\delta \epsilon} H_{\gamma}{}^{\varepsilon \mu} H_{\delta \epsilon}{}^{\zeta} H_{\varepsilon \mu \zeta} \nabla^{\beta}\nabla^{\alpha}\Phi \nabla^{\gamma}\nabla_{\alpha}\Phi\nn\\&& + c_{266} H_{\beta}{}^{\delta \epsilon} H_{\gamma \delta}{}^{\varepsilon} H_{\epsilon}{}^{\mu \zeta} H_{\varepsilon \mu \zeta} \nabla^{\beta}\nabla^{\alpha}\Phi \nabla^{\gamma}\nabla_{\alpha}\Phi + c_{267} H_{\beta}{}^{\delta \epsilon} H_{\gamma \delta \epsilon} H_{\varepsilon \mu \zeta} H^{\varepsilon \mu \zeta} \nabla^{\beta}\nabla^{\alpha}\Phi \nabla^{\gamma}\nabla_{\alpha}\Phi\nn\\&& + c_{331} H_{\alpha}{}^{\epsilon \varepsilon} H_{\beta}{}^{\mu \zeta} H_{\gamma \epsilon \mu} H_{\delta \varepsilon \zeta} \nabla^{\beta}\nabla^{\alpha}\Phi \nabla^{\delta}\nabla^{\gamma}\Phi + c_{332} H_{\alpha}{}^{\epsilon \varepsilon} H_{\beta}{}^{\mu \zeta} H_{\gamma \epsilon \varepsilon} H_{\delta \mu \zeta} \nabla^{\beta}\nabla^{\alpha}\Phi \nabla^{\delta}\nabla^{\gamma}\Phi \nn\\&&+ c_{333} H_{\alpha}{}^{\epsilon \varepsilon} H_{\beta \epsilon}{}^{\mu} H_{\gamma \varepsilon}{}^{\zeta} H_{\delta \mu \zeta} \nabla^{\beta}\nabla^{\alpha}\Phi \nabla^{\delta}\nabla^{\gamma}\Phi + c_{334} H_{\alpha}{}^{\epsilon \varepsilon} H_{\beta \epsilon \varepsilon} H_{\gamma}{}^{\mu \zeta} H_{\delta \mu \zeta} \nabla^{\beta}\nabla^{\alpha}\Phi \nabla^{\delta}\nabla^{\gamma}\Phi\nn\\&& + c_{335} H_{\alpha \gamma}{}^{\epsilon} H_{\beta}{}^{\varepsilon \mu} H_{\delta \varepsilon}{}^{\zeta} H_{\epsilon \mu \zeta} \nabla^{\beta}\nabla^{\alpha}\Phi \nabla^{\delta}\nabla^{\gamma}\Phi + c_{336} H_{\alpha \gamma}{}^{\epsilon} H_{\beta \epsilon}{}^{\varepsilon} H_{\delta}{}^{\mu \zeta} H_{\varepsilon \mu \zeta} \nabla^{\beta}\nabla^{\alpha}\Phi \nabla^{\delta}\nabla^{\gamma}\Phi\nn\\&& + c_{337} H_{\alpha \gamma}{}^{\epsilon} H_{\beta \delta}{}^{\varepsilon} H_{\epsilon}{}^{\mu \zeta} H_{\varepsilon \mu \zeta} \nabla^{\beta}\nabla^{\alpha}\Phi \nabla^{\delta}\nabla^{\gamma}\Phi + c_{338} H_{\alpha \gamma}{}^{\epsilon} H_{\beta \delta \epsilon} H_{\varepsilon \mu \zeta} H^{\varepsilon \mu \zeta} \nabla^{\beta}\nabla^{\alpha}\Phi \nabla^{\delta}\nabla^{\gamma}\Phi\labell{T20}\nn
\eeqa
\beqa
{\cal L}_3^{R^2(\prt\prt\Phi)^2}&\!\!\!\!\!=\!\!\!\!\! & c_{254} R_{\gamma \epsilon \delta \varepsilon} R^{\gamma \delta \epsilon \varepsilon} \nabla_{\beta}\nabla_{\alpha}\Phi \nabla^{\beta}\nabla^{\alpha}\Phi + c_{269} R_{\beta}{}^{\delta \epsilon \varepsilon} R_{\gamma \epsilon \delta \varepsilon} \nabla^{\beta}\nabla^{\alpha}\Phi \nabla^{\gamma}\nabla_{\alpha}\Phi \nn\\&&+ c_{341} R_{\alpha}{}^{\epsilon}{}_{\gamma}{}^{\varepsilon} R_{\beta \epsilon \delta \varepsilon} \nabla^{\beta}\nabla^{\alpha}\Phi \nabla^{\delta}\nabla^{\gamma}\Phi + c_{342} R_{\alpha}{}^{\epsilon}{}_{\gamma}{}^{\varepsilon} R_{\beta \varepsilon \delta \epsilon} \nabla^{\beta}\nabla^{\alpha}\Phi \nabla^{\delta}\nabla^{\gamma}\Phi\nn\\&& + c_{346} R_{\alpha}{}^{\epsilon}{}_{\beta}{}^{\varepsilon} R_{\gamma \epsilon \delta \varepsilon} \nabla^{\beta}\nabla^{\alpha}\Phi \nabla^{\delta}\nabla^{\gamma}\Phi\labell{T21}
\eeqa
\beqa
{\cal L}_3^{H^2R(\prt\prt\Phi)^2}&\!\!\!\!\!=\!\!\!\!\! & c_{255} H_{\gamma}{}^{\varepsilon\mu} H^{\gamma \delta \epsilon} R_{\delta \varepsilon \epsilon\mu} \nabla_{\beta}\nabla_{\alpha}\Phi \nabla^{\beta}\nabla^{\alpha}\Phi + c_{268} H_{\delta \epsilon}{}^{\mu} H^{\delta \epsilon \varepsilon} R_{\beta \varepsilon \gamma\mu} \nabla^{\beta}\nabla^{\alpha}\Phi \nabla^{\gamma}\nabla_{\alpha}\Phi \nn\\&&+ c_{270} H_{\beta}{}^{\delta \epsilon} H_{\delta}{}^{\varepsilon\mu} R_{\gamma \varepsilon \epsilon\mu} \nabla^{\beta}\nabla^{\alpha}\Phi \nabla^{\gamma}\nabla_{\alpha}\Phi + c_{271} H_{\beta}{}^{\delta \epsilon} H_{\gamma}{}^{\varepsilon\mu} R_{\delta \varepsilon \epsilon\mu} \nabla^{\beta}\nabla^{\alpha}\Phi \nabla^{\gamma}\nabla_{\alpha}\Phi \nn\\&&+ c_{339} H_{\epsilon \varepsilon\mu} H^{\epsilon \varepsilon\mu} R_{\alpha \gamma \beta \delta} \nabla^{\beta}\nabla^{\alpha}\Phi \nabla^{\delta}\nabla^{\gamma}\Phi + c_{340} H_{\alpha}{}^{\epsilon \varepsilon} H_{\epsilon \varepsilon}{}^{\mu} R_{\beta \gamma \delta\mu} \nabla^{\beta}\nabla^{\alpha}\Phi \nabla^{\delta}\nabla^{\gamma}\Phi\nn\\&&+ c_{343} H_{\alpha}{}^{\epsilon \varepsilon} H_{\gamma \epsilon}{}^{\mu} R_{\beta \varepsilon \delta\mu} \nabla^{\beta}\nabla^{\alpha}\Phi \nabla^{\delta}\nabla^{\gamma}\Phi + c_{344} H_{\alpha \gamma}{}^{\epsilon} H_{\epsilon}{}^{\varepsilon\mu} R_{\beta \varepsilon \delta\mu} \nabla^{\beta}\nabla^{\alpha}\Phi \nabla^{\delta}\nabla^{\gamma}\Phi \nn\\&&+ c_{345} H_{\alpha}{}^{\epsilon \varepsilon} H_{\gamma \epsilon}{}^{\mu} R_{\beta\mu \delta \varepsilon} \nabla^{\beta}\nabla^{\alpha}\Phi \nabla^{\delta}\nabla^{\gamma}\Phi + c_{347} H_{\alpha}{}^{\epsilon \varepsilon} H_{\beta \epsilon}{}^{\mu} R_{\gamma \varepsilon \delta\mu} \nabla^{\beta}\nabla^{\alpha}\Phi \nabla^{\delta}\nabla^{\gamma}\Phi \nn\\&&+ c_{348} H_{\alpha \gamma}{}^{\epsilon} H_{\beta}{}^{\varepsilon\mu} R_{\delta \varepsilon \epsilon\mu} \nabla^{\beta}\nabla^{\alpha}\Phi \nabla^{\delta}\nabla^{\gamma}\Phi\labell{T22}
\eeqa
\beqa
&&{\cal L}_3^{H^2(\prt\Phi)^4\prt\prt\Phi}=\nn\\&& c_{258} H_{\delta \epsilon \varepsilon} H^{\delta \epsilon \varepsilon} \nabla_{\alpha}\Phi \nabla^{\alpha}\Phi \nabla^{\beta}\Phi \nabla_{\gamma}\nabla_{\beta}\Phi \nabla^{\gamma}\Phi + c_{290} H_{\gamma}{}^{\epsilon \varepsilon} H_{\delta \epsilon \varepsilon} \nabla^{\alpha}\Phi \nabla_{\beta}\nabla_{\alpha}\Phi \nabla^{\beta}\Phi \nabla^{\gamma}\Phi \nabla^{\delta}\Phi \nn\\&& + c_{298} H_{\gamma}{}^{\epsilon \varepsilon} H_{\delta \epsilon \varepsilon} \nabla_{\alpha}\Phi \nabla^{\alpha}\Phi \nabla^{\beta}\Phi \nabla^{\gamma}\Phi \nabla^{\delta}\nabla_{\beta}\Phi + c_{329} H_{\gamma}{}^{\epsilon \varepsilon} H_{\delta \epsilon \varepsilon} \nabla_{\alpha}\Phi \nabla^{\alpha}\Phi \nabla_{\beta}\Phi \nabla^{\beta}\Phi \nabla^{\delta}\nabla^{\gamma}\Phi  \nn\\&&+ c_{373} H_{\beta \delta}{}^{\varepsilon} H_{\gamma \epsilon \varepsilon} \nabla_{\alpha}\Phi \nabla^{\alpha}\Phi \nabla^{\beta}\Phi \nabla^{\gamma}\Phi \nabla^{\epsilon}\nabla^{\delta}\Phi\labell{T23}
\eeqa
\beqa
&&{\cal L}_3^{H^2(\prt\Phi)^2(\prt\prt\Phi)^2}=\nn\\&& c_{272} H_{\delta \epsilon \varepsilon} H^{\delta \epsilon \varepsilon} \nabla^{\alpha}\Phi \nabla^{\beta}\Phi \nabla_{\gamma}\nabla_{\beta}\Phi \nabla^{\gamma}\nabla_{\alpha}\Phi + c_{280} H_{\delta \epsilon \varepsilon} H^{\delta \epsilon \varepsilon} \nabla_{\alpha}\Phi \nabla^{\alpha}\Phi \nabla_{\gamma}\nabla_{\beta}\Phi \nabla^{\gamma}\nabla^{\beta}\Phi  \nn\\&&+ c_{299} H_{\gamma}{}^{\epsilon \varepsilon} H_{\delta \epsilon \varepsilon} \nabla^{\alpha}\Phi \nabla^{\beta}\Phi \nabla^{\gamma}\nabla_{\alpha}\Phi \nabla^{\delta}\nabla_{\beta}\Phi + c_{301} H_{\gamma}{}^{\epsilon \varepsilon} H_{\delta \epsilon \varepsilon} \nabla_{\alpha}\Phi \nabla^{\alpha}\Phi \nabla^{\gamma}\nabla^{\beta}\Phi \nabla^{\delta}\nabla_{\beta}\Phi  \nn\\&&+ c_{308} H_{\beta}{}^{\epsilon \varepsilon} H_{\delta \epsilon \varepsilon} \nabla^{\alpha}\Phi \nabla^{\beta}\Phi \nabla^{\gamma}\nabla_{\alpha}\Phi \nabla^{\delta}\nabla_{\gamma}\Phi + c_{330} H_{\gamma}{}^{\epsilon \varepsilon} H_{\delta \epsilon \varepsilon} \nabla^{\alpha}\Phi \nabla_{\beta}\nabla_{\alpha}\Phi \nabla^{\beta}\Phi \nabla^{\delta}\nabla^{\gamma}\Phi \nn\\&& + c_{358} H_{\alpha}{}^{\epsilon \varepsilon} H_{\beta \epsilon \varepsilon} \nabla^{\alpha}\Phi \nabla^{\beta}\Phi \nabla_{\delta}\nabla_{\gamma}\Phi \nabla^{\delta}\nabla^{\gamma}\Phi + c_{371} H_{\alpha \delta}{}^{\varepsilon} H_{\beta \epsilon \varepsilon} \nabla^{\alpha}\Phi \nabla^{\beta}\Phi \nabla^{\delta}\nabla^{\gamma}\Phi \nabla^{\epsilon}\nabla_{\gamma}\Phi \nn\\&& + c_{376} H_{\beta \delta}{}^{\varepsilon} H_{\gamma \epsilon \varepsilon} \nabla^{\alpha}\Phi \nabla^{\beta}\Phi \nabla^{\gamma}\nabla_{\alpha}\Phi \nabla^{\epsilon}\nabla^{\delta}\Phi + c_{380} H_{\beta \delta}{}^{\varepsilon} H_{\gamma \epsilon \varepsilon} \nabla_{\alpha}\Phi \nabla^{\alpha}\Phi \nabla^{\gamma}\nabla^{\beta}\Phi \nabla^{\epsilon}\nabla^{\delta}\Phi  \nn\\&&+ c_{436} H_{\alpha \gamma \epsilon} H_{\beta \delta \varepsilon} \nabla^{\alpha}\Phi \nabla^{\beta}\Phi \nabla^{\delta}\nabla^{\gamma}\Phi \nabla^{\varepsilon}\nabla^{\epsilon}\Phi\labell{T24}
\eeqa
\beqa
{\cal L}_3^{H^2(\prt\prt\Phi)^3}&\!\!\!\!\!=\!\!\!\!\! & c_{273} H_{\delta \epsilon \varepsilon} H^{\delta \epsilon \varepsilon} \nabla^{\beta}\nabla^{\alpha}\Phi \nabla_{\gamma}\nabla_{\beta}\Phi \nabla^{\gamma}\nabla_{\alpha}\Phi + c_{300} H_{\gamma}{}^{\epsilon \varepsilon} H_{\delta \epsilon \varepsilon} \nabla^{\beta}\nabla^{\alpha}\Phi \nabla^{\gamma}\nabla_{\alpha}\Phi \nabla^{\delta}\nabla_{\beta}\Phi  \nn\\&&+ c_{349} H_{\gamma}{}^{\epsilon \varepsilon} H_{\delta \epsilon \varepsilon} \nabla_{\beta}\nabla_{\alpha}\Phi \nabla^{\beta}\nabla^{\alpha}\Phi \nabla^{\delta}\nabla^{\gamma}\Phi + c_{378} H_{\beta \delta}{}^{\varepsilon} H_{\gamma \epsilon \varepsilon} \nabla^{\beta}\nabla^{\alpha}\Phi \nabla^{\gamma}\nabla_{\alpha}\Phi \nabla^{\epsilon}\nabla^{\delta}\Phi  \nn\\&&+ c_{437} H_{\alpha \gamma \epsilon} H_{\beta \delta \varepsilon} \nabla^{\beta}\nabla^{\alpha}\Phi \nabla^{\delta}\nabla^{\gamma}\Phi \nabla^{\varepsilon}\nabla^{\epsilon}\Phi\labell{T25}
\eeqa
\beqa
{\cal L}_3^{H\prt H R\prt\Phi\prt\prt\Phi}&\!\!\!\!\!=\!\!\!\!\! & c_{281} H_{\alpha}{}^{\delta \epsilon} R_{\gamma \varepsilon \epsilon \mu} \nabla^{\alpha}\Phi \nabla^{\gamma}\nabla^{\beta}\Phi \nabla_{\delta}H_{\beta}{}^{\varepsilon \mu} + c_{363} H^{\delta \epsilon \varepsilon} R_{\alpha \mu \gamma \varepsilon} \nabla^{\alpha}\Phi \nabla^{\gamma}\nabla^{\beta}\Phi \nabla_{\epsilon}H_{\beta \delta}{}^{\mu}\nn\\&& + c_{471} H_{\beta}{}^{\delta \epsilon} R_{\gamma \varepsilon \epsilon \mu} \nabla^{\alpha}\Phi \nabla^{\gamma}\nabla^{\beta}\Phi \nabla^{\mu}H_{\alpha \delta}{}^{\varepsilon} + c_{479} H^{\gamma \delta \epsilon} R_{\delta \varepsilon \epsilon \mu} \nabla^{\alpha}\Phi \nabla^{\beta}\nabla_{\alpha}\Phi \nabla^{\mu}H_{\beta \gamma}{}^{\varepsilon}\nn\\&& + c_{482} H^{\delta \epsilon \varepsilon} R_{\alpha \mu \gamma \varepsilon} \nabla^{\alpha}\Phi \nabla^{\gamma}\nabla^{\beta}\Phi \nabla^{\mu}H_{\beta \delta \epsilon} + c_{483} H_{\alpha}{}^{\delta \epsilon} R_{\gamma \varepsilon \epsilon \mu} \nabla^{\alpha}\Phi \nabla^{\gamma}\nabla^{\beta}\Phi \nabla^{\mu}H_{\beta \delta}{}^{\varepsilon} \nn\\&&+ c_{484} H_{\alpha}{}^{\delta \epsilon} R_{\gamma \mu \epsilon \varepsilon} \nabla^{\alpha}\Phi \nabla^{\gamma}\nabla^{\beta}\Phi \nabla^{\mu}H_{\beta \delta}{}^{\varepsilon} + c_{495} H^{\gamma \delta \epsilon} R_{\beta \varepsilon \epsilon \mu} \nabla^{\alpha}\Phi \nabla^{\beta}\nabla_{\alpha}\Phi \nabla^{\mu}H_{\gamma \delta}{}^{\varepsilon}\nn\\&& + c_{496} H^{\gamma \delta \epsilon} R_{\beta \mu \epsilon \varepsilon} \nabla^{\alpha}\Phi \nabla^{\beta}\nabla_{\alpha}\Phi \nabla^{\mu}H_{\gamma \delta}{}^{\varepsilon} + c_{497} H_{\beta}{}^{\delta \epsilon} R_{\alpha \varepsilon \epsilon \mu} \nabla^{\alpha}\Phi \nabla^{\gamma}\nabla^{\beta}\Phi \nabla^{\mu}H_{\gamma \delta}{}^{\varepsilon} \nn\\&&+ c_{498} H_{\beta}{}^{\delta \epsilon} R_{\alpha \mu \epsilon \varepsilon} \nabla^{\alpha}\Phi \nabla^{\gamma}\nabla^{\beta}\Phi \nabla^{\mu}H_{\gamma \delta}{}^{\varepsilon} + c_{505} H_{\beta}{}^{\gamma \delta} R_{\delta \mu \epsilon \varepsilon} \nabla^{\alpha}\Phi \nabla^{\beta}\nabla_{\alpha}\Phi \nabla^{\mu}H_{\gamma}{}^{\epsilon \varepsilon}\nn\\&& + c_{507} H_{\alpha \beta}{}^{\delta} R_{\delta \mu \epsilon \varepsilon} \nabla^{\alpha}\Phi \nabla^{\gamma}\nabla^{\beta}\Phi \nabla^{\mu}H_{\gamma}{}^{\epsilon \varepsilon} + c_{510} H_{\beta}{}^{\delta \epsilon} R_{\alpha \varepsilon \gamma \mu} \nabla^{\alpha}\Phi \nabla^{\gamma}\nabla^{\beta}\Phi \nabla^{\mu}H_{\delta \epsilon}{}^{\varepsilon} \nn\\&&+ c_{511} H_{\beta}{}^{\delta \epsilon} R_{\alpha \mu \gamma \varepsilon} \nabla^{\alpha}\Phi \nabla^{\gamma}\nabla^{\beta}\Phi \nabla^{\mu}H_{\delta \epsilon}{}^{\varepsilon} + c_{512} H_{\alpha}{}^{\delta \epsilon} R_{\beta \varepsilon \gamma \mu} \nabla^{\alpha}\Phi \nabla^{\gamma}\nabla^{\beta}\Phi \nabla^{\mu}H_{\delta \epsilon}{}^{\varepsilon}\nn\\&& + c_{513} H_{\alpha \beta}{}^{\delta} R_{\gamma \mu \epsilon \varepsilon} \nabla^{\alpha}\Phi \nabla^{\gamma}\nabla^{\beta}\Phi \nabla^{\mu}H_{\delta}{}^{\epsilon \varepsilon}\labell{T26}
\eeqa
\beqa
{\cal L}_3^{(\prt H)^2 R^2}&\!\!\!\!\!=\!\!\!\!\! & c_{282} R_{\epsilon \mu \varepsilon \zeta} R^{\epsilon \varepsilon \mu \zeta} \nabla_{\delta}H_{\alpha \beta \gamma} \nabla^{\delta}H^{\alpha \beta \gamma} + c_{283} R_{\gamma}{}^{\varepsilon \mu \zeta} R_{\epsilon \mu \varepsilon \zeta} \nabla_{\delta}H_{\alpha \beta}{}^{\epsilon} \nabla^{\delta}H^{\alpha \beta \gamma} \nn\\&&+ c_{284} R_{\beta}{}^{\mu}{}_{\epsilon}{}^{\zeta} R_{\gamma \mu \varepsilon \zeta} \nabla_{\delta}H_{\alpha}{}^{\epsilon \varepsilon} \nabla^{\delta}H^{\alpha \beta \gamma} + c_{285} R_{\beta}{}^{\mu}{}_{\epsilon}{}^{\zeta} R_{\gamma \zeta \varepsilon \mu} \nabla_{\delta}H_{\alpha}{}^{\epsilon \varepsilon} \nabla^{\delta}H^{\alpha \beta \gamma} \nn\\&&+ c_{286} R_{\beta}{}^{\mu}{}_{\gamma}{}^{\zeta} R_{\epsilon \mu \varepsilon \zeta} \nabla_{\delta}H_{\alpha}{}^{\epsilon \varepsilon} \nabla^{\delta}H^{\alpha \beta \gamma} + c_{369} R_{\gamma}{}^{\varepsilon \mu \zeta} R_{\epsilon \mu \varepsilon \zeta} \nabla^{\delta}H^{\alpha \beta \gamma} \nabla^{\epsilon}H_{\alpha \beta \delta} \nn\\&&+ c_{407} R_{\gamma}{}^{\mu}{}_{\epsilon}{}^{\zeta} R_{\delta \mu \varepsilon \zeta} \nabla^{\delta}H^{\alpha \beta \gamma} \nabla^{\varepsilon}H_{\alpha \beta}{}^{\epsilon} + c_{408} R_{\gamma}{}^{\mu}{}_{\epsilon}{}^{\zeta} R_{\delta \zeta \varepsilon \mu} \nabla^{\delta}H^{\alpha \beta \gamma} \nabla^{\varepsilon}H_{\alpha \beta}{}^{\epsilon} \nn\\&&+ c_{409} R_{\gamma}{}^{\mu}{}_{\delta}{}^{\zeta} R_{\epsilon \zeta \varepsilon \mu} \nabla^{\delta}H^{\alpha \beta \gamma} \nabla^{\varepsilon}H_{\alpha \beta}{}^{\epsilon} + c_{416} R_{\beta}{}^{\mu}{}_{\epsilon}{}^{\zeta} R_{\gamma \mu \varepsilon \zeta} \nabla^{\delta}H^{\alpha \beta \gamma} \nabla^{\varepsilon}H_{\alpha \delta}{}^{\epsilon} \nn\\&&+ c_{417} R_{\beta}{}^{\mu}{}_{\epsilon}{}^{\zeta} R_{\gamma \zeta \varepsilon \mu} \nabla^{\delta}H^{\alpha \beta \gamma} \nabla^{\varepsilon}H_{\alpha \delta}{}^{\epsilon} + c_{418} R_{\beta}{}^{\mu}{}_{\gamma}{}^{\zeta} R_{\epsilon \mu \varepsilon \zeta} \nabla^{\delta}H^{\alpha \beta \gamma} \nabla^{\varepsilon}H_{\alpha \delta}{}^{\epsilon} \nn\\&&+ c_{475} R_{\beta \mu \epsilon}{}^{\zeta} R_{\gamma \zeta \delta \varepsilon} \nabla^{\delta}H^{\alpha \beta \gamma} \nabla^{\mu}H_{\alpha}{}^{\epsilon \varepsilon} + c_{476} R_{\beta \epsilon \gamma}{}^{\zeta} R_{\delta \mu \varepsilon \zeta} \nabla^{\delta}H^{\alpha \beta \gamma} \nabla^{\mu}H_{\alpha}{}^{\epsilon \varepsilon} \nn\\&&+ c_{477} R_{\beta \gamma \epsilon}{}^{\zeta} R_{\delta \zeta \varepsilon \mu} \nabla^{\delta}H^{\alpha \beta \gamma} \nabla^{\mu}H_{\alpha}{}^{\epsilon \varepsilon} + c_{478} R_{\beta \delta \gamma}{}^{\zeta} R_{\epsilon \mu \varepsilon \zeta} \nabla^{\delta}H^{\alpha \beta \gamma} \nabla^{\mu}H_{\alpha}{}^{\epsilon \varepsilon}\nn\\&& + c_{514} R_{\alpha \epsilon \beta}{}^{\zeta} R_{\gamma \mu \varepsilon \zeta} \nabla^{\delta}H^{\alpha \beta \gamma} \nabla^{\mu}H_{\delta}{}^{\epsilon \varepsilon} + c_{515} R_{\alpha \mu \beta}{}^{\zeta} R_{\gamma \zeta \epsilon \varepsilon} \nabla^{\delta}H^{\alpha \beta \gamma} \nabla^{\mu}H_{\delta}{}^{\epsilon \varepsilon}\nn\\&& + c_{516} R_{\alpha \epsilon \beta}{}^{\zeta} R_{\gamma \zeta \varepsilon \mu} \nabla^{\delta}H^{\alpha \beta \gamma} \nabla^{\mu}H_{\delta}{}^{\epsilon \varepsilon} + c_{697} R_{\alpha \epsilon \beta \varepsilon} R_{\gamma \mu \delta \zeta} \nabla^{\delta}H^{\alpha \beta \gamma} \nabla^{\zeta}H^{\epsilon \varepsilon \mu}\nn\\&& + c_{698} R_{\alpha \epsilon \beta \varepsilon} R_{\gamma \zeta \delta \mu} \nabla^{\delta}H^{\alpha \beta \gamma} \nabla^{\zeta}H^{\epsilon \varepsilon \mu} + c_{699} R_{\alpha \beta \delta \epsilon} R_{\gamma \zeta \varepsilon \mu} \nabla^{\delta}H^{\alpha \beta \gamma} \nabla^{\zeta}H^{\epsilon \varepsilon \mu}\labell{T27}
\eeqa
\beqa
{\cal L}_3^{H\prt H(\prt\Phi)^5}&\!\!\!\!\!=\!\!\!\!\! & c_{291} H_{\beta}{}^{\epsilon \varepsilon} \nabla_{\alpha}\Phi \nabla^{\alpha}\Phi \nabla^{\beta}\Phi \nabla^{\gamma}\Phi \nabla_{\delta}H_{\gamma \epsilon \varepsilon} \nabla^{\delta}\Phi\labell{T28}
\eeqa
\beqa
{\cal L}_3^{(\prt H)^2(\prt\Phi)^4}&\!\!\!\!\!=\!\!\!\!\! & c_{292} \nabla^{\alpha}\Phi \nabla_{\beta}H_{\alpha}{}^{\epsilon \varepsilon} \nabla^{\beta}\Phi \nabla^{\gamma}\Phi \nabla_{\delta}H_{\gamma \epsilon \varepsilon} \nabla^{\delta}\Phi + c_{425} \nabla_{\alpha}\Phi \nabla^{\alpha}\Phi \nabla^{\beta}\Phi \nabla^{\gamma}\Phi \nabla_{\epsilon}H_{\gamma \delta \varepsilon} \nabla^{\varepsilon}H_{\beta}{}^{\delta \epsilon}\nn\\&& + c_{428} \nabla_{\alpha}\Phi \nabla^{\alpha}\Phi \nabla^{\beta}\Phi \nabla^{\gamma}\Phi \nabla_{\varepsilon}H_{\gamma \delta \epsilon} \nabla^{\varepsilon}H_{\beta}{}^{\delta \epsilon} + c_{433} \nabla_{\alpha}\Phi \nabla^{\alpha}\Phi \nabla_{\beta}\Phi \nabla^{\beta}\Phi \nabla_{\varepsilon}H_{\gamma \delta \epsilon} \nabla^{\varepsilon}H^{\gamma \delta \epsilon}\labell{T29}\nn
\eeqa
\beqa
{\cal L}_3^{(\prt\Phi)^8
}&\!\!\!\!\!=\!\!\!\!\! & c_{293} \nabla_{\alpha}\Phi \nabla^{\alpha}\Phi \nabla_{\beta}\Phi \nabla^{\beta}\Phi \nabla_{\gamma}\Phi \nabla^{\gamma}\Phi \nabla_{\delta}\Phi \nabla^{\delta}\Phi\labell{T30}
\eeqa
\beqa
{\cal L}_3^{(\prt\Phi)^6\prt\prt\Phi}&\!\!\!\!\!=\!\!\!\!\! & c_{294} \nabla_{\alpha}\Phi \nabla^{\alpha}\Phi \nabla_{\beta}\Phi \nabla^{\beta}\Phi \nabla^{\gamma}\Phi \nabla_{\delta}\nabla_{\gamma}\Phi \nabla^{\delta}\Phi\labell{T31}
\eeqa
\beqa
{\cal L}_3^{(\prt\Phi)^4(\prt\prt\Phi)^2}&\!\!\!\!\!=\!\!\!\!\! &   c_{295} \nabla^{\alpha}\Phi \nabla_{\beta}\nabla_{\alpha}\Phi \nabla^{\beta}\Phi \nabla^{\gamma}\Phi \nabla_{\delta}\nabla_{\gamma}\Phi \nabla^{\delta}\Phi + c_{304} \nabla_{\alpha}\Phi \nabla^{\alpha}\Phi \nabla^{\beta}\Phi \nabla^{\gamma}\Phi \nabla_{\delta}\nabla_{\gamma}\Phi \nabla^{\delta}\nabla_{\beta}\Phi \nn\\&&+ c_{359} \nabla_{\alpha}\Phi \nabla^{\alpha}\Phi \nabla_{\beta}\Phi \nabla^{\beta}\Phi \nabla_{\delta}\nabla_{\gamma}\Phi \nabla^{\delta}\nabla^{\gamma}\Phi\labell{T32}
\eeqa
\beqa
&&{\cal L}_3^{H\prt H(\prt\Phi)^3\prt\prt\Phi}=\nn\\&&  c_{296} H_{\delta}{}^{\epsilon \varepsilon} \nabla^{\alpha}\Phi \nabla^{\beta}\Phi \nabla_{\gamma}H_{\beta \epsilon \varepsilon} \nabla^{\gamma}\Phi \nabla^{\delta}\nabla_{\alpha}\Phi + c_{297} H_{\beta}{}^{\epsilon \varepsilon} \nabla^{\alpha}\Phi \nabla^{\beta}\Phi \nabla^{\gamma}\Phi \nabla_{\delta}H_{\gamma \epsilon \varepsilon} \nabla^{\delta}\nabla_{\alpha}\Phi\nn\\&& + c_{350} H_{\gamma}{}^{\epsilon \varepsilon} \nabla_{\alpha}\Phi \nabla^{\alpha}\Phi \nabla^{\beta}\Phi \nabla_{\delta}H_{\beta \epsilon \varepsilon} \nabla^{\delta}\nabla^{\gamma}\Phi + c_{354} H_{\beta}{}^{\epsilon \varepsilon} \nabla_{\alpha}\Phi \nabla^{\alpha}\Phi \nabla^{\beta}\Phi \nabla_{\delta}H_{\gamma \epsilon \varepsilon} \nabla^{\delta}\nabla^{\gamma}\Phi \nn\\&&+ c_{375} H_{\alpha \delta}{}^{\varepsilon} \nabla^{\alpha}\Phi \nabla^{\beta}\Phi \nabla_{\gamma}H_{\beta \epsilon \varepsilon} \nabla^{\gamma}\Phi \nabla^{\epsilon}\nabla^{\delta}\Phi + c_{387} H_{\gamma}{}^{\epsilon \varepsilon} \nabla_{\alpha}\Phi \nabla^{\alpha}\Phi \nabla^{\beta}\Phi \nabla^{\delta}\nabla^{\gamma}\Phi \nabla_{\varepsilon}H_{\beta \delta \epsilon} \nn\\&&+ c_{392} H_{\beta}{}^{\epsilon \varepsilon} \nabla^{\alpha}\Phi \nabla^{\beta}\Phi \nabla^{\gamma}\Phi \nabla^{\delta}\nabla_{\alpha}\Phi \nabla_{\varepsilon}H_{\gamma \delta \epsilon}\labell{T33}
\eeqa
\beqa
{\cal L}_3^{H\prt H\prt\Phi(\prt\prt\Phi)^2}&\!\!\!\!\!=\!\!\!\!\! & c_{302} H_{\gamma}{}^{\epsilon \varepsilon} \nabla^{\alpha}\Phi \nabla^{\gamma}\nabla^{\beta}\Phi \nabla_{\delta}H_{\alpha \epsilon \varepsilon} \nabla^{\delta}\nabla_{\beta}\Phi + c_{303} H_{\alpha}{}^{\epsilon \varepsilon} \nabla^{\alpha}\Phi \nabla^{\gamma}\nabla^{\beta}\Phi \nabla_{\delta}H_{\gamma \epsilon \varepsilon} \nabla^{\delta}\nabla_{\beta}\Phi\nn\\&& + c_{351} H_{\gamma}{}^{\epsilon \varepsilon} \nabla^{\alpha}\Phi \nabla^{\beta}\nabla_{\alpha}\Phi \nabla_{\delta}H_{\beta \epsilon \varepsilon} \nabla^{\delta}\nabla^{\gamma}\Phi + c_{356} H_{\beta}{}^{\epsilon \varepsilon} \nabla^{\alpha}\Phi \nabla^{\beta}\nabla_{\alpha}\Phi \nabla_{\delta}H_{\gamma \epsilon \varepsilon} \nabla^{\delta}\nabla^{\gamma}\Phi\nn\\&& + c_{382} H_{\beta \delta}{}^{\varepsilon} \nabla^{\alpha}\Phi \nabla^{\gamma}\nabla^{\beta}\Phi \nabla_{\epsilon}H_{\alpha \gamma \varepsilon} \nabla^{\epsilon}\nabla^{\delta}\Phi + c_{383} H_{\alpha \beta}{}^{\varepsilon} \nabla^{\alpha}\Phi \nabla^{\gamma}\nabla^{\beta}\Phi \nabla_{\epsilon}H_{\gamma \delta \varepsilon} \nabla^{\epsilon}\nabla^{\delta}\Phi \nn\\&&+ c_{384} H_{\beta \delta}{}^{\varepsilon} \nabla^{\alpha}\Phi \nabla^{\gamma}\nabla^{\beta}\Phi \nabla^{\epsilon}\nabla^{\delta}\Phi \nabla_{\varepsilon}H_{\alpha \gamma \epsilon} + c_{385} H_{\gamma}{}^{\epsilon \varepsilon} \nabla^{\alpha}\Phi \nabla^{\gamma}\nabla^{\beta}\Phi \nabla^{\delta}\nabla_{\beta}\Phi \nabla_{\varepsilon}H_{\alpha \delta \epsilon}\nn\\&& + c_{388} H_{\gamma}{}^{\epsilon \varepsilon} \nabla^{\alpha}\Phi \nabla^{\beta}\nabla_{\alpha}\Phi \nabla^{\delta}\nabla^{\gamma}\Phi \nabla_{\varepsilon}H_{\beta \delta \epsilon} + c_{391} H^{\delta \epsilon \varepsilon} \nabla^{\alpha}\Phi \nabla^{\beta}\nabla_{\alpha}\Phi \nabla^{\gamma}\nabla_{\beta}\Phi \nabla_{\varepsilon}H_{\gamma \delta \epsilon}\labell{T34}\nn
\eeqa
\beqa
{\cal L}_3^{(\prt\Phi)^2(\prt\prt\Phi)^3}&\!\!\!\!\!=\!\!\!\!\! & c_{305} \nabla^{\alpha}\Phi \nabla^{\beta}\Phi \nabla^{\gamma}\nabla_{\alpha}\Phi \nabla_{\delta}\nabla_{\gamma}\Phi \nabla^{\delta}\nabla_{\beta}\Phi + c_{307} \nabla_{\alpha}\Phi \nabla^{\alpha}\Phi \nabla^{\gamma}\nabla^{\beta}\Phi \nabla_{\delta}\nabla_{\gamma}\Phi \nabla^{\delta}\nabla_{\beta}\Phi \nn\\&& + c_{360} \nabla^{\alpha}\Phi \nabla_{\beta}\nabla_{\alpha}\Phi \nabla^{\beta}\Phi \nabla_{\delta}\nabla_{\gamma}\Phi \nabla^{\delta}\nabla^{\gamma}\Phi\labell{T35}
\eeqa
\beqa
{\cal L}_3^{(\prt\prt\Phi)^4}&\!\!\!\!\!=\!\!\!\!\! & c_{306} \nabla^{\beta}\nabla^{\alpha}\Phi \nabla^{\gamma}\nabla_{\alpha}\Phi \nabla_{\delta}\nabla_{\gamma}\Phi \nabla^{\delta}\nabla_{\beta}\Phi + c_{361} \nabla_{\beta}\nabla_{\alpha}\Phi \nabla^{\beta}\nabla^{\alpha}\Phi \nabla_{\delta}\nabla_{\gamma}\Phi \nabla^{\delta}\nabla^{\gamma}\Phi\labell{T36}
\eeqa
\beqa
{\cal L}_3^{(\prt H)^2(\prt\Phi)^2\prt\prt\Phi}&\!\!\!\!\!=\!\!\!\!\! & c_{352} \nabla^{\alpha}\Phi \nabla^{\beta}\Phi \nabla_{\gamma}H_{\alpha}{}^{\epsilon \varepsilon} \nabla_{\delta}H_{\beta \epsilon \varepsilon} \nabla^{\delta}\nabla^{\gamma}\Phi + c_{355} \nabla^{\alpha}\Phi \nabla_{\beta}H_{\alpha}{}^{\epsilon \varepsilon} \nabla^{\beta}\Phi \nabla_{\delta}H_{\gamma \epsilon \varepsilon} \nabla^{\delta}\nabla^{\gamma}\Phi\nn\\&& + c_{411} \nabla^{\alpha}\Phi \nabla^{\beta}\Phi \nabla_{\delta}H_{\beta \epsilon \varepsilon} \nabla^{\delta}\nabla^{\gamma}\Phi \nabla^{\varepsilon}H_{\alpha \gamma}{}^{\epsilon} + c_{412} \nabla^{\alpha}\Phi \nabla^{\beta}\Phi \nabla^{\delta}\nabla^{\gamma}\Phi \nabla_{\epsilon}H_{\beta \delta \varepsilon} \nabla^{\varepsilon}H_{\alpha \gamma}{}^{\epsilon}\nn\\&& + c_{414} \nabla^{\alpha}\Phi \nabla^{\beta}\Phi \nabla^{\delta}\nabla^{\gamma}\Phi \nabla_{\varepsilon}H_{\beta \delta \epsilon} \nabla^{\varepsilon}H_{\alpha \gamma}{}^{\epsilon} + c_{427} \nabla_{\alpha}\Phi \nabla^{\alpha}\Phi \nabla^{\gamma}\nabla^{\beta}\Phi \nabla_{\epsilon}H_{\gamma \delta \varepsilon} \nabla^{\varepsilon}H_{\beta}{}^{\delta \epsilon} \nn\\&&+ c_{429} \nabla^{\alpha}\Phi \nabla^{\beta}\Phi \nabla^{\gamma}\nabla_{\alpha}\Phi \nabla_{\varepsilon}H_{\gamma \delta \epsilon} \nabla^{\varepsilon}H_{\beta}{}^{\delta \epsilon} + c_{431} \nabla_{\alpha}\Phi \nabla^{\alpha}\Phi \nabla^{\gamma}\nabla^{\beta}\Phi \nabla_{\varepsilon}H_{\gamma \delta \epsilon} \nabla^{\varepsilon}H_{\beta}{}^{\delta \epsilon} \nn\\&&+ c_{432} \nabla^{\alpha}\Phi \nabla_{\beta}H_{\delta \epsilon \varepsilon} \nabla^{\beta}\Phi \nabla^{\gamma}\nabla_{\alpha}\Phi \nabla^{\varepsilon}H_{\gamma}{}^{\delta \epsilon} + c_{434} \nabla^{\alpha}\Phi \nabla_{\beta}\nabla_{\alpha}\Phi \nabla^{\beta}\Phi \nabla_{\varepsilon}H_{\gamma \delta \epsilon} \nabla^{\varepsilon}H^{\gamma \delta \epsilon}\labell{T37}\nn
\eeqa
\beqa
{\cal L}_3^{(\prt H)^2(\prt\prt\Phi)^2}&\!\!\!\!\!=\!\!\!\!\! & c_{353} \nabla^{\beta}\nabla^{\alpha}\Phi \nabla_{\gamma}H_{\alpha}{}^{\epsilon \varepsilon} \nabla_{\delta}H_{\beta \epsilon \varepsilon} \nabla^{\delta}\nabla^{\gamma}\Phi + c_{357} \nabla_{\beta}H_{\alpha}{}^{\epsilon \varepsilon} \nabla^{\beta}\nabla^{\alpha}\Phi \nabla_{\delta}H_{\gamma \epsilon \varepsilon} \nabla^{\delta}\nabla^{\gamma}\Phi \nn\\&&+ c_{413} \nabla^{\beta}\nabla^{\alpha}\Phi \nabla^{\delta}\nabla^{\gamma}\Phi \nabla_{\epsilon}H_{\beta \delta \varepsilon} \nabla^{\varepsilon}H_{\alpha \gamma}{}^{\epsilon} + c_{415} \nabla^{\beta}\nabla^{\alpha}\Phi \nabla^{\delta}\nabla^{\gamma}\Phi \nabla_{\varepsilon}H_{\beta \delta \epsilon} \nabla^{\varepsilon}H_{\alpha \gamma}{}^{\epsilon}\nn\\&& + c_{426} \nabla^{\beta}\nabla^{\alpha}\Phi \nabla^{\gamma}\nabla_{\alpha}\Phi \nabla_{\epsilon}H_{\gamma \delta \varepsilon} \nabla^{\varepsilon}H_{\beta}{}^{\delta \epsilon} + c_{430} \nabla^{\beta}\nabla^{\alpha}\Phi \nabla^{\gamma}\nabla_{\alpha}\Phi \nabla_{\varepsilon}H_{\gamma \delta \epsilon} \nabla^{\varepsilon}H_{\beta}{}^{\delta \epsilon} \nn\\&&+ c_{435} \nabla_{\beta}\nabla_{\alpha}\Phi \nabla^{\beta}\nabla^{\alpha}\Phi \nabla_{\varepsilon}H_{\gamma \delta \epsilon} \nabla^{\varepsilon}H^{\gamma \delta \epsilon}\labell{T38}
\eeqa
\beqa
{\cal L}_3^{H^3\prt H R\prt\Phi}&\!\!\!\!\!=\!\!\!\!\! & c_{362} H_{\alpha}{}^{\beta \gamma} H_{\delta}{}^{\mu \zeta} H^{\delta \epsilon \varepsilon} R_{\varepsilon \eta \mu \zeta} \nabla^{\alpha}\Phi \nabla_{\epsilon}H_{\beta \gamma}{}^{\eta} + c_{366} H_{\alpha}{}^{\beta \gamma} H_{\beta}{}^{\delta \epsilon} H_{\delta}{}^{\varepsilon \mu} R_{\varepsilon \zeta \mu \eta} \nabla^{\alpha}\Phi \nabla_{\epsilon}H_{\gamma}{}^{\zeta \eta} \nn\\&&+ c_{367} H_{\beta \gamma}{}^{\epsilon} H^{\beta \gamma \delta} H^{\varepsilon \mu \zeta} R_{\alpha \eta \mu \zeta} \nabla^{\alpha}\Phi \nabla_{\epsilon}H_{\delta \varepsilon}{}^{\eta} + c_{386} H_{\beta \gamma}{}^{\epsilon} H^{\beta \gamma \delta} H_{\delta}{}^{\varepsilon \mu} R_{\epsilon \zeta \mu \eta} \nabla^{\alpha}\Phi \nabla_{\varepsilon}H_{\alpha}{}^{\zeta \eta}\nn\\&& + c_{389} H_{\alpha}{}^{\beta \gamma} H_{\delta}{}^{\mu \zeta} H^{\delta \epsilon \varepsilon} R_{\gamma \eta \mu \zeta} \nabla^{\alpha}\Phi \nabla_{\varepsilon}H_{\beta \epsilon}{}^{\eta} + c_{390} H_{\alpha}{}^{\beta \gamma} H_{\delta \epsilon}{}^{\mu} H^{\delta \epsilon \varepsilon} R_{\gamma \zeta \mu \eta} \nabla^{\alpha}\Phi \nabla_{\varepsilon}H_{\beta}{}^{\zeta \eta}\nn\\&& + c_{393} H_{\alpha}{}^{\beta \gamma} H_{\beta}{}^{\delta \epsilon} H_{\delta}{}^{\varepsilon \mu} R_{\epsilon \zeta \mu \eta} \nabla^{\alpha}\Phi \nabla_{\varepsilon}H_{\gamma}{}^{\zeta \eta} + c_{394} H_{\alpha}{}^{\beta \gamma} H_{\beta}{}^{\delta \epsilon} H^{\varepsilon \mu \zeta} R_{\gamma \eta \mu \zeta} \nabla^{\alpha}\Phi \nabla_{\varepsilon}H_{\delta \epsilon}{}^{\eta} \nn\\&&+ c_{397} H_{\alpha}{}^{\beta \gamma} H_{\beta}{}^{\delta \epsilon} H_{\gamma}{}^{\varepsilon \mu} R_{\epsilon \zeta \mu \eta} \nabla^{\alpha}\Phi \nabla_{\varepsilon}H_{\delta}{}^{\zeta \eta} + c_{402} H_{\beta \gamma}{}^{\epsilon} H^{\beta \gamma \delta} H_{\delta}{}^{\varepsilon \mu} R_{\alpha \zeta \mu \eta} \nabla^{\alpha}\Phi \nabla_{\varepsilon}H_{\epsilon}{}^{\zeta \eta} \nn\\&&+ c_{403} H_{\alpha}{}^{\beta \gamma} H_{\beta}{}^{\delta \epsilon} H_{\delta}{}^{\varepsilon \mu} R_{\gamma \zeta \mu \eta} \nabla^{\alpha}\Phi \nabla_{\varepsilon}H_{\epsilon}{}^{\zeta \eta} + c_{438} H_{\beta}{}^{\epsilon \varepsilon} H^{\beta \gamma \delta} H^{\mu \zeta \eta} R_{\epsilon \zeta \varepsilon \eta} \nabla^{\alpha}\Phi \nabla_{\mu}H_{\alpha \gamma \delta}\nn\\&& + c_{439} H_{\beta}{}^{\epsilon \varepsilon} H^{\beta \gamma \delta} H^{\mu \zeta \eta} R_{\delta \zeta \varepsilon \eta} \nabla^{\alpha}\Phi \nabla_{\mu}H_{\alpha \gamma \epsilon} + c_{440} H_{\beta}{}^{\epsilon \varepsilon} H^{\beta \gamma \delta} H_{\gamma}{}^{\mu \zeta} R_{\delta \zeta \varepsilon \eta} \nabla^{\alpha}\Phi \nabla_{\mu}H_{\alpha \epsilon}{}^{\eta} \nn\\&&+ c_{441} H_{\beta}{}^{\epsilon \varepsilon} H^{\beta \gamma \delta} H_{\gamma}{}^{\mu \zeta} R_{\delta \eta \varepsilon \zeta} \nabla^{\alpha}\Phi \nabla_{\mu}H_{\alpha \epsilon}{}^{\eta} + c_{442} H_{\beta \gamma}{}^{\epsilon} H^{\beta \gamma \delta} H^{\varepsilon \mu \zeta} R_{\delta \zeta \epsilon \eta} \nabla^{\alpha}\Phi \nabla_{\mu}H_{\alpha \varepsilon}{}^{\eta} \nn\\&&+ c_{444} H_{\alpha}{}^{\beta \gamma} H^{\delta \epsilon \varepsilon} H^{\mu \zeta \eta} R_{\epsilon \zeta \varepsilon \eta} \nabla^{\alpha}\Phi \nabla_{\mu}H_{\beta \gamma \delta} + c_{445} H_{\alpha}{}^{\beta \gamma} H^{\delta \epsilon \varepsilon} H^{\mu \zeta \eta} R_{\gamma \zeta \varepsilon \eta} \nabla^{\alpha}\Phi \nabla_{\mu}H_{\beta \delta \epsilon} \nn\\&&+ c_{446} H_{\alpha}{}^{\beta \gamma} H_{\delta}{}^{\mu \zeta} H^{\delta \epsilon \varepsilon} R_{\gamma \zeta \varepsilon \eta} \nabla^{\alpha}\Phi \nabla_{\mu}H_{\beta \epsilon}{}^{\eta} + c_{447} H_{\alpha}{}^{\beta \gamma} H_{\delta}{}^{\mu \zeta} H^{\delta \epsilon \varepsilon} R_{\gamma \eta \varepsilon \zeta} \nabla^{\alpha}\Phi \nabla_{\mu}H_{\beta \epsilon}{}^{\eta} \nn\\&&+ c_{448} H_{\beta}{}^{\epsilon \varepsilon} H^{\beta \gamma \delta} H^{\mu \zeta \eta} R_{\alpha \zeta \varepsilon \eta} \nabla^{\alpha}\Phi \nabla_{\mu}H_{\gamma \delta \epsilon} + c_{451} H_{\alpha}{}^{\beta \gamma} H_{\beta}{}^{\delta \epsilon} H^{\varepsilon \mu \zeta} R_{\delta \zeta \epsilon \eta} \nabla^{\alpha}\Phi \nabla_{\mu}H_{\gamma \varepsilon}{}^{\eta} \nn\\&&+ c_{455} H_{\alpha}{}^{\beta \gamma} H^{\delta \epsilon \varepsilon} H^{\mu \zeta \eta} R_{\beta \zeta \gamma \eta} \nabla^{\alpha}\Phi \nabla_{\mu}H_{\delta \epsilon \varepsilon} + c_{456} H_{\beta}{}^{\epsilon \varepsilon} H^{\beta \gamma \delta} H_{\gamma}{}^{\mu \zeta} R_{\alpha \zeta \varepsilon \eta} \nabla^{\alpha}\Phi \nabla_{\mu}H_{\delta \epsilon}{}^{\eta}\nn\\&& + c_{457} H_{\beta}{}^{\epsilon \varepsilon} H^{\beta \gamma \delta} H_{\gamma}{}^{\mu \zeta} R_{\alpha \eta \varepsilon \zeta} \nabla^{\alpha}\Phi \nabla_{\mu}H_{\delta \epsilon}{}^{\eta} + c_{458} H_{\beta \gamma}{}^{\epsilon} H^{\beta \gamma \delta} H^{\varepsilon \mu \zeta} R_{\alpha \zeta \epsilon \eta} \nabla^{\alpha}\Phi \nabla_{\mu}H_{\delta \varepsilon}{}^{\eta} \nn\\&&+ c_{459} H_{\beta \gamma}{}^{\epsilon} H^{\beta \gamma \delta} H^{\varepsilon \mu \zeta} R_{\alpha \eta \epsilon \zeta} \nabla^{\alpha}\Phi \nabla_{\mu}H_{\delta \varepsilon}{}^{\eta} + c_{460} H_{\alpha}{}^{\beta \gamma} H_{\beta}{}^{\delta \epsilon} H^{\varepsilon \mu \zeta} R_{\gamma \zeta \epsilon \eta} \nabla^{\alpha}\Phi \nabla_{\mu}H_{\delta \varepsilon}{}^{\eta} \nn\\&&+ c_{461} H_{\alpha}{}^{\beta \gamma} H_{\beta}{}^{\delta \epsilon} H^{\varepsilon \mu \zeta} R_{\gamma \eta \epsilon \zeta} \nabla^{\alpha}\Phi \nabla_{\mu}H_{\delta \varepsilon}{}^{\eta} + c_{462} H_{\beta}{}^{\epsilon \varepsilon} H^{\beta \gamma \delta} H_{\gamma}{}^{\mu \zeta} R_{\alpha \zeta \delta \eta} \nabla^{\alpha}\Phi \nabla_{\mu}H_{\epsilon \varepsilon}{}^{\eta} \nn\\&&+ c_{463} H_{\beta}{}^{\epsilon \varepsilon} H^{\beta \gamma \delta} H_{\gamma}{}^{\mu \zeta} R_{\alpha \eta \delta \zeta} \nabla^{\alpha}\Phi \nabla_{\mu}H_{\epsilon \varepsilon}{}^{\eta} + c_{464} H_{\alpha}{}^{\beta \gamma} H_{\delta}{}^{\mu \zeta} H^{\delta \epsilon \varepsilon} R_{\beta \zeta \gamma \eta} \nabla^{\alpha}\Phi \nabla_{\mu}H_{\epsilon \varepsilon}{}^{\eta} \nn\\&&+ c_{470} H_{\alpha}{}^{\beta \gamma} H_{\delta \epsilon}{}^{\mu} H^{\delta \epsilon \varepsilon} R_{\beta \zeta \gamma \eta} \nabla^{\alpha}\Phi \nabla_{\mu}H_{\varepsilon}{}^{\zeta \eta} + c_{524} H_{\beta}{}^{\epsilon \varepsilon} H^{\beta \gamma \delta} H^{\mu \zeta \eta} R_{\delta \eta \epsilon \varepsilon} \nabla^{\alpha}\Phi \nabla_{\zeta}H_{\alpha \gamma \mu} \nn\\&&+ c_{525} H_{\beta}{}^{\epsilon \varepsilon} H^{\beta \gamma \delta} H^{\mu \zeta \eta} R_{\alpha \eta \epsilon \varepsilon} \nabla^{\alpha}\Phi \nabla_{\zeta}H_{\gamma \delta \mu} + c_{526} H_{\beta}{}^{\epsilon \varepsilon} H^{\beta \gamma \delta} H^{\mu \zeta \eta} R_{\alpha \eta \delta \varepsilon} \nabla^{\alpha}\Phi \nabla_{\zeta}H_{\gamma \epsilon \mu}\nn\\&& + c_{527} H_{\alpha}{}^{\beta \gamma} H^{\delta \epsilon \varepsilon} H^{\mu \zeta \eta} R_{\beta \varepsilon \gamma \eta} \nabla^{\alpha}\Phi \nabla_{\zeta}H_{\delta \epsilon \mu} + c_{528} H_{\beta}{}^{\epsilon \varepsilon} H^{\beta \gamma \delta} H^{\mu \zeta \eta} R_{\gamma \epsilon \delta \varepsilon} \nabla^{\alpha}\Phi \nabla_{\eta}H_{\alpha \mu \zeta}\nn\\&& + c_{529} H_{\beta}{}^{\epsilon \varepsilon} H^{\beta \gamma \delta} H^{\mu \zeta \eta} R_{\alpha \epsilon \delta \varepsilon} \nabla^{\alpha}\Phi \nabla_{\eta}H_{\gamma \mu \zeta} + c_{530} H_{\alpha}{}^{\beta \gamma} H_{\delta}{}^{\mu \zeta} H^{\delta \epsilon \varepsilon} R_{\varepsilon \eta \mu \zeta} \nabla^{\alpha}\Phi \nabla^{\eta}H_{\beta \gamma \epsilon}\nn\\&& + c_{531} H_{\alpha}{}^{\beta \gamma} H_{\delta \epsilon}{}^{\mu} H^{\delta \epsilon \varepsilon} R_{\varepsilon \zeta \mu \eta} \nabla^{\alpha}\Phi \nabla^{\eta}H_{\beta \gamma}{}^{\zeta} + c_{532} H_{\alpha}{}^{\beta \gamma} H_{\delta}{}^{\mu \zeta} H^{\delta \epsilon \varepsilon} R_{\gamma \eta \mu \zeta} \nabla^{\alpha}\Phi \nabla^{\eta}H_{\beta \epsilon \varepsilon} \nn\\&&+ c_{533} H_{\alpha}{}^{\beta \gamma} H_{\delta}{}^{\mu \zeta} H^{\delta \epsilon \varepsilon} R_{\gamma \eta \varepsilon \zeta} \nabla^{\alpha}\Phi \nabla^{\eta}H_{\beta \epsilon \mu} + c_{534} H_{\alpha}{}^{\beta \gamma} H_{\delta \epsilon}{}^{\mu} H^{\delta \epsilon \varepsilon} R_{\gamma \zeta \mu \eta} \nabla^{\alpha}\Phi \nabla^{\eta}H_{\beta \varepsilon}{}^{\zeta} \nn\\&&+ c_{535} H_{\alpha}{}^{\beta \gamma} H_{\delta \epsilon}{}^{\mu} H^{\delta \epsilon \varepsilon} R_{\gamma \eta \mu \zeta} \nabla^{\alpha}\Phi \nabla^{\eta}H_{\beta \varepsilon}{}^{\zeta} + c_{536} H_{\alpha}{}^{\beta \gamma} H_{\delta \epsilon \varepsilon} H^{\delta \epsilon \varepsilon} R_{\gamma \eta \mu \zeta} \nabla^{\alpha}\Phi \nabla^{\eta}H_{\beta}{}^{\mu \zeta}\nn\\&& + c_{537} H_{\alpha}{}^{\beta \gamma} H_{\beta}{}^{\delta \epsilon} H^{\varepsilon \mu \zeta} R_{\epsilon \eta \mu \zeta} \nabla^{\alpha}\Phi \nabla^{\eta}H_{\gamma \delta \varepsilon} + c_{538} H_{\alpha}{}^{\beta \gamma} H_{\beta}{}^{\delta \epsilon} H_{\delta}{}^{\varepsilon \mu} R_{\varepsilon \zeta \mu \eta} \nabla^{\alpha}\Phi \nabla^{\eta}H_{\gamma \epsilon}{}^{\zeta} \nn\\&&+ c_{539} H_{\alpha}{}^{\beta \gamma} H_{\beta}{}^{\delta \epsilon} H^{\varepsilon \mu \zeta} R_{\delta \zeta \epsilon \eta} \nabla^{\alpha}\Phi \nabla^{\eta}H_{\gamma \varepsilon \mu} + c_{540} H_{\alpha}{}^{\beta \gamma} H_{\beta}{}^{\delta \epsilon} H_{\delta}{}^{\varepsilon \mu} R_{\epsilon \zeta \mu \eta} \nabla^{\alpha}\Phi \nabla^{\eta}H_{\gamma \varepsilon}{}^{\zeta} \nn\\&&+ c_{541} H_{\alpha}{}^{\beta \gamma} H_{\beta}{}^{\delta \epsilon} H_{\delta}{}^{\varepsilon \mu} R_{\epsilon \eta \mu \zeta} \nabla^{\alpha}\Phi \nabla^{\eta}H_{\gamma \varepsilon}{}^{\zeta} + c_{542} H_{\alpha}{}^{\beta \gamma} H_{\beta}{}^{\delta \epsilon} H_{\delta \epsilon}{}^{\varepsilon} R_{\varepsilon \eta \mu \zeta} \nabla^{\alpha}\Phi \nabla^{\eta}H_{\gamma}{}^{\mu \zeta} \nn\\&&+ c_{543} H_{\beta}{}^{\epsilon \varepsilon} H^{\beta \gamma \delta} H_{\gamma}{}^{\mu \zeta} R_{\alpha \eta \mu \zeta} \nabla^{\alpha}\Phi \nabla^{\eta}H_{\delta \epsilon \varepsilon} + c_{544} H_{\alpha}{}^{\beta \gamma} H_{\beta}{}^{\delta \epsilon} H^{\varepsilon \mu \zeta} R_{\gamma \eta \mu \zeta} \nabla^{\alpha}\Phi \nabla^{\eta}H_{\delta \epsilon \varepsilon}\nn\\&& + c_{545} H_{\alpha}{}^{\beta \gamma} H_{\beta}{}^{\delta \epsilon} H_{\gamma}{}^{\varepsilon \mu} R_{\varepsilon \zeta \mu \eta} \nabla^{\alpha}\Phi \nabla^{\eta}H_{\delta \epsilon}{}^{\zeta} + c_{546} H_{\alpha}{}^{\beta \gamma} H_{\beta \gamma}{}^{\delta} H^{\epsilon \varepsilon \mu} R_{\varepsilon \zeta \mu \eta} \nabla^{\alpha}\Phi \nabla^{\eta}H_{\delta \epsilon}{}^{\zeta}\nn\\&& + c_{547} H_{\beta \gamma}{}^{\epsilon} H^{\beta \gamma \delta} H^{\varepsilon \mu \zeta} R_{\alpha \eta \epsilon \zeta} \nabla^{\alpha}\Phi \nabla^{\eta}H_{\delta \varepsilon \mu} + c_{548} H_{\alpha}{}^{\beta \gamma} H_{\beta}{}^{\delta \epsilon} H^{\varepsilon \mu \zeta} R_{\gamma \zeta \epsilon \eta} \nabla^{\alpha}\Phi \nabla^{\eta}H_{\delta \varepsilon \mu}\nn\\&& + c_{549} H_{\alpha}{}^{\beta \gamma} H_{\beta}{}^{\delta \epsilon} H^{\varepsilon \mu \zeta} R_{\gamma \eta \epsilon \zeta} \nabla^{\alpha}\Phi \nabla^{\eta}H_{\delta \varepsilon \mu} + c_{550} H_{\beta}{}^{\epsilon \varepsilon} H^{\beta \gamma \delta} H_{\gamma \epsilon}{}^{\mu} R_{\alpha \zeta \mu \eta} \nabla^{\alpha}\Phi \nabla^{\eta}H_{\delta \varepsilon}{}^{\zeta}\nn\\&& + c_{551} H_{\beta}{}^{\epsilon \varepsilon} H^{\beta \gamma \delta} H_{\gamma \epsilon}{}^{\mu} R_{\alpha \eta \mu \zeta} \nabla^{\alpha}\Phi \nabla^{\eta}H_{\delta \varepsilon}{}^{\zeta} + c_{552} H_{\alpha}{}^{\beta \gamma} H_{\beta}{}^{\delta \epsilon} H_{\gamma}{}^{\varepsilon \mu} R_{\epsilon \zeta \mu \eta} \nabla^{\alpha}\Phi \nabla^{\eta}H_{\delta \varepsilon}{}^{\zeta} \nn\\&&+ c_{553} H_{\beta}{}^{\epsilon \varepsilon} H^{\beta \gamma \delta} H_{\gamma}{}^{\mu \zeta} R_{\alpha \eta \delta \zeta} \nabla^{\alpha}\Phi \nabla^{\eta}H_{\epsilon \varepsilon \mu} + c_{554} H_{\alpha}{}^{\beta \gamma} H_{\delta}{}^{\mu \zeta} H^{\delta \epsilon \varepsilon} R_{\beta \zeta \gamma \eta} \nabla^{\alpha}\Phi \nabla^{\eta}H_{\epsilon \varepsilon \mu}\nn\\&& + c_{555} H_{\beta \gamma}{}^{\epsilon} H^{\beta \gamma \delta} H_{\delta}{}^{\varepsilon \mu} R_{\alpha \zeta \mu \eta} \nabla^{\alpha}\Phi \nabla^{\eta}H_{\epsilon \varepsilon}{}^{\zeta} + c_{556} H_{\beta \gamma}{}^{\epsilon} H^{\beta \gamma \delta} H_{\delta}{}^{\varepsilon \mu} R_{\alpha \eta \mu \zeta} \nabla^{\alpha}\Phi \nabla^{\eta}H_{\epsilon \varepsilon}{}^{\zeta} \nn\\&&+ c_{557} H_{\beta \gamma \delta} H^{\beta \gamma \delta} H^{\epsilon \varepsilon \mu} R_{\alpha \eta \mu \zeta} \nabla^{\alpha}\Phi \nabla^{\eta}H_{\epsilon \varepsilon}{}^{\zeta} + c_{558} H_{\alpha}{}^{\beta \gamma} H_{\beta}{}^{\delta \epsilon} H_{\delta}{}^{\varepsilon \mu} R_{\gamma \zeta \mu \eta} \nabla^{\alpha}\Phi \nabla^{\eta}H_{\epsilon \varepsilon}{}^{\zeta} \nn\\&&+ c_{559} H_{\alpha}{}^{\beta \gamma} H_{\beta}{}^{\delta \epsilon} H_{\delta}{}^{\varepsilon \mu} R_{\gamma \eta \mu \zeta} \nabla^{\alpha}\Phi \nabla^{\eta}H_{\epsilon \varepsilon}{}^{\zeta} + c_{560} H_{\alpha}{}^{\beta \gamma} H_{\beta \gamma}{}^{\delta} H^{\epsilon \varepsilon \mu} R_{\delta \zeta \mu \eta} \nabla^{\alpha}\Phi \nabla^{\eta}H_{\epsilon \varepsilon}{}^{\zeta} \nn\\&&+ c_{561} H_{\alpha}{}^{\beta \gamma} H_{\beta \gamma}{}^{\delta} H^{\epsilon \varepsilon \mu} R_{\delta \eta \mu \zeta} \nabla^{\alpha}\Phi \nabla^{\eta}H_{\epsilon \varepsilon}{}^{\zeta} + c_{562} H_{\alpha}{}^{\beta \gamma} H_{\beta}{}^{\delta \epsilon} H_{\gamma \delta}{}^{\varepsilon} R_{\varepsilon \eta \mu \zeta} \nabla^{\alpha}\Phi \nabla^{\eta}H_{\epsilon}{}^{\mu \zeta} \nn\\&&+ c_{563} H_{\alpha}{}^{\beta \gamma} H_{\beta \gamma}{}^{\delta} H_{\delta}{}^{\epsilon \varepsilon} R_{\varepsilon \eta \mu \zeta} \nabla^{\alpha}\Phi \nabla^{\eta}H_{\epsilon}{}^{\mu \zeta} + c_{564} H_{\beta \gamma}{}^{\epsilon} H^{\beta \gamma \delta} H^{\varepsilon \mu \zeta} R_{\alpha \delta \epsilon \eta} \nabla^{\alpha}\Phi \nabla^{\eta}H_{\varepsilon \mu \zeta} \nn\\&&+ c_{565} H_{\alpha}{}^{\beta \gamma} H_{\beta}{}^{\delta \epsilon} H^{\varepsilon \mu \zeta} R_{\gamma \eta \delta \epsilon} \nabla^{\alpha}\Phi \nabla^{\eta}H_{\varepsilon \mu \zeta} + c_{566} H_{\beta \gamma}{}^{\epsilon} H^{\beta \gamma \delta} H_{\delta}{}^{\varepsilon \mu} R_{\alpha \zeta \epsilon \eta} \nabla^{\alpha}\Phi \nabla^{\eta}H_{\varepsilon \mu}{}^{\zeta} \nn\\&&+ c_{567} H_{\beta \gamma}{}^{\epsilon} H^{\beta \gamma \delta} H_{\delta}{}^{\varepsilon \mu} R_{\alpha \eta \epsilon \zeta} \nabla^{\alpha}\Phi \nabla^{\eta}H_{\varepsilon \mu}{}^{\zeta} + c_{568} H_{\alpha}{}^{\beta \gamma} H_{\beta}{}^{\delta \epsilon} H_{\delta}{}^{\varepsilon \mu} R_{\gamma \zeta \epsilon \eta} \nabla^{\alpha}\Phi \nabla^{\eta}H_{\varepsilon \mu}{}^{\zeta}\nn\\&& + c_{569} H_{\alpha}{}^{\beta \gamma} H_{\beta}{}^{\delta \epsilon} H_{\delta}{}^{\varepsilon \mu} R_{\gamma \eta \epsilon \zeta} \nabla^{\alpha}\Phi \nabla^{\eta}H_{\varepsilon \mu}{}^{\zeta} + c_{570} H_{\alpha}{}^{\beta \gamma} H_{\beta}{}^{\delta \epsilon} H_{\delta \epsilon}{}^{\varepsilon} R_{\gamma \eta \mu \zeta} \nabla^{\alpha}\Phi \nabla^{\eta}H_{\varepsilon}{}^{\mu \zeta}\labell{T39}\nn
\eeqa
\beqa
&&{\cal L}_3^{H^3\prt H \prt\Phi\prt\prt\Phi}=\nn\\&&  c_{364}H_{\alpha}{}^{\delta \epsilon} H_{\beta}{}^{\varepsilon \mu} H_{\delta \varepsilon}{}^{\zeta} \nabla^{\alpha}\Phi \nabla^{\gamma}\nabla^{\beta}\Phi \nabla_{\epsilon}H_{\gamma \mu \zeta} + c_{365}H_{\alpha}{}^{\delta \epsilon} H_{\beta \delta}{}^{\varepsilon} H_{\varepsilon}{}^{\mu \zeta} \nabla^{\alpha}\Phi \nabla^{\gamma}\nabla^{\beta}\Phi \nabla_{\epsilon}H_{\gamma \mu \zeta}\nn\\&& + c_{395}H_{\beta}{}^{\gamma \delta} H_{\gamma}{}^{\epsilon \varepsilon} H_{\epsilon}{}^{\mu \zeta} \nabla^{\alpha}\Phi \nabla^{\beta}\nabla_{\alpha}\Phi \nabla_{\varepsilon}H_{\delta \mu \zeta} + c_{396}H_{\alpha \beta}{}^{\delta} H_{\gamma}{}^{\epsilon \varepsilon} H_{\epsilon}{}^{\mu \zeta} \nabla^{\alpha}\Phi \nabla^{\gamma}\nabla^{\beta}\Phi \nabla_{\varepsilon}H_{\delta \mu \zeta}\nn\\&& + c_{401}H_{\alpha}{}^{\delta \epsilon} H_{\beta \delta}{}^{\varepsilon} H_{\gamma}{}^{\mu \zeta} \nabla^{\alpha}\Phi \nabla^{\gamma}\nabla^{\beta}\Phi \nabla_{\varepsilon}H_{\epsilon \mu \zeta} + c_{443}H_{\beta}{}^{\delta \epsilon} H_{\gamma}{}^{\varepsilon \mu} H_{\delta \epsilon}{}^{\zeta} \nabla^{\alpha}\Phi \nabla^{\gamma}\nabla^{\beta}\Phi \nabla_{\mu}H_{\alpha \varepsilon \zeta} \nn\\&&+ c_{450}H_{\alpha}{}^{\delta \epsilon} H_{\beta}{}^{\varepsilon \mu} H_{\delta \varepsilon}{}^{\zeta} \nabla^{\alpha}\Phi \nabla^{\gamma}\nabla^{\beta}\Phi \nabla_{\mu}H_{\gamma \epsilon \zeta} + c_{453}H_{\alpha}{}^{\delta \epsilon} H_{\beta}{}^{\varepsilon \mu} H_{\delta \epsilon}{}^{\zeta} \nabla^{\alpha}\Phi \nabla^{\gamma}\nabla^{\beta}\Phi \nabla_{\mu}H_{\gamma \varepsilon \zeta}\nn\\&& + c_{679}H_{\beta}{}^{\delta \epsilon} H_{\delta}{}^{\varepsilon \mu} H_{\varepsilon \mu}{}^{\zeta} \nabla^{\alpha}\Phi \nabla^{\gamma}\nabla^{\beta}\Phi \nabla_{\zeta}H_{\alpha \gamma \epsilon} + c_{680}H_{\beta}{}^{\delta \epsilon} H_{\delta \epsilon}{}^{\varepsilon} H_{\varepsilon}{}^{\mu \zeta} \nabla^{\alpha}\Phi \nabla^{\gamma}\nabla^{\beta}\Phi \nabla_{\zeta}H_{\alpha \gamma \mu} \nn\\&&+ c_{681}H_{\beta}{}^{\delta \epsilon} H_{\gamma \delta}{}^{\varepsilon} H_{\epsilon}{}^{\mu \zeta} \nabla^{\alpha}\Phi \nabla^{\gamma}\nabla^{\beta}\Phi \nabla_{\zeta}H_{\alpha \varepsilon \mu} + c_{682}H_{\gamma \delta}{}^{\varepsilon} H^{\gamma \delta \epsilon} H_{\epsilon}{}^{\mu \zeta} \nabla^{\alpha}\Phi \nabla^{\beta}\nabla_{\alpha}\Phi \nabla_{\zeta}H_{\beta \varepsilon \mu}\nn\\&& + c_{683}H_{\alpha}{}^{\delta \epsilon} H_{\beta}{}^{\varepsilon \mu} H_{\varepsilon \mu}{}^{\zeta} \nabla^{\alpha}\Phi \nabla^{\gamma}\nabla^{\beta}\Phi \nabla_{\zeta}H_{\gamma \delta \epsilon} + c_{684}H_{\beta}{}^{\gamma \delta} H_{\epsilon \varepsilon}{}^{\zeta} H^{\epsilon \varepsilon \mu} \nabla^{\alpha}\Phi \nabla^{\beta}\nabla_{\alpha}\Phi \nabla_{\zeta}H_{\gamma \delta \mu} \nn\\&&+ c_{685}H_{\alpha \beta}{}^{\delta} H_{\epsilon \varepsilon}{}^{\zeta} H^{\epsilon \varepsilon \mu} \nabla^{\alpha}\Phi \nabla^{\gamma}\nabla^{\beta}\Phi \nabla_{\zeta}H_{\gamma \delta \mu} + c_{686}H_{\alpha}{}^{\delta \epsilon} H_{\beta}{}^{\varepsilon \mu} H_{\delta \varepsilon}{}^{\zeta} \nabla^{\alpha}\Phi \nabla^{\gamma}\nabla^{\beta}\Phi \nabla_{\zeta}H_{\gamma \epsilon \mu}\nn\\&& + c_{687}H_{\alpha}{}^{\delta \epsilon} H_{\beta \delta}{}^{\varepsilon} H_{\varepsilon}{}^{\mu \zeta} \nabla^{\alpha}\Phi \nabla^{\gamma}\nabla^{\beta}\Phi \nabla_{\zeta}H_{\gamma \epsilon \mu} + c_{688}H_{\alpha}{}^{\delta \epsilon} H_{\beta}{}^{\varepsilon \mu} H_{\delta \epsilon}{}^{\zeta} \nabla^{\alpha}\Phi \nabla^{\gamma}\nabla^{\beta}\Phi \nabla_{\zeta}H_{\gamma \varepsilon \mu} \nn\\&&+ c_{689}H_{\alpha}{}^{\delta \epsilon} H_{\beta \delta}{}^{\varepsilon} H_{\epsilon}{}^{\mu \zeta} \nabla^{\alpha}\Phi \nabla^{\gamma}\nabla^{\beta}\Phi \nabla_{\zeta}H_{\gamma \varepsilon \mu} + c_{690}H_{\alpha \beta}{}^{\delta} H_{\delta}{}^{\epsilon \varepsilon} H_{\epsilon}{}^{\mu \zeta} \nabla^{\alpha}\Phi \nabla^{\gamma}\nabla^{\beta}\Phi \nabla_{\zeta}H_{\gamma \varepsilon \mu}\nn\\&& + c_{691}H_{\alpha}{}^{\delta \epsilon} H_{\beta}{}^{\varepsilon \mu} H_{\gamma \varepsilon}{}^{\zeta} \nabla^{\alpha}\Phi \nabla^{\gamma}\nabla^{\beta}\Phi \nabla_{\zeta}H_{\delta \epsilon \mu} + c_{692}H_{\beta}{}^{\gamma \delta} H_{\gamma}{}^{\epsilon \varepsilon} H_{\epsilon}{}^{\mu \zeta} \nabla^{\alpha}\Phi \nabla^{\beta}\nabla_{\alpha}\Phi \nabla_{\zeta}H_{\delta \varepsilon \mu}\nn\\&& + c_{693}H_{\alpha \beta}{}^{\delta} H_{\gamma}{}^{\epsilon \varepsilon} H_{\epsilon}{}^{\mu \zeta} \nabla^{\alpha}\Phi \nabla^{\gamma}\nabla^{\beta}\Phi \nabla_{\zeta}H_{\delta \varepsilon \mu} + c_{694}H_{\beta}{}^{\gamma \delta} H_{\gamma}{}^{\epsilon \varepsilon} H_{\delta}{}^{\mu \zeta} \nabla^{\alpha}\Phi \nabla^{\beta}\nabla_{\alpha}\Phi \nabla_{\zeta}H_{\epsilon \varepsilon \mu} \nn\\&&+ c_{695}H_{\alpha}{}^{\delta \epsilon} H_{\beta \delta}{}^{\varepsilon} H_{\gamma}{}^{\mu \zeta} \nabla^{\alpha}\Phi \nabla^{\gamma}\nabla^{\beta}\Phi \nabla_{\zeta}H_{\epsilon \varepsilon \mu} + c_{696}H_{\alpha \beta}{}^{\delta} H_{\gamma}{}^{\epsilon \varepsilon} H_{\delta}{}^{\mu \zeta} \nabla^{\alpha}\Phi \nabla^{\gamma}\nabla^{\beta}\Phi \nabla_{\zeta}H_{\epsilon \varepsilon \mu}\labell{T40}\nn
\eeqa
\beqa
{\cal L}_3^{H\prt H R^2\prt\Phi}&\!\!\!\!\!=\!\!\!\!\! & c_{368}H^{\beta \gamma \delta} R_{\delta}{}^{\varepsilon \mu \zeta} R_{\epsilon \mu \varepsilon \zeta} \nabla^{\alpha}\Phi \nabla^{\epsilon}H_{\alpha \beta \gamma} + c_{370}H_{\alpha}{}^{\beta \gamma} R_{\delta}{}^{\varepsilon \mu \zeta} R_{\epsilon \mu \varepsilon \zeta} \nabla^{\alpha}\Phi \nabla^{\epsilon}H_{\beta \gamma}{}^{\delta}\nn\\&& + c_{404}H^{\beta \gamma \delta} R_{\gamma}{}^{\mu}{}_{\epsilon}{}^{\zeta} R_{\delta \mu \varepsilon \zeta} \nabla^{\alpha}\Phi \nabla^{\varepsilon}H_{\alpha \beta}{}^{\epsilon} + c_{405}H^{\beta \gamma \delta} R_{\gamma}{}^{\mu}{}_{\epsilon}{}^{\zeta} R_{\delta \zeta \varepsilon \mu} \nabla^{\alpha}\Phi \nabla^{\varepsilon}H_{\alpha \beta}{}^{\epsilon} \nn\\&&+ c_{406}H^{\beta \gamma \delta} R_{\gamma}{}^{\mu}{}_{\delta}{}^{\zeta} R_{\epsilon \mu \varepsilon \zeta} \nabla^{\alpha}\Phi \nabla^{\varepsilon}H_{\alpha \beta}{}^{\epsilon} + c_{419}H^{\beta \gamma \delta} R_{\alpha}{}^{\mu}{}_{\epsilon}{}^{\zeta} R_{\delta \mu \varepsilon \zeta} \nabla^{\alpha}\Phi \nabla^{\varepsilon}H_{\beta \gamma}{}^{\epsilon} \nn\\&&+ c_{420}H^{\beta \gamma \delta} R_{\alpha}{}^{\mu}{}_{\varepsilon}{}^{\zeta} R_{\delta \zeta \epsilon \mu} \nabla^{\alpha}\Phi \nabla^{\varepsilon}H_{\beta \gamma}{}^{\epsilon} + c_{421}H^{\beta \gamma \delta} R_{\alpha}{}^{\mu}{}_{\epsilon}{}^{\zeta} R_{\delta \zeta \varepsilon \mu} \nabla^{\alpha}\Phi \nabla^{\varepsilon}H_{\beta \gamma}{}^{\epsilon} \nn\\&&+ c_{422}H^{\beta \gamma \delta} R_{\alpha}{}^{\mu}{}_{\delta}{}^{\zeta} R_{\epsilon \zeta \varepsilon \mu} \nabla^{\alpha}\Phi \nabla^{\varepsilon}H_{\beta \gamma}{}^{\epsilon} + c_{423}H_{\alpha}{}^{\beta \gamma} R_{\gamma}{}^{\mu}{}_{\delta}{}^{\zeta} R_{\epsilon \mu \varepsilon \zeta} \nabla^{\alpha}\Phi \nabla^{\varepsilon}H_{\beta}{}^{\delta \epsilon} \nn\\&&+ c_{424}H_{\alpha}{}^{\beta \gamma} R_{\gamma}{}^{\mu}{}_{\delta}{}^{\zeta} R_{\epsilon \zeta \varepsilon \mu} \nabla^{\alpha}\Phi \nabla^{\varepsilon}H_{\beta}{}^{\delta \epsilon} + c_{472}H^{\beta \gamma \delta} R_{\beta \epsilon \gamma}{}^{\zeta} R_{\delta \mu \varepsilon \zeta} \nabla^{\alpha}\Phi \nabla^{\mu}H_{\alpha}{}^{\epsilon \varepsilon} \nn\\&&+ c_{473}H^{\beta \gamma \delta} R_{\beta \mu \gamma}{}^{\zeta} R_{\delta \zeta \epsilon \varepsilon} \nabla^{\alpha}\Phi \nabla^{\mu}H_{\alpha}{}^{\epsilon \varepsilon} + c_{474}H^{\beta \gamma \delta} R_{\beta \epsilon \gamma}{}^{\zeta} R_{\delta \zeta \varepsilon \mu} \nabla^{\alpha}\Phi \nabla^{\mu}H_{\alpha}{}^{\epsilon \varepsilon} \nn\\&&+ c_{487}H^{\beta \gamma \delta} R_{\alpha \mu \epsilon}{}^{\zeta} R_{\gamma \varepsilon \delta \zeta} \nabla^{\alpha}\Phi \nabla^{\mu}H_{\beta}{}^{\epsilon \varepsilon} + c_{488}H^{\beta \gamma \delta} R_{\alpha}{}^{\zeta}{}_{\epsilon \varepsilon} R_{\gamma \mu \delta \zeta} \nabla^{\alpha}\Phi \nabla^{\mu}H_{\beta}{}^{\epsilon \varepsilon}\nn\\&& + c_{489}H^{\beta \gamma \delta} R_{\alpha \epsilon \gamma}{}^{\zeta} R_{\delta \mu \varepsilon \zeta} \nabla^{\alpha}\Phi \nabla^{\mu}H_{\beta}{}^{\epsilon \varepsilon} + c_{490}H^{\beta \gamma \delta} R_{\alpha}{}^{\zeta}{}_{\gamma \epsilon} R_{\delta \mu \varepsilon \zeta} \nabla^{\alpha}\Phi \nabla^{\mu}H_{\beta}{}^{\epsilon \varepsilon} \nn\\&&+ c_{491}H^{\beta \gamma \delta} R_{\alpha}{}^{\zeta}{}_{\gamma \mu} R_{\delta \zeta \epsilon \varepsilon} \nabla^{\alpha}\Phi \nabla^{\mu}H_{\beta}{}^{\epsilon \varepsilon} + c_{492}H^{\beta \gamma \delta} R_{\alpha \epsilon \gamma}{}^{\zeta} R_{\delta \zeta \varepsilon \mu} \nabla^{\alpha}\Phi \nabla^{\mu}H_{\beta}{}^{\epsilon \varepsilon} \nn\\&&+ c_{493}H^{\beta \gamma \delta} R_{\alpha}{}^{\zeta}{}_{\gamma \epsilon} R_{\delta \zeta \varepsilon \mu} \nabla^{\alpha}\Phi \nabla^{\mu}H_{\beta}{}^{\epsilon \varepsilon} + c_{494}H^{\beta \gamma \delta} R_{\alpha}{}^{\zeta}{}_{\gamma \delta} R_{\epsilon \mu \varepsilon \zeta} \nabla^{\alpha}\Phi \nabla^{\mu}H_{\beta}{}^{\epsilon \varepsilon} \nn\\&&+ c_{517}H_{\alpha}{}^{\beta \gamma} R_{\beta \mu \delta}{}^{\zeta} R_{\gamma \zeta \epsilon \varepsilon} \nabla^{\alpha}\Phi \nabla^{\mu}H^{\delta \epsilon \varepsilon} + c_{518}H_{\alpha}{}^{\beta \gamma} R_{\beta}{}^{\zeta}{}_{\delta \epsilon} R_{\gamma \zeta \varepsilon \mu} \nabla^{\alpha}\Phi \nabla^{\mu}H^{\delta \epsilon \varepsilon} \nn\\&&+ c_{519}H_{\alpha}{}^{\beta \gamma} R_{\beta \delta \gamma}{}^{\zeta} R_{\epsilon \mu \varepsilon \zeta} \nabla^{\alpha}\Phi \nabla^{\mu}H^{\delta \epsilon \varepsilon} + c_{521}H^{\beta \gamma \delta} R_{\alpha \zeta \beta \epsilon} R_{\gamma \varepsilon \delta \mu} \nabla^{\alpha}\Phi \nabla^{\zeta}H^{\epsilon \varepsilon \mu} \nn\\&&+ c_{522}H^{\beta \gamma \delta} R_{\alpha \epsilon \beta \varepsilon} R_{\gamma \mu \delta \zeta} \nabla^{\alpha}\Phi \nabla^{\zeta}H^{\epsilon \varepsilon \mu} + c_{523}H^{\beta \gamma \delta} R_{\alpha \epsilon \beta \gamma} R_{\delta \zeta \varepsilon \mu} \nabla^{\alpha}\Phi \nabla^{\zeta}H^{\epsilon \varepsilon \mu}\labell{T41}
\eeqa
\beqa
{\cal L}_3^{R(\prt\Phi)^2(\prt\prt\Phi)^2}&\!\!\!\!\!=\!\!\!\!\! & c_{372} R_{\alpha \delta \beta \epsilon} \nabla^{\alpha}\Phi \nabla^{\beta}\Phi \nabla^{\delta}\nabla^{\gamma}\Phi \nabla^{\epsilon}\nabla_{\gamma}\Phi + c_{377} R_{\beta \delta \gamma \epsilon} \nabla^{\alpha}\Phi \nabla^{\beta}\Phi \nabla^{\gamma}\nabla_{\alpha}\Phi \nabla^{\epsilon}\nabla^{\delta}\Phi\nn\\&& + c_{381} R_{\beta \delta \gamma \epsilon} \nabla_{\alpha}\Phi \nabla^{\alpha}\Phi \nabla^{\gamma}\nabla^{\beta}\Phi \nabla^{\epsilon}\nabla^{\delta}\Phi\labell{T42}
\eeqa
\beqa
{\cal L}_3^{R(\prt\Phi)^4\prt\prt\Phi}&\!\!\!\!\!=\!\!\!\!\! &  c_{374} R_{\beta \delta \gamma \epsilon} \nabla_{\alpha}\Phi \nabla^{\alpha}\Phi \nabla^{\beta}\Phi \nabla^{\gamma}\Phi \nabla^{\epsilon}\nabla^{\delta}\Phi\labell{T43}
\eeqa
\beqa
{\cal L}_3^{R(\prt\prt\Phi)^3}&\!\!\!\!\!=\!\!\!\!\! & c_{379} R_{\beta \delta \gamma \epsilon} \nabla^{\beta}\nabla^{\alpha}\Phi \nabla^{\gamma}\nabla_{\alpha}\Phi \nabla^{\epsilon}\nabla^{\delta}\Phi\labell{T44}
\eeqa
\beqa
&&{\cal L}_3^{H^2(\prt H)^2(\prt\Phi)^2}=\nn\\&& c_{398} H_{\gamma \delta}{}^{\varepsilon} H^{\gamma \delta \epsilon} \nabla^{\alpha}\Phi \nabla_{\beta}H_{\alpha}{}^{\mu \zeta} \nabla^{\beta}\Phi \nabla_{\varepsilon}H_{\epsilon\mu \zeta} + c_{466} H_{\gamma}{}^{\varepsilon\mu} H^{\gamma \delta \epsilon} \nabla^{\alpha}\Phi \nabla_{\beta}H_{\alpha \delta}{}^{\zeta} \nabla^{\beta}\Phi \nabla_{\mu}H_{\epsilon \varepsilon \zeta} \nn\\&&+ c_{630} H^{\gamma \delta \epsilon} H^{\varepsilon\mu \zeta} \nabla^{\alpha}\Phi \nabla_{\beta}H_{\alpha \gamma \varepsilon} \nabla^{\beta}\Phi \nabla_{\zeta}H_{\delta \epsilon\mu} + c_{631} H_{\gamma \delta}{}^{\varepsilon} H^{\gamma \delta \epsilon} \nabla^{\alpha}\Phi \nabla^{\beta}\Phi \nabla_{\varepsilon}H_{\beta\mu \zeta} \nabla^{\zeta}H_{\alpha \epsilon}{}^{\mu}\nn\\&& + c_{632} H_{\gamma \delta}{}^{\varepsilon} H^{\gamma \delta \epsilon} \nabla^{\alpha}\Phi \nabla^{\beta}\Phi \nabla_{\mu}H_{\beta \varepsilon \zeta} \nabla^{\zeta}H_{\alpha \epsilon}{}^{\mu} + c_{633} H_{\alpha}{}^{\gamma \delta} H_{\gamma}{}^{\epsilon \varepsilon} \nabla^{\alpha}\Phi \nabla^{\beta}\Phi \nabla_{\mu}H_{\epsilon \varepsilon \zeta} \nabla^{\zeta}H_{\beta \delta}{}^{\mu} \nn\\&&+ c_{634} H_{\alpha}{}^{\gamma \delta} H_{\gamma}{}^{\epsilon \varepsilon} \nabla^{\alpha}\Phi \nabla^{\beta}\Phi \nabla_{\zeta}H_{\epsilon \varepsilon\mu} \nabla^{\zeta}H_{\beta \delta}{}^{\mu} + c_{635} H_{\alpha}{}^{\gamma \delta} H_{\gamma}{}^{\epsilon \varepsilon} \nabla^{\alpha}\Phi \nabla^{\beta}\Phi \nabla_{\mu}H_{\delta \varepsilon \zeta} \nabla^{\zeta}H_{\beta \epsilon}{}^{\mu}\nn\\&& + c_{636} H_{\alpha}{}^{\gamma \delta} H_{\gamma \delta}{}^{\epsilon} \nabla^{\alpha}\Phi \nabla^{\beta}\Phi \nabla_{\zeta}H_{\epsilon \varepsilon\mu} \nabla^{\zeta}H_{\beta}{}^{\varepsilon\mu} + c_{637} H_{\alpha}{}^{\gamma \delta} H^{\epsilon \varepsilon\mu} \nabla^{\alpha}\Phi \nabla_{\beta}H_{\varepsilon\mu \zeta} \nabla^{\beta}\Phi \nabla^{\zeta}H_{\gamma \delta \epsilon} \nn\\&&+ c_{638} H_{\alpha}{}^{\gamma \delta} H_{\beta}{}^{\epsilon \varepsilon} \nabla^{\alpha}\Phi \nabla^{\beta}\Phi \nabla_{\mu}H_{\epsilon \varepsilon \zeta} \nabla^{\zeta}H_{\gamma \delta}{}^{\mu} + c_{639} H_{\beta}{}^{\epsilon \varepsilon} H^{\beta \gamma \delta} \nabla_{\alpha}\Phi \nabla^{\alpha}\Phi \nabla_{\zeta}H_{\epsilon \varepsilon\mu} \nabla^{\zeta}H_{\gamma \delta}{}^{\mu} \nn\\&&+ c_{640} H_{\alpha}{}^{\gamma \delta} H_{\beta}{}^{\epsilon \varepsilon} \nabla^{\alpha}\Phi \nabla^{\beta}\Phi \nabla_{\zeta}H_{\epsilon \varepsilon\mu} \nabla^{\zeta}H_{\gamma \delta}{}^{\mu} + c_{641} H_{\alpha}{}^{\gamma \delta} H^{\epsilon \varepsilon\mu} \nabla^{\alpha}\Phi \nabla_{\beta}H_{\delta\mu \zeta} \nabla^{\beta}\Phi \nabla^{\zeta}H_{\gamma \epsilon \varepsilon}\nn\\&& + c_{642} H_{\beta}{}^{\epsilon \varepsilon} H^{\beta \gamma \delta} \nabla_{\alpha}\Phi \nabla^{\alpha}\Phi \nabla_{\mu}H_{\delta \varepsilon \zeta} \nabla^{\zeta}H_{\gamma \epsilon}{}^{\mu} + c_{643} H_{\alpha}{}^{\gamma \delta} H_{\beta}{}^{\epsilon \varepsilon} \nabla^{\alpha}\Phi \nabla^{\beta}\Phi \nabla_{\mu}H_{\delta \varepsilon \zeta} \nabla^{\zeta}H_{\gamma \epsilon}{}^{\mu}\nn\\&& + c_{644} H_{\alpha}{}^{\gamma \delta} H_{\beta}{}^{\epsilon \varepsilon} \nabla^{\alpha}\Phi \nabla^{\beta}\Phi \nabla_{\zeta}H_{\delta \varepsilon\mu} \nabla^{\zeta}H_{\gamma \epsilon}{}^{\mu} + c_{645} H_{\alpha}{}^{\gamma \delta} H_{\gamma}{}^{\epsilon \varepsilon} \nabla^{\alpha}\Phi \nabla_{\beta}H_{\varepsilon\mu \zeta} \nabla^{\beta}\Phi \nabla^{\zeta}H_{\delta \epsilon}{}^{\mu}\nn\\&& + c_{646} H_{\beta \gamma}{}^{\epsilon} H^{\beta \gamma \delta} \nabla_{\alpha}\Phi \nabla^{\alpha}\Phi \nabla_{\mu}H_{\epsilon \varepsilon \zeta} \nabla^{\zeta}H_{\delta}{}^{\varepsilon\mu} + c_{647} H_{\alpha}{}^{\gamma \delta} H_{\beta \gamma}{}^{\epsilon} \nabla^{\alpha}\Phi \nabla^{\beta}\Phi \nabla_{\mu}H_{\epsilon \varepsilon \zeta} \nabla^{\zeta}H_{\delta}{}^{\varepsilon\mu}\nn\\&& + c_{648} H_{\beta \gamma}{}^{\epsilon} H^{\beta \gamma \delta} \nabla_{\alpha}\Phi \nabla^{\alpha}\Phi \nabla_{\zeta}H_{\epsilon \varepsilon\mu} \nabla^{\zeta}H_{\delta}{}^{\varepsilon\mu} + c_{649} H_{\alpha}{}^{\gamma \delta} H_{\beta \gamma}{}^{\epsilon} \nabla^{\alpha}\Phi \nabla^{\beta}\Phi \nabla_{\zeta}H_{\epsilon \varepsilon\mu} \nabla^{\zeta}H_{\delta}{}^{\varepsilon\mu} \nn\\&&+ c_{650} H_{\alpha}{}^{\gamma \delta} H_{\gamma \delta}{}^{\epsilon} \nabla^{\alpha}\Phi \nabla_{\beta}H_{\varepsilon\mu \zeta} \nabla^{\beta}\Phi \nabla^{\zeta}H_{\epsilon}{}^{\varepsilon\mu}\labell{T45}
\eeqa
\beqa
{\cal L}_3^{H^2(\prt H)^2\prt\prt\Phi}&\!\!\!\!\!=\!\!\!\!\! & c_{400} H_{\gamma \delta}{}^{\varepsilon} H^{\gamma \delta \epsilon} \nabla_{\beta}H_{\alpha}{}^{\mu \zeta} \nabla^{\beta}\nabla^{\alpha}\Phi \nabla_{\varepsilon}H_{\epsilon\mu \zeta} + c_{467} H_{\gamma}{}^{\varepsilon\mu} H^{\gamma \delta \epsilon} \nabla_{\beta}H_{\alpha \delta}{}^{\zeta} \nabla^{\beta}\nabla^{\alpha}\Phi \nabla_{\mu}H_{\epsilon \varepsilon \zeta}\nn\\&& + c_{654} H^{\gamma \delta \epsilon} H^{\varepsilon\mu \zeta} \nabla_{\beta}H_{\alpha \gamma \varepsilon} \nabla^{\beta}\nabla^{\alpha}\Phi \nabla_{\zeta}H_{\delta \epsilon\mu} + c_{655} H^{\gamma \delta \epsilon} H^{\varepsilon\mu \zeta} \nabla_{\beta}H_{\alpha \gamma \delta} \nabla^{\beta}\nabla^{\alpha}\Phi \nabla_{\zeta}H_{\epsilon \varepsilon\mu}\nn\\&& + c_{656} H_{\gamma \delta}{}^{\varepsilon} H^{\gamma \delta \epsilon} \nabla^{\beta}\nabla^{\alpha}\Phi \nabla_{\varepsilon}H_{\beta\mu \zeta} \nabla^{\zeta}H_{\alpha \epsilon}{}^{\mu} + c_{657} H_{\gamma \delta}{}^{\varepsilon} H^{\gamma \delta \epsilon} \nabla^{\beta}\nabla^{\alpha}\Phi \nabla_{\mu}H_{\beta \varepsilon \zeta} \nabla^{\zeta}H_{\alpha \epsilon}{}^{\mu}\nn\\&& + c_{658} H_{\alpha}{}^{\gamma \delta} H^{\epsilon \varepsilon\mu} \nabla^{\beta}\nabla^{\alpha}\Phi \nabla_{\mu}H_{\delta \varepsilon \zeta} \nabla^{\zeta}H_{\beta \gamma \epsilon} + c_{659} H_{\alpha}{}^{\gamma \delta} H^{\epsilon \varepsilon\mu} \nabla^{\beta}\nabla^{\alpha}\Phi \nabla_{\zeta}H_{\delta \varepsilon\mu} \nabla^{\zeta}H_{\beta \gamma \epsilon}\nn\\&& + c_{660} H_{\alpha}{}^{\gamma \delta} H_{\gamma}{}^{\epsilon \varepsilon} \nabla^{\beta}\nabla^{\alpha}\Phi \nabla_{\mu}H_{\epsilon \varepsilon \zeta} \nabla^{\zeta}H_{\beta \delta}{}^{\mu} + c_{661} H_{\alpha}{}^{\gamma \delta} H_{\gamma}{}^{\epsilon \varepsilon} \nabla^{\beta}\nabla^{\alpha}\Phi \nabla_{\zeta}H_{\epsilon \varepsilon\mu} \nabla^{\zeta}H_{\beta \delta}{}^{\mu} \nn\\&&+ c_{662} H_{\alpha}{}^{\gamma \delta} H^{\epsilon \varepsilon\mu} \nabla^{\beta}\nabla^{\alpha}\Phi \nabla_{\mu}H_{\gamma \delta \zeta} \nabla^{\zeta}H_{\beta \epsilon \varepsilon} + c_{663} H_{\alpha}{}^{\gamma \delta} H^{\epsilon \varepsilon\mu} \nabla^{\beta}\nabla^{\alpha}\Phi \nabla_{\zeta}H_{\gamma \delta\mu} \nabla^{\zeta}H_{\beta \epsilon \varepsilon} \nn\\&&+ c_{664} H_{\alpha}{}^{\gamma \delta} H_{\gamma}{}^{\epsilon \varepsilon} \nabla^{\beta}\nabla^{\alpha}\Phi \nabla_{\varepsilon}H_{\delta\mu \zeta} \nabla^{\zeta}H_{\beta \epsilon}{}^{\mu} + c_{665} H_{\alpha}{}^{\gamma \delta} H_{\gamma}{}^{\epsilon \varepsilon} \nabla^{\beta}\nabla^{\alpha}\Phi \nabla_{\mu}H_{\delta \varepsilon \zeta} \nabla^{\zeta}H_{\beta \epsilon}{}^{\mu} \nn\\&&+ c_{666} H_{\alpha}{}^{\gamma \delta} H_{\gamma}{}^{\epsilon \varepsilon} \nabla^{\beta}\nabla^{\alpha}\Phi \nabla_{\zeta}H_{\delta \varepsilon\mu} \nabla^{\zeta}H_{\beta \epsilon}{}^{\mu} + c_{667} H_{\alpha}{}^{\gamma \delta} H_{\gamma \delta}{}^{\epsilon} \nabla^{\beta}\nabla^{\alpha}\Phi \nabla_{\zeta}H_{\epsilon \varepsilon\mu} \nabla^{\zeta}H_{\beta}{}^{\varepsilon\mu}\nn\\&& + c_{668} H_{\alpha}{}^{\gamma \delta} H^{\epsilon \varepsilon\mu} \nabla_{\beta}H_{\varepsilon\mu \zeta} \nabla^{\beta}\nabla^{\alpha}\Phi \nabla^{\zeta}H_{\gamma \delta \epsilon} + c_{669} H_{\alpha}{}^{\gamma \delta} H_{\beta}{}^{\epsilon \varepsilon} \nabla^{\beta}\nabla^{\alpha}\Phi \nabla_{\mu}H_{\epsilon \varepsilon \zeta} \nabla^{\zeta}H_{\gamma \delta}{}^{\mu} \nn\\&&+ c_{670} H_{\alpha}{}^{\gamma \delta} H_{\beta}{}^{\epsilon \varepsilon} \nabla^{\beta}\nabla^{\alpha}\Phi \nabla_{\zeta}H_{\epsilon \varepsilon\mu} \nabla^{\zeta}H_{\gamma \delta}{}^{\mu} + c_{671} H_{\alpha}{}^{\gamma \delta} H^{\epsilon \varepsilon\mu} \nabla_{\beta}H_{\delta\mu \zeta} \nabla^{\beta}\nabla^{\alpha}\Phi \nabla^{\zeta}H_{\gamma \epsilon \varepsilon} \nn\\&&+ c_{672} H_{\alpha}{}^{\gamma \delta} H_{\beta}{}^{\epsilon \varepsilon} \nabla^{\beta}\nabla^{\alpha}\Phi \nabla_{\mu}H_{\delta \varepsilon \zeta} \nabla^{\zeta}H_{\gamma \epsilon}{}^{\mu} + c_{673} H_{\alpha}{}^{\gamma \delta} H_{\beta}{}^{\epsilon \varepsilon} \nabla^{\beta}\nabla^{\alpha}\Phi \nabla_{\zeta}H_{\delta \varepsilon\mu} \nabla^{\zeta}H_{\gamma \epsilon}{}^{\mu} \nn\\&&+ c_{674} H_{\alpha}{}^{\gamma \delta} H_{\gamma}{}^{\epsilon \varepsilon} \nabla_{\beta}H_{\varepsilon\mu \zeta} \nabla^{\beta}\nabla^{\alpha}\Phi \nabla^{\zeta}H_{\delta \epsilon}{}^{\mu} + c_{675} H_{\alpha}{}^{\gamma \delta} H_{\beta \gamma}{}^{\epsilon} \nabla^{\beta}\nabla^{\alpha}\Phi \nabla_{\mu}H_{\epsilon \varepsilon \zeta} \nabla^{\zeta}H_{\delta}{}^{\varepsilon\mu} \nn\\&&+ c_{676} H_{\alpha}{}^{\gamma \delta} H_{\beta \gamma}{}^{\epsilon} \nabla^{\beta}\nabla^{\alpha}\Phi \nabla_{\zeta}H_{\epsilon \varepsilon\mu} \nabla^{\zeta}H_{\delta}{}^{\varepsilon\mu} + c_{677} H_{\alpha}{}^{\gamma \delta} H_{\gamma \delta}{}^{\epsilon} \nabla_{\beta}H_{\varepsilon\mu \zeta} \nabla^{\beta}\nabla^{\alpha}\Phi \nabla^{\zeta}H_{\epsilon}{}^{\varepsilon\mu}\nn\\&& + c_{678} H_{\alpha}{}^{\gamma \delta} H_{\beta \gamma \delta} \nabla^{\beta}\nabla^{\alpha}\Phi \nabla_{\zeta}H_{\epsilon \varepsilon\mu} \nabla^{\zeta}H^{\epsilon \varepsilon\mu}\labell{T46}
\eeqa
\beqa
&&{\cal L}_3^{H^3\prt H(\prt\Phi)^3}=\nn\\&& c_{399} H_{\alpha}{}^{\delta \epsilon} H_{\beta \delta}{}^{\varepsilon} H_{\gamma}{}^{\mu \zeta} \nabla^{\alpha}\Phi \nabla^{\beta}\Phi \nabla^{\gamma}\Phi \nabla_{\varepsilon}H_{\epsilon \mu \zeta} + c_{449} H_{\alpha}{}^{\delta \epsilon} H_{\beta}{}^{\varepsilon \mu} H_{\delta \varepsilon}{}^{\zeta} \nabla^{\alpha}\Phi \nabla^{\beta}\Phi \nabla^{\gamma}\Phi \nabla_{\mu}H_{\gamma \epsilon \zeta}  \nn\\&&+ c_{452} H_{\alpha}{}^{\delta \epsilon} H_{\beta}{}^{\varepsilon \mu} H_{\delta \epsilon}{}^{\zeta} \nabla^{\alpha}\Phi \nabla^{\beta}\Phi \nabla^{\gamma}\Phi \nabla_{\mu}H_{\gamma \varepsilon \zeta} + c_{651} H_{\beta}{}^{\gamma \delta} H_{\epsilon \varepsilon}{}^{\zeta} H^{\epsilon \varepsilon \mu} \nabla_{\alpha}\Phi \nabla^{\alpha}\Phi \nabla^{\beta}\Phi \nabla_{\zeta}H_{\gamma \delta \mu} \nn\\&& + c_{652} H_{\alpha}{}^{\delta \epsilon} H_{\beta \delta}{}^{\varepsilon} H_{\epsilon}{}^{\mu \zeta} \nabla^{\alpha}\Phi \nabla^{\beta}\Phi \nabla^{\gamma}\Phi \nabla_{\zeta}H_{\gamma \varepsilon \mu} + c_{653} H_{\beta}{}^{\gamma \delta} H_{\gamma}{}^{\epsilon \varepsilon} H_{\delta}{}^{\mu \zeta} \nabla_{\alpha}\Phi \nabla^{\alpha}\Phi \nabla^{\beta}\Phi \nabla_{\zeta}H_{\epsilon \varepsilon \mu}\labell{T47}\nn
\eeqa
\beqa
{\cal L}_3^{(\prt H)^4}&\!\!\!\!\!=\!\!\!\!\! & c_{410} \nabla_{\delta}H_{\gamma}{}^{\mu \zeta} \nabla^{\delta}H^{\alpha \beta \gamma} \nabla_{\varepsilon}H_{\epsilon \mu \zeta} \nabla^{\varepsilon}H_{\alpha \beta}{}^{\epsilon} + c_{615} \nabla^{\delta}H^{\alpha \beta \gamma} \nabla^{\varepsilon}H_{\alpha \beta}{}^{\epsilon} \nabla_{\zeta}H_{\epsilon \varepsilon \mu} \nabla^{\zeta}H_{\gamma \delta}{}^{\mu} \nn\\&& + c_{616} \nabla^{\delta}H^{\alpha \beta \gamma} \nabla_{\epsilon}H_{\delta \mu \zeta} \nabla^{\varepsilon}H_{\alpha \beta}{}^{\epsilon} \nabla^{\zeta}H_{\gamma \varepsilon}{}^{\mu} + c_{617} \nabla^{\delta}H^{\alpha \beta \gamma} \nabla^{\varepsilon}H_{\alpha \beta}{}^{\epsilon} \nabla_{\mu}H_{\delta \epsilon \zeta} \nabla^{\zeta}H_{\gamma \varepsilon}{}^{\mu}  \nn\\&&+ c_{618} \nabla^{\delta}H^{\alpha \beta \gamma} \nabla^{\varepsilon}H_{\alpha \beta}{}^{\epsilon} \nabla_{\zeta}H_{\delta \epsilon \mu} \nabla^{\zeta}H_{\gamma \varepsilon}{}^{\mu} + c_{619} \nabla_{\delta}H_{\alpha \beta}{}^{\epsilon} \nabla^{\delta}H^{\alpha \beta \gamma} \nabla_{\zeta}H_{\epsilon \varepsilon \mu} \nabla^{\zeta}H_{\gamma}{}^{\varepsilon \mu}  \nn\\&&+ c_{620} \nabla^{\delta}H^{\alpha \beta \gamma} \nabla^{\epsilon}H_{\alpha \beta \gamma} \nabla_{\zeta}H_{\epsilon \varepsilon \mu} \nabla^{\zeta}H_{\delta}{}^{\varepsilon \mu} + c_{621} \nabla_{\gamma}H_{\varepsilon \mu \zeta} \nabla^{\delta}H^{\alpha \beta \gamma} \nabla^{\epsilon}H_{\alpha \beta \delta} \nabla^{\zeta}H_{\epsilon}{}^{\varepsilon \mu}  \nn\\&&+ c_{622} \nabla_{\delta}H_{\alpha \beta \gamma} \nabla^{\delta}H^{\alpha \beta \gamma} \nabla_{\zeta}H_{\epsilon \varepsilon \mu} \nabla^{\zeta}H^{\epsilon \varepsilon \mu}\labell{T48}
\eeqa
\beqa
&&{\cal L}_3^{H^5\prt H\prt\Phi}=\nn\\&& c_{454} H_{\alpha}{}^{\beta \gamma} H_{\beta}{}^{\delta \epsilon} H_{\gamma}{}^{\varepsilon \mu} H_{\zeta \eta \theta} H^{\zeta \eta \theta} \nabla^{\alpha}\Phi \nabla_{\mu}H_{\delta \epsilon \varepsilon} + c_{465} H_{\alpha}{}^{\beta \gamma} H_{\beta}{}^{\delta \epsilon} H_{\gamma}{}^{\varepsilon \mu} H_{\delta}{}^{\zeta \eta} H_{\zeta \eta}{}^{\theta} \nabla^{\alpha}\Phi \nabla_{\mu}H_{\epsilon \varepsilon \theta}\nn\\&& + c_{468} H_{\alpha}{}^{\beta \gamma} H_{\beta}{}^{\delta \epsilon} H_{\gamma}{}^{\varepsilon \mu} H_{\delta}{}^{\zeta \eta} H_{\varepsilon \zeta}{}^{\theta} \nabla^{\alpha}\Phi \nabla_{\mu}H_{\epsilon \eta \theta} + c_{469} H_{\alpha}{}^{\beta \gamma} H_{\beta}{}^{\delta \epsilon} H_{\gamma}{}^{\varepsilon \mu} H_{\delta \varepsilon}{}^{\zeta} H_{\zeta}{}^{\eta \theta} \nabla^{\alpha}\Phi \nabla_{\mu}H_{\epsilon \eta \theta}\nn\\&& + c_{571} H_{\alpha}{}^{\beta \gamma} H_{\beta}{}^{\delta \epsilon} H_{\delta \epsilon}{}^{\varepsilon} H_{\varepsilon}{}^{\mu \zeta} H_{\mu}{}^{\eta \theta} \nabla^{\alpha}\Phi \nabla_{\zeta}H_{\gamma \eta \theta} + c_{572} H_{\alpha}{}^{\beta \gamma} H_{\beta}{}^{\delta \epsilon} H_{\gamma \delta}{}^{\varepsilon} H_{\epsilon}{}^{\mu \zeta} H_{\mu}{}^{\eta \theta} \nabla^{\alpha}\Phi \nabla_{\zeta}H_{\varepsilon \eta \theta} \nn\\&&+ c_{573} H_{\alpha}{}^{\beta \gamma} H_{\beta \gamma}{}^{\delta} H_{\delta}{}^{\epsilon \varepsilon} H_{\epsilon}{}^{\mu \zeta} H_{\mu}{}^{\eta \theta} \nabla^{\alpha}\Phi \nabla_{\zeta}H_{\varepsilon \eta \theta} + c_{574} H_{\alpha}{}^{\beta \gamma} H_{\beta}{}^{\delta \epsilon} H_{\gamma}{}^{\varepsilon \mu} H_{\delta \epsilon}{}^{\zeta} H_{\varepsilon}{}^{\eta \theta} \nabla^{\alpha}\Phi \nabla_{\zeta}H_{\mu \eta \theta} \nn\\&&+ c_{575} H_{\beta \gamma}{}^{\epsilon} H^{\beta \gamma \delta} H_{\delta}{}^{\varepsilon \mu} H_{\epsilon}{}^{\zeta \eta} H_{\varepsilon \zeta}{}^{\theta} \nabla^{\alpha}\Phi \nabla_{\eta}H_{\alpha \mu \theta} + c_{576} H_{\beta \gamma}{}^{\epsilon} H^{\beta \gamma \delta} H_{\delta}{}^{\varepsilon \mu} H_{\epsilon}{}^{\zeta \eta} H_{\varepsilon \mu}{}^{\theta} \nabla^{\alpha}\Phi \nabla_{\eta}H_{\alpha \zeta \theta} \nn\\&&+ c_{577} H_{\alpha}{}^{\beta \gamma} H_{\beta}{}^{\delta \epsilon} H_{\delta}{}^{\varepsilon \mu} H_{\epsilon}{}^{\zeta \eta} H_{\varepsilon \mu}{}^{\theta} \nabla^{\alpha}\Phi \nabla_{\eta}H_{\gamma \zeta \theta} + c_{578} H_{\beta \gamma}{}^{\epsilon} H^{\beta \gamma \delta} H_{\delta}{}^{\varepsilon \mu} H_{\varepsilon}{}^{\zeta \eta} H_{\mu \zeta}{}^{\theta} \nabla^{\alpha}\Phi \nabla_{\theta}H_{\alpha \epsilon \eta} \nn\\&&+ c_{579} H_{\beta}{}^{\epsilon \varepsilon} H^{\beta \gamma \delta} H_{\gamma \epsilon}{}^{\mu} H_{\delta}{}^{\zeta \eta} H_{\varepsilon \zeta}{}^{\theta} \nabla^{\alpha}\Phi \nabla_{\theta}H_{\alpha \mu \eta} + c_{580} H_{\beta \gamma}{}^{\epsilon} H^{\beta \gamma \delta} H_{\delta}{}^{\varepsilon \mu} H_{\epsilon}{}^{\zeta \eta} H_{\varepsilon \zeta}{}^{\theta} \nabla^{\alpha}\Phi \nabla_{\theta}H_{\alpha \mu \eta}\nn\\&& + c_{581} H_{\beta \gamma \delta} H^{\beta \gamma \delta} H_{\epsilon}{}^{\zeta \eta} H^{\epsilon \varepsilon \mu} H_{\varepsilon \zeta}{}^{\theta} \nabla^{\alpha}\Phi \nabla_{\theta}H_{\alpha \mu \eta} + c_{582} H_{\beta}{}^{\epsilon \varepsilon} H^{\beta \gamma \delta} H_{\gamma \epsilon}{}^{\mu} H_{\delta \varepsilon}{}^{\zeta} H_{\mu}{}^{\eta \theta} \nabla^{\alpha}\Phi \nabla_{\theta}H_{\alpha \zeta \eta} \nn\\&&+ c_{583} H_{\beta \gamma}{}^{\epsilon} H^{\beta \gamma \delta} H_{\delta}{}^{\varepsilon \mu} H_{\epsilon \varepsilon}{}^{\zeta} H_{\mu}{}^{\eta \theta} \nabla^{\alpha}\Phi \nabla_{\theta}H_{\alpha \zeta \eta} + c_{584} H_{\beta \gamma \delta} H^{\beta \gamma \delta} H_{\epsilon \varepsilon}{}^{\zeta} H^{\epsilon \varepsilon \mu} H_{\mu}{}^{\eta \theta} \nabla^{\alpha}\Phi \nabla_{\theta}H_{\alpha \zeta \eta} \nn\\&&+ c_{585} H_{\beta \gamma}{}^{\epsilon} H^{\beta \gamma \delta} H_{\delta}{}^{\varepsilon \mu} H_{\epsilon \varepsilon \mu} H^{\zeta \eta \theta} \nabla^{\alpha}\Phi \nabla_{\theta}H_{\alpha \zeta \eta} + c_{586} H_{\alpha}{}^{\beta \gamma} H_{\delta \epsilon}{}^{\mu} H^{\delta \epsilon \varepsilon} H_{\varepsilon}{}^{\zeta \eta} H_{\zeta \eta}{}^{\theta} \nabla^{\alpha}\Phi \nabla_{\theta}H_{\beta \gamma \mu} \nn\\&&+ c_{587} H_{\alpha}{}^{\beta \gamma} H_{\delta}{}^{\mu \zeta} H^{\delta \epsilon \varepsilon} H_{\epsilon \mu}{}^{\eta} H_{\varepsilon \zeta}{}^{\theta} \nabla^{\alpha}\Phi \nabla_{\theta}H_{\beta \gamma \eta} + c_{588} H_{\alpha}{}^{\beta \gamma} H_{\delta \epsilon}{}^{\mu} H^{\delta \epsilon \varepsilon} H_{\varepsilon}{}^{\zeta \eta} H_{\mu \zeta}{}^{\theta} \nabla^{\alpha}\Phi \nabla_{\theta}H_{\beta \gamma \eta} \nn\\&&+ c_{589} H_{\alpha}{}^{\beta \gamma} H_{\delta \epsilon \varepsilon} H^{\delta \epsilon \varepsilon} H_{\mu \zeta}{}^{\theta} H^{\mu \zeta \eta} \nabla^{\alpha}\Phi \nabla_{\theta}H_{\beta \gamma \eta} + c_{590} H_{\alpha}{}^{\beta \gamma} H_{\beta}{}^{\delta \epsilon} H_{\delta}{}^{\varepsilon \mu} H_{\varepsilon}{}^{\zeta \eta} H_{\zeta \eta}{}^{\theta} \nabla^{\alpha}\Phi \nabla_{\theta}H_{\gamma \epsilon \mu} \nn\\&&+ c_{591} H_{\alpha}{}^{\beta \gamma} H_{\beta}{}^{\delta \epsilon} H_{\delta}{}^{\varepsilon \mu} H_{\varepsilon}{}^{\zeta \eta} H_{\mu \zeta}{}^{\theta} \nabla^{\alpha}\Phi \nabla_{\theta}H_{\gamma \epsilon \eta} + c_{592} H_{\alpha}{}^{\beta \gamma} H_{\beta}{}^{\delta \epsilon} H_{\delta}{}^{\varepsilon \mu} H_{\varepsilon \mu}{}^{\zeta} H_{\zeta}{}^{\eta \theta} \nabla^{\alpha}\Phi \nabla_{\theta}H_{\gamma \epsilon \eta} \nn\\&&+ c_{593} H_{\alpha}{}^{\beta \gamma} H_{\beta}{}^{\delta \epsilon} H_{\delta \epsilon}{}^{\varepsilon} H_{\mu \zeta}{}^{\theta} H^{\mu \zeta \eta} \nabla^{\alpha}\Phi \nabla_{\theta}H_{\gamma \varepsilon \eta} + c_{594} H_{\alpha}{}^{\beta \gamma} H_{\beta}{}^{\delta \epsilon} H_{\delta}{}^{\varepsilon \mu} H_{\epsilon}{}^{\zeta \eta} H_{\varepsilon \mu}{}^{\theta} \nabla^{\alpha}\Phi \nabla_{\theta}H_{\gamma \zeta \eta} \nn\\&&+ c_{595} H_{\alpha}{}^{\beta \gamma} H_{\beta}{}^{\delta \epsilon} H_{\delta}{}^{\varepsilon \mu} H_{\epsilon \varepsilon}{}^{\zeta} H_{\mu}{}^{\eta \theta} \nabla^{\alpha}\Phi \nabla_{\theta}H_{\gamma \zeta \eta} + c_{596} H_{\alpha}{}^{\beta \gamma} H_{\beta}{}^{\delta \epsilon} H_{\delta \epsilon}{}^{\varepsilon} H_{\varepsilon}{}^{\mu \zeta} H_{\mu}{}^{\eta \theta} \nabla^{\alpha}\Phi \nabla_{\theta}H_{\gamma \zeta \eta}\nn\\&& + c_{597} H_{\alpha}{}^{\beta \gamma} H_{\beta \gamma}{}^{\delta} H_{\epsilon}{}^{\zeta \eta} H^{\epsilon \varepsilon \mu} H_{\varepsilon \zeta}{}^{\theta} \nabla^{\alpha}\Phi \nabla_{\theta}H_{\delta \mu \eta} + c_{598} H_{\alpha}{}^{\beta \gamma} H_{\beta \gamma}{}^{\delta} H_{\epsilon \varepsilon}{}^{\zeta} H^{\epsilon \varepsilon \mu} H_{\mu}{}^{\eta \theta} \nabla^{\alpha}\Phi \nabla_{\theta}H_{\delta \zeta \eta}\nn\\&& + c_{599} H_{\alpha}{}^{\beta \gamma} H_{\beta \gamma}{}^{\delta} H_{\epsilon \varepsilon \mu} H^{\epsilon \varepsilon \mu} H^{\zeta \eta \theta} \nabla^{\alpha}\Phi \nabla_{\theta}H_{\delta \zeta \eta} + c_{600} H_{\alpha}{}^{\beta \gamma} H_{\beta}{}^{\delta \epsilon} H_{\gamma}{}^{\varepsilon \mu} H_{\delta}{}^{\zeta \eta} H_{\zeta \eta}{}^{\theta} \nabla^{\alpha}\Phi \nabla_{\theta}H_{\epsilon \varepsilon \mu}\nn\\&& + c_{601} H_{\alpha}{}^{\beta \gamma} H_{\beta}{}^{\delta \epsilon} H_{\gamma \delta}{}^{\varepsilon} H_{\mu \zeta}{}^{\theta} H^{\mu \zeta \eta} \nabla^{\alpha}\Phi \nabla_{\theta}H_{\epsilon \varepsilon \eta} + c_{602} H_{\alpha}{}^{\beta \gamma} H_{\beta \gamma}{}^{\delta} H_{\delta}{}^{\epsilon \varepsilon} H_{\mu \zeta}{}^{\theta} H^{\mu \zeta \eta} \nabla^{\alpha}\Phi \nabla_{\theta}H_{\epsilon \varepsilon \eta} \nn\\&&+ c_{603} H_{\alpha}{}^{\beta \gamma} H_{\beta}{}^{\delta \epsilon} H_{\gamma}{}^{\varepsilon \mu} H_{\delta}{}^{\zeta \eta} H_{\varepsilon \zeta}{}^{\theta} \nabla^{\alpha}\Phi \nabla_{\theta}H_{\epsilon \mu \eta} + c_{604} H_{\alpha}{}^{\beta \gamma} H_{\beta}{}^{\delta \epsilon} H_{\gamma}{}^{\varepsilon \mu} H_{\delta}{}^{\zeta \eta} H_{\epsilon \zeta}{}^{\theta} \nabla^{\alpha}\Phi \nabla_{\theta}H_{\varepsilon \mu \eta} \nn\\&&+ c_{605} H_{\alpha}{}^{\beta \gamma} H_{\beta}{}^{\delta \epsilon} H_{\gamma}{}^{\varepsilon \mu} H_{\delta \epsilon}{}^{\zeta} H_{\zeta}{}^{\eta \theta} \nabla^{\alpha}\Phi \nabla_{\theta}H_{\varepsilon \mu \eta} + c_{606} H_{\alpha}{}^{\beta \gamma} H_{\beta}{}^{\delta \epsilon} H_{\gamma \delta}{}^{\varepsilon} H_{\epsilon}{}^{\mu \zeta} H_{\mu}{}^{\eta \theta} \nabla^{\alpha}\Phi \nabla_{\theta}H_{\varepsilon \zeta \eta} \nn\\&&+ c_{607} H_{\alpha}{}^{\beta \gamma} H_{\beta \gamma}{}^{\delta} H_{\delta}{}^{\epsilon \varepsilon} H_{\epsilon}{}^{\mu \zeta} H_{\mu}{}^{\eta \theta} \nabla^{\alpha}\Phi \nabla_{\theta}H_{\varepsilon \zeta \eta} + c_{608} H_{\alpha}{}^{\beta \gamma} H_{\beta}{}^{\delta \epsilon} H_{\gamma}{}^{\varepsilon \mu} H_{\delta \varepsilon}{}^{\zeta} H_{\epsilon}{}^{\eta \theta} \nabla^{\alpha}\Phi \nabla_{\theta}H_{\mu \zeta \eta} \nn\\&&+ c_{609} H_{\alpha}{}^{\beta \gamma} H_{\beta}{}^{\delta \epsilon} H_{\gamma}{}^{\varepsilon \mu} H_{\delta \epsilon}{}^{\zeta} H_{\varepsilon}{}^{\eta \theta} \nabla^{\alpha}\Phi \nabla_{\theta}H_{\mu \zeta \eta} + c_{610} H_{\alpha}{}^{\beta \gamma} H_{\beta}{}^{\delta \epsilon} H_{\gamma \delta}{}^{\varepsilon} H_{\epsilon}{}^{\mu \zeta} H_{\varepsilon}{}^{\eta \theta} \nabla^{\alpha}\Phi \nabla_{\theta}H_{\mu \zeta \eta} \nn\\&&+ c_{611} H_{\alpha}{}^{\beta \gamma} H_{\beta \gamma}{}^{\delta} H_{\delta}{}^{\epsilon \varepsilon} H_{\epsilon}{}^{\mu \zeta} H_{\varepsilon}{}^{\eta \theta} \nabla^{\alpha}\Phi \nabla_{\theta}H_{\mu \zeta \eta} + c_{612} H_{\alpha}{}^{\beta \gamma} H_{\beta}{}^{\delta \epsilon} H_{\gamma}{}^{\varepsilon \mu} H_{\delta \epsilon \varepsilon} H^{\zeta \eta \theta} \nabla^{\alpha}\Phi \nabla_{\theta}H_{\mu \zeta \eta} \nn\\&&+ c_{613} H_{\alpha}{}^{\beta \gamma} H_{\beta}{}^{\delta \epsilon} H_{\gamma \delta}{}^{\varepsilon} H_{\epsilon \varepsilon}{}^{\mu} H^{\zeta \eta \theta} \nabla^{\alpha}\Phi \nabla_{\theta}H_{\mu \zeta \eta} + c_{614} H_{\alpha}{}^{\beta \gamma} H_{\beta \gamma}{}^{\delta} H_{\delta}{}^{\epsilon \varepsilon} H_{\epsilon \varepsilon}{}^{\mu} H^{\zeta \eta \theta} \nabla^{\alpha}\Phi \nabla_{\theta}H_{\mu \zeta \eta}\labell{T49}\nn
\eeqa
\beqa
{\cal L}_3^{R(\prt H)^2\prt\prt\Phi}&\!\!\!\!\!=\!\!\!\!\! & c_{480} R_{\delta \mu \epsilon \varepsilon} \nabla^{\beta}\nabla^{\alpha}\Phi \nabla^{\epsilon}H_{\alpha}{}^{\gamma \delta} \nabla^{\mu}H_{\beta \gamma}{}^{\varepsilon} + c_{486} R_{\gamma \varepsilon \delta \mu} \nabla^{\beta}\nabla^{\alpha}\Phi \nabla^{\epsilon}H_{\alpha}{}^{\gamma \delta} \nabla^{\mu}H_{\beta \epsilon}{}^{\varepsilon}  \nn\\&&+ c_{499} R_{\beta \varepsilon \epsilon \mu} \nabla^{\beta}\nabla^{\alpha}\Phi \nabla^{\epsilon}H_{\alpha}{}^{\gamma \delta} \nabla^{\mu}H_{\gamma \delta}{}^{\varepsilon} + c_{500} R_{\beta \mu \epsilon \varepsilon} \nabla^{\beta}\nabla^{\alpha}\Phi \nabla^{\epsilon}H_{\alpha}{}^{\gamma \delta} \nabla^{\mu}H_{\gamma \delta}{}^{\varepsilon}  \nn\\&&+ c_{502} R_{\beta \varepsilon \delta \mu} \nabla^{\beta}\nabla^{\alpha}\Phi \nabla^{\epsilon}H_{\alpha}{}^{\gamma \delta} \nabla^{\mu}H_{\gamma \epsilon}{}^{\varepsilon} + c_{503} R_{\beta \mu \delta \varepsilon} \nabla^{\beta}\nabla^{\alpha}\Phi \nabla^{\epsilon}H_{\alpha}{}^{\gamma \delta} \nabla^{\mu}H_{\gamma \epsilon}{}^{\varepsilon} \nn\\&& + c_{506} R_{\delta \mu \epsilon \varepsilon} \nabla_{\beta}H_{\alpha}{}^{\gamma \delta} \nabla^{\beta}\nabla^{\alpha}\Phi \nabla^{\mu}H_{\gamma}{}^{\epsilon \varepsilon}\labell{T50}
\eeqa
\beqa
{\cal L}_3^{H^2R(\prt H)^2}&\!\!\!\!\!=\!\!\!\!\! & c_{481} H_{\alpha}{}^{\delta \epsilon} H^{\alpha \beta \gamma} R_{\delta \zeta \epsilon \eta} \nabla_{\mu}H_{\varepsilon}{}^{\zeta \eta} \nabla^{\mu}H_{\beta \gamma}{}^{\varepsilon} + c_{485} H_{\alpha}{}^{\delta \epsilon} H^{\alpha \beta \gamma} R_{\epsilon \zeta\mu \eta} \nabla_{\varepsilon}H_{\gamma}{}^{\zeta \eta} \nabla^{\mu}H_{\beta \delta}{}^{\varepsilon} \nn\\&& + c_{508} H_{\alpha \beta}{}^{\delta} H^{\alpha \beta \gamma} R_{\epsilon \zeta \varepsilon \eta} \nabla_{\mu}H_{\delta}{}^{\zeta \eta} \nabla^{\mu}H_{\gamma}{}^{\epsilon \varepsilon} + c_{509} H_{\alpha \beta}{}^{\delta} H^{\alpha \beta \gamma} R_{\delta \zeta \varepsilon \eta} \nabla_{\mu}H_{\epsilon}{}^{\zeta \eta} \nabla^{\mu}H_{\gamma}{}^{\epsilon \varepsilon}  \nn\\&&+ c_{700} H^{\alpha \beta \gamma} H^{\delta \epsilon \varepsilon} R_{\beta \varepsilon \gamma \eta} \nabla_{\delta}H_{\alpha}{}^{\mu \zeta} \nabla_{\zeta}H_{\epsilon\mu}{}^{\eta} + c_{701} H^{\alpha \beta \gamma} H^{\delta \epsilon \varepsilon} R_{\epsilon\mu \varepsilon \eta} \nabla_{\delta}H_{\gamma \zeta}{}^{\eta} \nabla^{\zeta}H_{\alpha \beta}{}^{\mu}  \nn\\&&+ c_{702} H^{\alpha \beta \gamma} H^{\delta \epsilon \varepsilon} R_{\gamma \eta \epsilon \varepsilon} \nabla_{\mu}H_{\delta \zeta}{}^{\eta} \nabla^{\zeta}H_{\alpha \beta}{}^{\mu} + c_{703} H^{\alpha \beta \gamma} H^{\delta \epsilon \varepsilon} R_{\epsilon\mu \varepsilon \eta} \nabla_{\zeta}H_{\gamma \delta}{}^{\eta} \nabla^{\zeta}H_{\alpha \beta}{}^{\mu}  \nn\\&&+ c_{704} H^{\alpha \beta \gamma} H^{\delta \epsilon \varepsilon} R_{\gamma\mu \varepsilon \eta} \nabla_{\zeta}H_{\delta \epsilon}{}^{\eta} \nabla^{\zeta}H_{\alpha \beta}{}^{\mu} + c_{705} H^{\alpha \beta \gamma} H^{\delta \epsilon \varepsilon} R_{\gamma \eta \varepsilon\mu} \nabla_{\zeta}H_{\delta \epsilon}{}^{\eta} \nabla^{\zeta}H_{\alpha \beta}{}^{\mu}  \nn\\&&+ c_{706} H^{\alpha \beta \gamma} H^{\delta \epsilon \varepsilon} R_{\gamma \eta \epsilon \varepsilon} \nabla_{\zeta}H_{\delta\mu}{}^{\eta} \nabla^{\zeta}H_{\alpha \beta}{}^{\mu} + c_{707} H^{\alpha \beta \gamma} H^{\delta \epsilon \varepsilon} R_{\epsilon\mu \varepsilon \eta} \nabla_{\gamma}H_{\beta \zeta}{}^{\eta} \nabla^{\zeta}H_{\alpha \delta}{}^{\mu}  \nn\\&&+ c_{708} H^{\alpha \beta \gamma} H^{\delta \epsilon \varepsilon} R_{\gamma \eta \varepsilon \zeta} \nabla_{\epsilon}H_{\beta\mu}{}^{\eta} \nabla^{\zeta}H_{\alpha \delta}{}^{\mu} + c_{709} H^{\alpha \beta \gamma} H^{\delta \epsilon \varepsilon} R_{\gamma \eta \varepsilon\mu} \nabla_{\epsilon}H_{\beta \zeta}{}^{\eta} \nabla^{\zeta}H_{\alpha \delta}{}^{\mu} \nn\\&& + c_{710} H^{\alpha \beta \gamma} H^{\delta \epsilon \varepsilon} R_{\gamma \eta \epsilon \varepsilon} \nabla_{\mu}H_{\beta \zeta}{}^{\eta} \nabla^{\zeta}H_{\alpha \delta}{}^{\mu} + c_{711} H^{\alpha \beta \gamma} H^{\delta \epsilon \varepsilon} R_{\gamma\mu \varepsilon \eta} \nabla_{\zeta}H_{\beta \epsilon}{}^{\eta} \nabla^{\zeta}H_{\alpha \delta}{}^{\mu} \nn\\&& + c_{712} H^{\alpha \beta \gamma} H^{\delta \epsilon \varepsilon} R_{\gamma \eta \epsilon \varepsilon} \nabla_{\zeta}H_{\beta\mu}{}^{\eta} \nabla^{\zeta}H_{\alpha \delta}{}^{\mu} + c_{713} H_{\alpha}{}^{\delta \epsilon} H^{\alpha \beta \gamma} R_{\delta\mu \epsilon \eta} \nabla_{\varepsilon}H_{\gamma \zeta}{}^{\eta} \nabla^{\zeta}H_{\beta}{}^{\varepsilon\mu}  \nn\\&&+ c_{714} H_{\alpha}{}^{\delta \epsilon} H^{\alpha \beta \gamma} R_{\gamma\mu \epsilon \eta} \nabla_{\varepsilon}H_{\delta \zeta}{}^{\eta} \nabla^{\zeta}H_{\beta}{}^{\varepsilon\mu} + c_{715} H_{\alpha}{}^{\delta \epsilon} H^{\alpha \beta \gamma} R_{\gamma \eta \epsilon\mu} \nabla_{\varepsilon}H_{\delta \zeta}{}^{\eta} \nabla^{\zeta}H_{\beta}{}^{\varepsilon\mu} \nn\\&& + c_{716} H_{\alpha}{}^{\delta \epsilon} H^{\alpha \beta \gamma} R_{\delta\mu \epsilon \eta} \nabla_{\zeta}H_{\gamma \varepsilon}{}^{\eta} \nabla^{\zeta}H_{\beta}{}^{\varepsilon\mu} + c_{717} H_{\alpha}{}^{\delta \epsilon} H^{\alpha \beta \gamma} R_{\gamma\mu \epsilon \eta} \nabla_{\zeta}H_{\delta \varepsilon}{}^{\eta} \nabla^{\zeta}H_{\beta}{}^{\varepsilon\mu}  \nn\\&&+ c_{718} H_{\alpha}{}^{\delta \epsilon} H^{\alpha \beta \gamma} R_{\gamma \eta \epsilon\mu} \nabla_{\zeta}H_{\delta \varepsilon}{}^{\eta} \nabla^{\zeta}H_{\beta}{}^{\varepsilon\mu} + c_{719} H_{\alpha}{}^{\delta \epsilon} H^{\alpha \beta \gamma} R_{\gamma \eta \delta \epsilon} \nabla_{\zeta}H_{\varepsilon\mu}{}^{\eta} \nabla^{\zeta}H_{\beta}{}^{\varepsilon\mu} \nn\\&& + c_{720} H_{\alpha \beta}{}^{\delta} H^{\alpha \beta \gamma} R_{\gamma\mu \delta \eta} \nabla_{\zeta}H_{\epsilon \varepsilon}{}^{\eta} \nabla^{\zeta}H^{\epsilon \varepsilon\mu} + c_{721} H^{\alpha \beta \gamma} H^{\delta \epsilon \varepsilon} R_{\beta \epsilon \gamma \varepsilon} \nabla_{\zeta}H_{\delta\mu \eta} \nabla^{\eta}H_{\alpha}{}^{\mu \zeta} \nn\\&& + c_{722} H^{\alpha \beta \gamma} H^{\delta \epsilon \varepsilon} R_{\beta \epsilon \gamma \varepsilon} \nabla_{\eta}H_{\delta\mu \zeta} \nabla^{\eta}H_{\alpha}{}^{\mu \zeta} + c_{723} H^{\alpha \beta \gamma} H^{\delta \epsilon \varepsilon} R_{\gamma \zeta \varepsilon \eta} \nabla^{\zeta}H_{\alpha \delta}{}^{\mu} \nabla^{\eta}H_{\beta \epsilon\mu}  \nn\\&&+ c_{724} H^{\alpha \beta \gamma} H^{\delta \epsilon \varepsilon} R_{\gamma\mu \varepsilon \eta} \nabla^{\zeta}H_{\alpha \delta}{}^{\mu} \nabla^{\eta}H_{\beta \epsilon \zeta} + c_{725} H^{\alpha \beta \gamma} H^{\delta \epsilon \varepsilon} R_{\gamma \eta \epsilon \varepsilon} \nabla^{\zeta}H_{\alpha \delta}{}^{\mu} \nabla^{\eta}H_{\beta\mu \zeta} \nn\\&& + c_{726} H^{\alpha \beta \gamma} H^{\delta \epsilon \varepsilon} R_{\epsilon\mu \varepsilon \eta} \nabla^{\zeta}H_{\alpha \beta}{}^{\mu} \nabla^{\eta}H_{\gamma \delta \zeta} + c_{727} H^{\alpha \beta \gamma} H^{\delta \epsilon \varepsilon} R_{\varepsilon \zeta\mu \eta} \nabla_{\delta}H_{\alpha \beta}{}^{\mu} \nabla^{\eta}H_{\gamma \epsilon}{}^{\zeta}  \nn\\&&+ c_{728} H^{\alpha \beta \gamma} H^{\delta \epsilon \varepsilon} R_{\varepsilon \eta\mu \zeta} \nabla_{\delta}H_{\alpha \beta}{}^{\mu} \nabla^{\eta}H_{\gamma \epsilon}{}^{\zeta} + c_{729} H^{\alpha \beta \gamma} H^{\delta \epsilon \varepsilon} R_{\varepsilon \eta\mu \zeta} \nabla^{\mu}H_{\alpha \beta \delta} \nabla^{\eta}H_{\gamma \epsilon}{}^{\zeta}  \nn\\&&+ c_{730} H_{\alpha}{}^{\delta \epsilon} H^{\alpha \beta \gamma} R_{\varepsilon \zeta\mu \eta} \nabla^{\mu}H_{\beta \delta}{}^{\varepsilon} \nabla^{\eta}H_{\gamma \epsilon}{}^{\zeta} + c_{731} H_{\alpha}{}^{\delta \epsilon} H^{\alpha \beta \gamma} R_{\varepsilon \eta\mu \zeta} \nabla^{\mu}H_{\beta \delta}{}^{\varepsilon} \nabla^{\eta}H_{\gamma \epsilon}{}^{\zeta}  \nn\\&&+ c_{732} H_{\alpha}{}^{\delta \epsilon} H^{\alpha \beta \gamma} R_{\delta \zeta \epsilon \eta} \nabla^{\zeta}H_{\beta}{}^{\varepsilon\mu} \nabla^{\eta}H_{\gamma \varepsilon\mu} + c_{733} H_{\alpha}{}^{\delta \epsilon} H^{\alpha \beta \gamma} R_{\delta\mu \epsilon \eta} \nabla^{\zeta}H_{\beta}{}^{\varepsilon\mu} \nabla^{\eta}H_{\gamma \varepsilon \zeta}  \nn\\&&+ c_{734} H_{\alpha}{}^{\delta \epsilon} H^{\alpha \beta \gamma} R_{\epsilon \zeta\mu \eta} \nabla^{\mu}H_{\beta \delta}{}^{\varepsilon} \nabla^{\eta}H_{\gamma \varepsilon}{}^{\zeta} + c_{735} H_{\alpha}{}^{\delta \epsilon} H^{\alpha \beta \gamma} R_{\epsilon \eta\mu \zeta} \nabla^{\mu}H_{\beta \delta}{}^{\varepsilon} \nabla^{\eta}H_{\gamma \varepsilon}{}^{\zeta} \nn\\&& + c_{736} H^{\alpha \beta \gamma} H^{\delta \epsilon \varepsilon} R_{\epsilon \zeta \varepsilon \eta} \nabla_{\delta}H_{\alpha \beta}{}^{\mu} \nabla^{\eta}H_{\gamma\mu}{}^{\zeta} + c_{737} H^{\alpha \beta \gamma} H^{\delta \epsilon \varepsilon} R_{\epsilon \zeta \varepsilon \eta} \nabla^{\mu}H_{\alpha \beta \delta} \nabla^{\eta}H_{\gamma\mu}{}^{\zeta} \nn\\&& + c_{738} H_{\alpha}{}^{\delta \epsilon} H^{\alpha \beta \gamma} R_{\epsilon \zeta \varepsilon \eta} \nabla^{\mu}H_{\beta \delta}{}^{\varepsilon} \nabla^{\eta}H_{\gamma\mu}{}^{\zeta} + c_{739} H_{\alpha}{}^{\delta \epsilon} H^{\alpha \beta \gamma} R_{\epsilon \eta \varepsilon \zeta} \nabla^{\mu}H_{\beta \delta}{}^{\varepsilon} \nabla^{\eta}H_{\gamma\mu}{}^{\zeta}  \nn\\&&+ c_{740} H^{\alpha \beta \gamma} H^{\delta \epsilon \varepsilon} R_{\gamma \zeta \varepsilon \eta} \nabla^{\zeta}H_{\alpha \beta}{}^{\mu} \nabla^{\eta}H_{\delta \epsilon\mu} + c_{741} H^{\alpha \beta \gamma} H^{\delta \epsilon \varepsilon} R_{\gamma \eta \varepsilon \zeta} \nabla^{\zeta}H_{\alpha \beta}{}^{\mu} \nabla^{\eta}H_{\delta \epsilon\mu}  \nn\\&&+ c_{742} H^{\alpha \beta \gamma} H^{\delta \epsilon \varepsilon} R_{\gamma\mu \varepsilon \eta} \nabla^{\zeta}H_{\alpha \beta}{}^{\mu} \nabla^{\eta}H_{\delta \epsilon \zeta} + c_{743} H^{\alpha \beta \gamma} H^{\delta \epsilon \varepsilon} R_{\gamma \eta \varepsilon\mu} \nabla^{\zeta}H_{\alpha \beta}{}^{\mu} \nabla^{\eta}H_{\delta \epsilon \zeta}  \nn\\&&+ c_{744} H_{\alpha}{}^{\delta \epsilon} H^{\alpha \beta \gamma} R_{\varepsilon \zeta\mu \eta} \nabla^{\mu}H_{\beta \gamma}{}^{\varepsilon} \nabla^{\eta}H_{\delta \epsilon}{}^{\zeta} + c_{745} H_{\alpha}{}^{\delta \epsilon} H^{\alpha \beta \gamma} R_{\varepsilon \eta\mu \zeta} \nabla^{\mu}H_{\beta \gamma}{}^{\varepsilon} \nabla^{\eta}H_{\delta \epsilon}{}^{\zeta} \nn\\&& + c_{746} H_{\alpha \beta}{}^{\delta} H^{\alpha \beta \gamma} R_{\varepsilon \zeta\mu \eta} \nabla^{\mu}H_{\gamma}{}^{\epsilon \varepsilon} \nabla^{\eta}H_{\delta \epsilon}{}^{\zeta} + c_{747} H_{\alpha \beta}{}^{\delta} H^{\alpha \beta \gamma} R_{\varepsilon \eta\mu \zeta} \nabla^{\mu}H_{\gamma}{}^{\epsilon \varepsilon} \nabla^{\eta}H_{\delta \epsilon}{}^{\zeta}  \nn\\&&+ c_{748} H_{\alpha \beta \gamma} H^{\alpha \beta \gamma} R_{\varepsilon \eta\mu \zeta} \nabla^{\mu}H^{\delta \epsilon \varepsilon} \nabla^{\eta}H_{\delta \epsilon}{}^{\zeta} + c_{749} H_{\alpha}{}^{\delta \epsilon} H^{\alpha \beta \gamma} R_{\gamma \zeta \epsilon \eta} \nabla^{\zeta}H_{\beta}{}^{\varepsilon\mu} \nabla^{\eta}H_{\delta \varepsilon\mu}  \nn\\&&+ c_{750} H_{\alpha}{}^{\delta \epsilon} H^{\alpha \beta \gamma} R_{\gamma \eta \epsilon \zeta} \nabla^{\zeta}H_{\beta}{}^{\varepsilon\mu} \nabla^{\eta}H_{\delta \varepsilon\mu} + c_{751} H_{\alpha}{}^{\delta \epsilon} H^{\alpha \beta \gamma} R_{\gamma\mu \epsilon \eta} \nabla^{\zeta}H_{\beta}{}^{\varepsilon\mu} \nabla^{\eta}H_{\delta \varepsilon \zeta}  \nn\\&&+ c_{752} H_{\alpha}{}^{\delta \epsilon} H^{\alpha \beta \gamma} R_{\gamma \eta \epsilon\mu} \nabla^{\zeta}H_{\beta}{}^{\varepsilon\mu} \nabla^{\eta}H_{\delta \varepsilon \zeta} + c_{753} H_{\alpha}{}^{\delta \epsilon} H^{\alpha \beta \gamma} R_{\epsilon \zeta\mu \eta} \nabla^{\mu}H_{\beta \gamma}{}^{\varepsilon} \nabla^{\eta}H_{\delta \varepsilon}{}^{\zeta} \nn\\&& + c_{754} H_{\alpha}{}^{\delta \epsilon} H^{\alpha \beta \gamma} R_{\epsilon \eta\mu \zeta} \nabla^{\mu}H_{\beta \gamma}{}^{\varepsilon} \nabla^{\eta}H_{\delta \varepsilon}{}^{\zeta} + c_{755} H^{\alpha \beta \gamma} H^{\delta \epsilon \varepsilon} R_{\gamma \eta \epsilon \varepsilon} \nabla^{\zeta}H_{\alpha \beta}{}^{\mu} \nabla^{\eta}H_{\delta\mu \zeta} \nn\\&& + c_{756} H^{\alpha \beta \gamma} H^{\delta \epsilon \varepsilon} R_{\epsilon \zeta \varepsilon \eta} \nabla^{\mu}H_{\alpha \beta \gamma} \nabla^{\eta}H_{\delta\mu}{}^{\zeta} + c_{757} H_{\alpha}{}^{\delta \epsilon} H^{\alpha \beta \gamma} R_{\epsilon \zeta \varepsilon \eta} \nabla^{\mu}H_{\beta \gamma}{}^{\varepsilon} \nabla^{\eta}H_{\delta\mu}{}^{\zeta}  \nn\\&&+ c_{758} H_{\alpha}{}^{\delta \epsilon} H^{\alpha \beta \gamma} R_{\epsilon \eta \varepsilon \zeta} \nabla^{\mu}H_{\beta \gamma}{}^{\varepsilon} \nabla^{\eta}H_{\delta\mu}{}^{\zeta} + c_{759} H_{\alpha \beta}{}^{\delta} H^{\alpha \beta \gamma} R_{\epsilon \zeta \varepsilon \eta} \nabla^{\mu}H_{\gamma}{}^{\epsilon \varepsilon} \nabla^{\eta}H_{\delta\mu}{}^{\zeta}  \nn\\&&+ c_{760} H_{\alpha \beta \gamma} H^{\alpha \beta \gamma} R_{\epsilon \zeta \varepsilon \eta} \nabla^{\mu}H^{\delta \epsilon \varepsilon} \nabla^{\eta}H_{\delta\mu}{}^{\zeta} + c_{761} H_{\alpha \beta}{}^{\delta} H^{\alpha \beta \gamma} R_{\gamma\mu \delta \eta} \nabla^{\zeta}H^{\epsilon \varepsilon\mu} \nabla^{\eta}H_{\epsilon \varepsilon \zeta}  \nn\\&&+ c_{762} H^{\alpha \beta \gamma} H^{\delta \epsilon \varepsilon} R_{\gamma \zeta\mu \eta} \nabla_{\delta}H_{\alpha \beta}{}^{\mu} \nabla^{\eta}H_{\epsilon \varepsilon}{}^{\zeta} + c_{763} H^{\alpha \beta \gamma} H^{\delta \epsilon \varepsilon} R_{\gamma \eta\mu \zeta} \nabla_{\delta}H_{\alpha \beta}{}^{\mu} \nabla^{\eta}H_{\epsilon \varepsilon}{}^{\zeta}  \nn\\&&+ c_{764} H_{\alpha}{}^{\delta \epsilon} H^{\alpha \beta \gamma} R_{\gamma \zeta\mu \eta} \nabla_{\delta}H_{\beta}{}^{\varepsilon\mu} \nabla^{\eta}H_{\epsilon \varepsilon}{}^{\zeta} + c_{765} H_{\alpha}{}^{\delta \epsilon} H^{\alpha \beta \gamma} R_{\gamma \eta\mu \zeta} \nabla_{\delta}H_{\beta}{}^{\varepsilon\mu} \nabla^{\eta}H_{\epsilon \varepsilon}{}^{\zeta}  \nn\\&&+ c_{766} H^{\alpha \beta \gamma} H^{\delta \epsilon \varepsilon} R_{\gamma \zeta\mu \eta} \nabla^{\mu}H_{\alpha \beta \delta} \nabla^{\eta}H_{\epsilon \varepsilon}{}^{\zeta} + c_{767} H^{\alpha \beta \gamma} H^{\delta \epsilon \varepsilon} R_{\gamma \eta\mu \zeta} \nabla^{\mu}H_{\alpha \beta \delta} \nabla^{\eta}H_{\epsilon \varepsilon}{}^{\zeta}  \nn\\&&+ c_{768} H_{\alpha \beta}{}^{\delta} H^{\alpha \beta \gamma} R_{\delta \zeta\mu \eta} \nabla^{\mu}H_{\gamma}{}^{\epsilon \varepsilon} \nabla^{\eta}H_{\epsilon \varepsilon}{}^{\zeta} + c_{769} H_{\alpha \beta}{}^{\delta} H^{\alpha \beta \gamma} R_{\delta \eta\mu \zeta} \nabla^{\mu}H_{\gamma}{}^{\epsilon \varepsilon} \nabla^{\eta}H_{\epsilon \varepsilon}{}^{\zeta}  \nn\\&&+ c_{770} H^{\alpha \beta \gamma} H^{\delta \epsilon \varepsilon} R_{\beta \varepsilon \gamma \eta} \nabla_{\delta}H_{\alpha}{}^{\mu \zeta} \nabla^{\eta}H_{\epsilon\mu \zeta} + c_{771} H^{\alpha \beta \gamma} H^{\delta \epsilon \varepsilon} R_{\gamma \zeta \varepsilon \eta} \nabla_{\delta}H_{\alpha \beta}{}^{\mu} \nabla^{\eta}H_{\epsilon\mu}{}^{\zeta} \nn\\&& + c_{772} H^{\alpha \beta \gamma} H^{\delta \epsilon \varepsilon} R_{\gamma \eta \varepsilon \zeta} \nabla_{\delta}H_{\alpha \beta}{}^{\mu} \nabla^{\eta}H_{\epsilon\mu}{}^{\zeta} + c_{773} H^{\alpha \beta \gamma} H^{\delta \epsilon \varepsilon} R_{\gamma \zeta \varepsilon \eta} \nabla^{\mu}H_{\alpha \beta \delta} \nabla^{\eta}H_{\epsilon\mu}{}^{\zeta}  \nn\\&&+ c_{774} H^{\alpha \beta \gamma} H^{\delta \epsilon \varepsilon} R_{\gamma \eta \varepsilon \zeta} \nabla^{\mu}H_{\alpha \beta \delta} \nabla^{\eta}H_{\epsilon\mu}{}^{\zeta} + c_{775} H_{\alpha \beta}{}^{\delta} H^{\alpha \beta \gamma} R_{\delta \zeta \varepsilon \eta} \nabla^{\mu}H_{\gamma}{}^{\epsilon \varepsilon} \nabla^{\eta}H_{\epsilon\mu}{}^{\zeta}  \nn\\&&+ c_{776} H_{\alpha \beta}{}^{\delta} H^{\alpha \beta \gamma} R_{\delta \eta \varepsilon \zeta} \nabla^{\mu}H_{\gamma}{}^{\epsilon \varepsilon} \nabla^{\eta}H_{\epsilon\mu}{}^{\zeta} + c_{777} H^{\alpha \beta \gamma} H^{\delta \epsilon \varepsilon} R_{\varepsilon \eta\mu \zeta} \nabla_{\delta}H_{\alpha \beta \gamma} \nabla^{\eta}H_{\epsilon}{}^{\mu \zeta} \nn\\&& + c_{778} H_{\alpha}{}^{\delta \epsilon} H^{\alpha \beta \gamma} R_{\varepsilon \eta\mu \zeta} \nabla_{\delta}H_{\beta \gamma}{}^{\varepsilon} \nabla^{\eta}H_{\epsilon}{}^{\mu \zeta} + c_{779} H_{\alpha \beta}{}^{\delta} H^{\alpha \beta \gamma} R_{\varepsilon \eta\mu \zeta} \nabla_{\delta}H_{\gamma}{}^{\epsilon \varepsilon} \nabla^{\eta}H_{\epsilon}{}^{\mu \zeta}  \nn\\&&+ c_{780} H_{\alpha}{}^{\delta \epsilon} H^{\alpha \beta \gamma} R_{\varepsilon \eta\mu \zeta} \nabla^{\varepsilon}H_{\beta \gamma \delta} \nabla^{\eta}H_{\epsilon}{}^{\mu \zeta} + c_{781} H_{\alpha}{}^{\delta \epsilon} H^{\alpha \beta \gamma} R_{\gamma \eta \delta \epsilon} \nabla^{\zeta}H_{\beta}{}^{\varepsilon\mu} \nabla^{\eta}H_{\varepsilon\mu \zeta} \nn\\&& + c_{782} H_{\alpha}{}^{\delta \epsilon} H^{\alpha \beta \gamma} R_{\gamma \eta \epsilon \zeta} \nabla_{\delta}H_{\beta}{}^{\varepsilon\mu} \nabla^{\eta}H_{\varepsilon\mu}{}^{\zeta} + c_{783} H_{\alpha}{}^{\delta \epsilon} H^{\alpha \beta \gamma} R_{\delta \zeta \epsilon \eta} \nabla^{\mu}H_{\beta \gamma}{}^{\varepsilon} \nabla^{\eta}H_{\varepsilon\mu}{}^{\zeta}  \nn\\&&+ c_{784} H_{\alpha}{}^{\delta \epsilon} H^{\alpha \beta \gamma} R_{\gamma \zeta \epsilon \eta} \nabla^{\mu}H_{\beta \delta}{}^{\varepsilon} \nabla^{\eta}H_{\varepsilon\mu}{}^{\zeta} + c_{785} H_{\alpha}{}^{\delta \epsilon} H^{\alpha \beta \gamma} R_{\epsilon \eta\mu \zeta} \nabla_{\delta}H_{\beta \gamma}{}^{\varepsilon} \nabla^{\eta}H_{\varepsilon}{}^{\mu \zeta}  \nn\\&&+ c_{786} H^{\alpha \beta \gamma} H^{\delta \epsilon \varepsilon} R_{\gamma \eta\mu \zeta} \nabla_{\epsilon}H_{\alpha \beta \delta} \nabla^{\eta}H_{\varepsilon}{}^{\mu \zeta} + c_{787} H_{\alpha}{}^{\delta \epsilon} H^{\alpha \beta \gamma} R_{\epsilon \eta\mu \zeta} \nabla^{\varepsilon}H_{\beta \gamma \delta} \nabla^{\eta}H_{\varepsilon}{}^{\mu \zeta}  \nn\\&&+ c_{788} H_{\alpha}{}^{\delta \epsilon} H^{\alpha \beta \gamma} R_{\beta \delta \gamma \epsilon} \nabla_{\eta}H_{\varepsilon\mu \zeta} \nabla^{\eta}H^{\varepsilon\mu \zeta}\labell{T51}
\eeqa
\beqa
{\cal L}_3^{R(\prt H)^2(\prt\Phi)^2}&\!\!\!\!\!=\!\!\!\!\! & c_{501} R_{\beta\mu \delta \varepsilon} \nabla^{\alpha}\Phi \nabla^{\beta}\Phi \nabla^{\epsilon}H_{\alpha}{}^{\gamma \delta} \nabla^{\varepsilon 1}H_{\gamma \epsilon}{}^{\varepsilon} + c_{504} R_{\delta\mu \epsilon \varepsilon} \nabla^{\alpha}\Phi \nabla_{\beta}H_{\alpha}{}^{\gamma \delta} \nabla^{\beta}\Phi \nabla^{\varepsilon 1}H_{\gamma}{}^{\epsilon \varepsilon}\labell{T52}\nn
\eeqa
\beqa
{\cal L}_3^{H^4(\prt H)^2}&\!\!\!\!\!=\!\!\!\!\! & c_{789} H_{\alpha \beta}{}^{\delta} H^{\alpha \beta \gamma} H_{\gamma}{}^{\epsilon \varepsilon} H_{\delta}{}^{\mu \zeta} \nabla_{\mu}H_{\epsilon}{}^{\eta \theta} \nabla_{\zeta}H_{\varepsilon \eta \theta} + c_{790} H_{\alpha \beta}{}^{\delta} H^{\alpha \beta \gamma} H_{\epsilon \varepsilon}{}^{\zeta} H^{\epsilon \varepsilon \mu} \nabla_{\delta}H_{\gamma}{}^{\eta \theta} \nabla_{\zeta}H_{\mu \eta \theta}  \nn\\&&+ c_{791} H_{\alpha}{}^{\delta \epsilon} H^{\alpha \beta \gamma} H_{\beta}{}^{\varepsilon \mu} H_{\delta \varepsilon}{}^{\zeta} \nabla_{\epsilon}H_{\gamma}{}^{\eta \theta} \nabla_{\zeta}H_{\mu \eta \theta} + c_{792} H_{\alpha}{}^{\delta \epsilon} H^{\alpha \beta \gamma} H_{\varepsilon}{}^{\eta \theta} H^{\varepsilon \mu \zeta} \nabla_{\zeta}H_{\gamma \epsilon \theta} \nabla_{\eta}H_{\beta \delta \mu}  \nn\\&&+ c_{793} H_{\alpha}{}^{\delta \epsilon} H^{\alpha \beta \gamma} H_{\beta}{}^{\varepsilon \mu} H_{\delta}{}^{\zeta \eta} \nabla_{\varepsilon}H_{\gamma \zeta}{}^{\theta} \nabla_{\eta}H_{\epsilon \mu \theta} + c_{794} H_{\alpha}{}^{\delta \epsilon} H^{\alpha \beta \gamma} H_{\beta \delta}{}^{\varepsilon} H^{\mu \zeta \eta} \nabla_{\mu}H_{\gamma \epsilon}{}^{\theta} \nabla_{\eta}H_{\varepsilon \zeta \theta} \nn\\&& + c_{795} H_{\alpha \beta}{}^{\delta} H^{\alpha \beta \gamma} H_{\epsilon}{}^{\zeta \eta} H^{\epsilon \varepsilon \mu} \nabla_{\delta}H_{\gamma \varepsilon}{}^{\theta} \nabla_{\eta}H_{\mu \zeta \theta} + c_{796} H_{\alpha}{}^{\delta \epsilon} H^{\alpha \beta \gamma} H_{\beta}{}^{\varepsilon \mu} H_{\delta}{}^{\zeta \eta} \nabla_{\epsilon}H_{\gamma \varepsilon}{}^{\theta} \nabla_{\eta}H_{\mu \zeta \theta} \nn\\&& + c_{797} H_{\alpha}{}^{\delta \epsilon} H^{\alpha \beta \gamma} H_{\varepsilon}{}^{\eta \theta} H^{\varepsilon \mu \zeta} \nabla_{\eta}H_{\beta \delta \mu} \nabla_{\theta}H_{\gamma \epsilon \zeta} + c_{798} H_{\alpha}{}^{\delta \epsilon} H^{\alpha \beta \gamma} H_{\varepsilon}{}^{\eta \theta} H^{\varepsilon \mu \zeta} \nabla_{\zeta}H_{\beta \delta \mu} \nabla_{\theta}H_{\gamma \epsilon \eta}  \nn\\&&+ c_{799} H_{\alpha}{}^{\delta \epsilon} H^{\alpha \beta \gamma} H_{\varepsilon}{}^{\eta \theta} H^{\varepsilon \mu \zeta} \nabla_{\eta}H_{\beta \gamma \mu} \nabla_{\theta}H_{\delta \epsilon \zeta} + c_{800} H_{\alpha}{}^{\delta \epsilon} H^{\alpha \beta \gamma} H_{\varepsilon}{}^{\eta \theta} H^{\varepsilon \mu \zeta} \nabla_{\zeta}H_{\beta \gamma \mu} \nabla_{\theta}H_{\delta \epsilon \eta}  \nn\\&&+ c_{801} H_{\alpha \beta}{}^{\delta} H^{\alpha \beta \gamma} H^{\epsilon \varepsilon \mu} H^{\zeta \eta \theta} \nabla_{\varepsilon}H_{\gamma \epsilon \zeta} \nabla_{\theta}H_{\delta \mu \eta} + c_{802} H_{\alpha \beta}{}^{\delta} H^{\alpha \beta \gamma} H^{\epsilon \varepsilon \mu} H^{\zeta \eta \theta} \nabla_{\zeta}H_{\gamma \epsilon \varepsilon} \nabla_{\theta}H_{\delta \mu \eta} \nn\\&& + c_{803} H_{\alpha \beta}{}^{\delta} H^{\alpha \beta \gamma} H^{\epsilon \varepsilon \mu} H^{\zeta \eta \theta} \nabla_{\mu}H_{\gamma \epsilon \varepsilon} \nabla_{\theta}H_{\delta \zeta \eta} + c_{804} H_{\alpha}{}^{\delta \epsilon} H^{\alpha \beta \gamma} H_{\beta}{}^{\varepsilon \mu} H^{\zeta \eta \theta} \nabla_{\delta}H_{\gamma \zeta \eta} \nabla_{\theta}H_{\epsilon \varepsilon \mu} \nn\\&& + c_{805} H_{\alpha}{}^{\delta \epsilon} H^{\alpha \beta \gamma} H_{\beta}{}^{\varepsilon \mu} H^{\zeta \eta \theta} \nabla_{\eta}H_{\gamma \delta \zeta} \nabla_{\theta}H_{\epsilon \varepsilon \mu} + c_{806} H_{\alpha}{}^{\delta \epsilon} H^{\alpha \beta \gamma} H_{\beta}{}^{\varepsilon \mu} H^{\zeta \eta \theta} \nabla_{\varepsilon}H_{\gamma \delta \zeta} \nabla_{\theta}H_{\epsilon \mu \eta} \nn\\&& + c_{807} H_{\alpha}{}^{\delta \epsilon} H^{\alpha \beta \gamma} H_{\varepsilon}{}^{\eta \theta} H^{\varepsilon \mu \zeta} \nabla_{\delta}H_{\beta \gamma \mu} \nabla_{\theta}H_{\epsilon \zeta \eta} + c_{808} H_{\alpha}{}^{\delta \epsilon} H^{\alpha \beta \gamma} H_{\varepsilon}{}^{\eta \theta} H^{\varepsilon \mu \zeta} \nabla_{\mu}H_{\beta \gamma \delta} \nabla_{\theta}H_{\epsilon \zeta \eta}  \nn\\&&+ c_{809} H_{\alpha \beta}{}^{\delta} H^{\alpha \beta \gamma} H^{\epsilon \varepsilon \mu} H^{\zeta \eta \theta} \nabla_{\delta}H_{\gamma \epsilon \zeta} \nabla_{\theta}H_{\varepsilon \mu \eta} + c_{810} H_{\alpha}{}^{\delta \epsilon} H^{\alpha \beta \gamma} H_{\beta}{}^{\varepsilon \mu} H^{\zeta \eta \theta} \nabla_{\epsilon}H_{\gamma \delta \zeta} \nabla_{\theta}H_{\varepsilon \mu \eta}  \nn\\&&+ c_{811} H_{\alpha}{}^{\delta \epsilon} H^{\alpha \beta \gamma} H_{\beta}{}^{\varepsilon \mu} H^{\zeta \eta \theta} \nabla_{\zeta}H_{\gamma \delta \epsilon} \nabla_{\theta}H_{\varepsilon \mu \eta} + c_{812} H_{\alpha \beta}{}^{\delta} H^{\alpha \beta \gamma} H^{\epsilon \varepsilon \mu} H^{\zeta \eta \theta} \nabla_{\delta}H_{\gamma \epsilon \varepsilon} \nabla_{\theta}H_{\mu \zeta \eta}  \nn\\&&+ c_{813} H_{\alpha \beta}{}^{\delta} H^{\alpha \beta \gamma} H_{\epsilon}{}^{\zeta \eta} H^{\epsilon \varepsilon \mu} \nabla_{\delta}H_{\gamma \varepsilon}{}^{\theta} \nabla_{\theta}H_{\mu \zeta \eta} + c_{814} H_{\alpha}{}^{\delta \epsilon} H^{\alpha \beta \gamma} H_{\beta}{}^{\varepsilon \mu} H^{\zeta \eta \theta} \nabla_{\epsilon}H_{\gamma \delta \varepsilon} \nabla_{\theta}H_{\mu \zeta \eta}  \nn\\&&+ c_{815} H_{\alpha}{}^{\delta \epsilon} H^{\alpha \beta \gamma} H_{\beta}{}^{\varepsilon \mu} H^{\zeta \eta \theta} \nabla_{\varepsilon}H_{\gamma \delta \epsilon} \nabla_{\theta}H_{\mu \zeta \eta} + c_{816} H_{\alpha}{}^{\delta \epsilon} H^{\alpha \beta \gamma} H_{\beta}{}^{\varepsilon \mu} H_{\delta}{}^{\zeta \eta} \nabla_{\varepsilon}H_{\gamma \epsilon}{}^{\theta} \nabla_{\theta}H_{\mu \zeta \eta}  \nn\\&&+ c_{817} H_{\alpha \beta}{}^{\delta} H^{\alpha \beta \gamma} H_{\gamma}{}^{\epsilon \varepsilon} H_{\epsilon}{}^{\mu \zeta} \nabla_{\varepsilon}H_{\delta}{}^{\eta \theta} \nabla_{\theta}H_{\mu \zeta \eta} + c_{818} H_{\alpha}{}^{\delta \epsilon} H^{\alpha \beta \gamma} H_{\beta \delta}{}^{\varepsilon} H^{\mu \zeta \eta} \nabla_{\eta}H_{\varepsilon \zeta \theta} \nabla^{\theta}H_{\gamma \epsilon \mu}  \nn\\&&+ c_{819} H_{\alpha}{}^{\delta \epsilon} H^{\alpha \beta \gamma} H_{\beta}{}^{\varepsilon \mu} H_{\delta \varepsilon}{}^{\zeta} \nabla_{\theta}H_{\mu \zeta \eta} \nabla^{\theta}H_{\gamma \epsilon}{}^{\eta} + c_{820} H_{\alpha}{}^{\delta \epsilon} H^{\alpha \beta \gamma} H_{\beta}{}^{\varepsilon \mu} H_{\delta}{}^{\zeta \eta} \nabla_{\eta}H_{\epsilon \zeta \theta} \nabla^{\theta}H_{\gamma \varepsilon \mu}  \nn\\&&+ c_{821} H_{\alpha \beta}{}^{\delta} H^{\alpha \beta \gamma} H_{\epsilon}{}^{\zeta \eta} H^{\epsilon \varepsilon \mu} \nabla_{\theta}H_{\delta \zeta \eta} \nabla^{\theta}H_{\gamma \varepsilon \mu} + c_{822} H_{\alpha}{}^{\delta \epsilon} H^{\alpha \beta \gamma} H_{\beta}{}^{\varepsilon \mu} H_{\delta}{}^{\zeta \eta} \nabla_{\theta}H_{\epsilon \zeta \eta} \nabla^{\theta}H_{\gamma \varepsilon \mu}  \nn\\&&+ c_{823} H_{\alpha}{}^{\delta \epsilon} H^{\alpha \beta \gamma} H_{\beta}{}^{\varepsilon \mu} H_{\delta}{}^{\zeta \eta} \nabla_{\mu}H_{\epsilon \eta \theta} \nabla^{\theta}H_{\gamma \varepsilon \zeta} + c_{824} H_{\alpha \beta}{}^{\delta} H^{\alpha \beta \gamma} H_{\epsilon}{}^{\zeta \eta} H^{\epsilon \varepsilon \mu} \nabla_{\eta}H_{\delta \mu \theta} \nabla^{\theta}H_{\gamma \varepsilon \zeta}  \nn\\&&+ c_{825} H_{\alpha}{}^{\delta \epsilon} H^{\alpha \beta \gamma} H_{\beta}{}^{\varepsilon \mu} H_{\delta}{}^{\zeta \eta} \nabla_{\eta}H_{\epsilon \mu \theta} \nabla^{\theta}H_{\gamma \varepsilon \zeta} + c_{826} H_{\alpha \beta}{}^{\delta} H^{\alpha \beta \gamma} H_{\epsilon}{}^{\zeta \eta} H^{\epsilon \varepsilon \mu} \nabla_{\theta}H_{\delta \mu \eta} \nabla^{\theta}H_{\gamma \varepsilon \zeta}  \nn\\&&+ c_{827} H_{\alpha}{}^{\delta \epsilon} H^{\alpha \beta \gamma} H_{\beta}{}^{\varepsilon \mu} H_{\delta}{}^{\zeta \eta} \nabla_{\theta}H_{\epsilon \mu \eta} \nabla^{\theta}H_{\gamma \varepsilon \zeta} + c_{828} H_{\alpha}{}^{\delta \epsilon} H^{\alpha \beta \gamma} H_{\beta \delta}{}^{\varepsilon} H^{\mu \zeta \eta} \nabla_{\varepsilon}H_{\epsilon \eta \theta} \nabla^{\theta}H_{\gamma \mu \zeta}  \nn\\&&+ c_{829} H_{\alpha \beta}{}^{\delta} H^{\alpha \beta \gamma} H_{\epsilon \varepsilon}{}^{\zeta} H^{\epsilon \varepsilon \mu} \nabla_{\eta}H_{\delta \zeta \theta} \nabla^{\theta}H_{\gamma \mu}{}^{\eta} + c_{830} H_{\alpha \beta}{}^{\delta} H^{\alpha \beta \gamma} H_{\epsilon \varepsilon}{}^{\zeta} H^{\epsilon \varepsilon \mu} \nabla_{\theta}H_{\delta \zeta \eta} \nabla^{\theta}H_{\gamma \mu}{}^{\eta}  \nn\\&&+ c_{831} H_{\alpha}{}^{\delta \epsilon} H^{\alpha \beta \gamma} H_{\beta}{}^{\varepsilon \mu} H_{\delta}{}^{\zeta \eta} \nabla_{\mu}H_{\epsilon \varepsilon \theta} \nabla^{\theta}H_{\gamma \zeta \eta} + c_{832} H_{\alpha}{}^{\delta \epsilon} H^{\alpha \beta \gamma} H_{\beta}{}^{\varepsilon \mu} H_{\delta}{}^{\zeta \eta} \nabla_{\theta}H_{\epsilon \varepsilon \mu} \nabla^{\theta}H_{\gamma \zeta \eta}  \nn\\&&+ c_{833} H_{\alpha}{}^{\delta \epsilon} H^{\alpha \beta \gamma} H_{\beta}{}^{\varepsilon \mu} H_{\gamma}{}^{\zeta \eta} \nabla_{\eta}H_{\mu \zeta \theta} \nabla^{\theta}H_{\delta \epsilon \varepsilon} + c_{834} H_{\alpha}{}^{\delta \epsilon} H^{\alpha \beta \gamma} H_{\beta}{}^{\varepsilon \mu} H_{\gamma}{}^{\zeta \eta} \nabla_{\theta}H_{\mu \zeta \eta} \nabla^{\theta}H_{\delta \epsilon \varepsilon}  \nn\\&&+ c_{835} H_{\alpha \beta}{}^{\delta} H^{\alpha \beta \gamma} H_{\gamma}{}^{\epsilon \varepsilon} H^{\mu \zeta \eta} \nabla_{\eta}H_{\varepsilon \zeta \theta} \nabla^{\theta}H_{\delta \epsilon \mu} + c_{836} H_{\alpha \beta \gamma} H^{\alpha \beta \gamma} H^{\delta \epsilon \varepsilon} H^{\mu \zeta \eta} \nabla_{\eta}H_{\varepsilon \zeta \theta} \nabla^{\theta}H_{\delta \epsilon \mu}  \nn\\&&+ c_{837} H_{\alpha \beta}{}^{\delta} H^{\alpha \beta \gamma} H_{\gamma}{}^{\epsilon \varepsilon} H^{\mu \zeta \eta} \nabla_{\theta}H_{\varepsilon \zeta \eta} \nabla^{\theta}H_{\delta \epsilon \mu} + c_{838} H_{\alpha \beta \gamma} H^{\alpha \beta \gamma} H^{\delta \epsilon \varepsilon} H^{\mu \zeta \eta} \nabla_{\theta}H_{\varepsilon \zeta \eta} \nabla^{\theta}H_{\delta \epsilon \mu} \nn\\&& + c_{839} H_{\alpha \beta}{}^{\delta} H^{\alpha \beta \gamma} H_{\gamma}{}^{\epsilon \varepsilon} H_{\epsilon}{}^{\mu \zeta} \nabla_{\eta}H_{\mu \zeta \theta} \nabla^{\theta}H_{\delta \varepsilon}{}^{\eta} + c_{840} H_{\alpha \beta}{}^{\delta} H^{\alpha \beta \gamma} H_{\gamma}{}^{\epsilon \varepsilon} H_{\epsilon}{}^{\mu \zeta} \nabla_{\theta}H_{\mu \zeta \eta} \nabla^{\theta}H_{\delta \varepsilon}{}^{\eta} \nn\\&& + c_{841} H_{\alpha \beta}{}^{\delta} H^{\alpha \beta \gamma} H_{\gamma}{}^{\epsilon \varepsilon} H^{\mu \zeta \eta} \nabla_{\eta}H_{\epsilon \varepsilon \theta} \nabla^{\theta}H_{\delta \mu \zeta} + c_{842} H_{\alpha \beta}{}^{\delta} H^{\alpha \beta \gamma} H_{\gamma}{}^{\epsilon \varepsilon} H^{\mu \zeta \eta} \nabla_{\theta}H_{\epsilon \varepsilon \eta} \nabla^{\theta}H_{\delta \mu \zeta}  \nn\\&&+ c_{843} H_{\alpha \beta}{}^{\delta} H^{\alpha \beta \gamma} H_{\gamma}{}^{\epsilon \varepsilon} H_{\epsilon}{}^{\mu \zeta} \nabla_{\zeta}H_{\varepsilon \eta \theta} \nabla^{\theta}H_{\delta \mu}{}^{\eta} + c_{844} H_{\alpha \beta}{}^{\delta} H^{\alpha \beta \gamma} H_{\gamma}{}^{\epsilon \varepsilon} H_{\epsilon}{}^{\mu \zeta} \nabla_{\eta}H_{\varepsilon \zeta \theta} \nabla^{\theta}H_{\delta \mu}{}^{\eta}  \nn\\&&+ c_{845} H_{\alpha \beta}{}^{\delta} H^{\alpha \beta \gamma} H_{\gamma}{}^{\epsilon \varepsilon} H_{\epsilon}{}^{\mu \zeta} \nabla_{\theta}H_{\varepsilon \zeta \eta} \nabla^{\theta}H_{\delta \mu}{}^{\eta} + c_{846} H_{\alpha \beta}{}^{\delta} H^{\alpha \beta \gamma} H_{\gamma}{}^{\epsilon \varepsilon} H_{\epsilon \varepsilon}{}^{\mu} \nabla_{\eta}H_{\mu \zeta \theta} \nabla^{\theta}H_{\delta}{}^{\zeta \eta}  \nn\\&&+ c_{847} H_{\alpha \beta}{}^{\delta} H^{\alpha \beta \gamma} H_{\gamma}{}^{\epsilon \varepsilon} H_{\epsilon \varepsilon}{}^{\mu} \nabla_{\theta}H_{\mu \zeta \eta} \nabla^{\theta}H_{\delta}{}^{\zeta \eta} + c_{848} H_{\alpha \beta}{}^{\delta} H^{\alpha \beta \gamma} H_{\gamma}{}^{\epsilon \varepsilon} H^{\mu \zeta \eta} \nabla_{\delta}H_{\zeta \eta \theta} \nabla^{\theta}H_{\epsilon \varepsilon \mu}  \nn\\&&+ c_{849} H_{\alpha}{}^{\delta \epsilon} H^{\alpha \beta \gamma} H_{\beta \delta}{}^{\varepsilon} H_{\gamma}{}^{\mu \zeta} \nabla_{\eta}H_{\mu \zeta \theta} \nabla^{\theta}H_{\epsilon \varepsilon}{}^{\eta} + c_{850} H_{\alpha \beta}{}^{\delta} H^{\alpha \beta \gamma} H_{\gamma}{}^{\epsilon \varepsilon} H_{\delta}{}^{\mu \zeta} \nabla_{\eta}H_{\mu \zeta \theta} \nabla^{\theta}H_{\epsilon \varepsilon}{}^{\eta}  \nn\\&&+ c_{851} H_{\alpha \beta \gamma} H^{\alpha \beta \gamma} H_{\delta}{}^{\mu \zeta} H^{\delta \epsilon \varepsilon} \nabla_{\eta}H_{\mu \zeta \theta} \nabla^{\theta}H_{\epsilon \varepsilon}{}^{\eta} + c_{852} H_{\alpha}{}^{\delta \epsilon} H^{\alpha \beta \gamma} H_{\beta \delta}{}^{\varepsilon} H_{\gamma}{}^{\mu \zeta} \nabla_{\theta}H_{\mu \zeta \eta} \nabla^{\theta}H_{\epsilon \varepsilon}{}^{\eta}  \nn\\&&+ c_{853} H_{\alpha \beta}{}^{\delta} H^{\alpha \beta \gamma} H_{\gamma}{}^{\epsilon \varepsilon} H_{\delta}{}^{\mu \zeta} \nabla_{\theta}H_{\mu \zeta \eta} \nabla^{\theta}H_{\epsilon \varepsilon}{}^{\eta} + c_{854} H_{\alpha \beta \gamma} H^{\alpha \beta \gamma} H_{\delta}{}^{\mu \zeta} H^{\delta \epsilon \varepsilon} \nabla_{\theta}H_{\mu \zeta \eta} \nabla^{\theta}H_{\epsilon \varepsilon}{}^{\eta} \nn\\&& + c_{855} H_{\alpha \beta}{}^{\delta} H^{\alpha \beta \gamma} H_{\gamma}{}^{\epsilon \varepsilon} H^{\mu \zeta \eta} \nabla_{\delta}H_{\varepsilon \eta \theta} \nabla^{\theta}H_{\epsilon \mu \zeta} + c_{856} H_{\alpha}{}^{\delta \epsilon} H^{\alpha \beta \gamma} H_{\beta \delta}{}^{\varepsilon} H_{\gamma}{}^{\mu \zeta} \nabla_{\zeta}H_{\varepsilon \eta \theta} \nabla^{\theta}H_{\epsilon \mu}{}^{\eta}  \nn\\&&+ c_{857} H_{\alpha}{}^{\delta \epsilon} H^{\alpha \beta \gamma} H_{\beta \delta}{}^{\varepsilon} H_{\gamma}{}^{\mu \zeta} \nabla_{\eta}H_{\varepsilon \zeta \theta} \nabla^{\theta}H_{\epsilon \mu}{}^{\eta} + c_{858} H_{\alpha \beta}{}^{\delta} H^{\alpha \beta \gamma} H_{\gamma}{}^{\epsilon \varepsilon} H_{\delta}{}^{\mu \zeta} \nabla_{\eta}H_{\varepsilon \zeta \theta} \nabla^{\theta}H_{\epsilon \mu}{}^{\eta}  \nn\\&&+ c_{859} H_{\alpha \beta \gamma} H^{\alpha \beta \gamma} H_{\delta}{}^{\mu \zeta} H^{\delta \epsilon \varepsilon} \nabla_{\eta}H_{\varepsilon \zeta \theta} \nabla^{\theta}H_{\epsilon \mu}{}^{\eta} + c_{860} H_{\alpha}{}^{\delta \epsilon} H^{\alpha \beta \gamma} H_{\beta \delta}{}^{\varepsilon} H_{\gamma}{}^{\mu \zeta} \nabla_{\theta}H_{\varepsilon \zeta \eta} \nabla^{\theta}H_{\epsilon \mu}{}^{\eta}  \nn\\&&+ c_{861} H_{\alpha \beta}{}^{\delta} H^{\alpha \beta \gamma} H_{\gamma}{}^{\epsilon \varepsilon} H_{\delta}{}^{\mu \zeta} \nabla_{\theta}H_{\varepsilon \zeta \eta} \nabla^{\theta}H_{\epsilon \mu}{}^{\eta} + c_{862} H_{\alpha \beta \gamma} H^{\alpha \beta \gamma} H_{\delta}{}^{\mu \zeta} H^{\delta \epsilon \varepsilon} \nabla_{\theta}H_{\varepsilon \zeta \eta} \nabla^{\theta}H_{\epsilon \mu}{}^{\eta}  \nn\\&&+ c_{863} H_{\alpha \beta}{}^{\delta} H^{\alpha \beta \gamma} H_{\gamma}{}^{\epsilon \varepsilon} H_{\epsilon}{}^{\mu \zeta} \nabla_{\delta}H_{\zeta \eta \theta} \nabla^{\theta}H_{\varepsilon \mu}{}^{\eta} + c_{864} H_{\alpha}{}^{\delta \epsilon} H^{\alpha \beta \gamma} H_{\beta \delta}{}^{\varepsilon} H_{\gamma \epsilon}{}^{\mu} \nabla_{\eta}H_{\mu \zeta \theta} \nabla^{\theta}H_{\varepsilon}{}^{\zeta \eta}  \nn\\&&+ c_{865} H_{\alpha \beta}{}^{\delta} H^{\alpha \beta \gamma} H_{\gamma}{}^{\epsilon \varepsilon} H_{\delta \epsilon}{}^{\mu} \nabla_{\eta}H_{\mu \zeta \theta} \nabla^{\theta}H_{\varepsilon}{}^{\zeta \eta} + c_{866} H_{\alpha \beta \gamma} H^{\alpha \beta \gamma} H_{\delta \epsilon}{}^{\mu} H^{\delta \epsilon \varepsilon} \nabla_{\eta}H_{\mu \zeta \theta} \nabla^{\theta}H_{\varepsilon}{}^{\zeta \eta}  \nn\\&&+ c_{867} H_{\alpha}{}^{\delta \epsilon} H^{\alpha \beta \gamma} H_{\beta \delta}{}^{\varepsilon} H_{\gamma \epsilon}{}^{\mu} \nabla_{\theta}H_{\mu \zeta \eta} \nabla^{\theta}H_{\varepsilon}{}^{\zeta \eta} + c_{868} H_{\alpha \beta}{}^{\delta} H^{\alpha \beta \gamma} H_{\gamma}{}^{\epsilon \varepsilon} H_{\delta \epsilon}{}^{\mu} \nabla_{\theta}H_{\mu \zeta \eta} \nabla^{\theta}H_{\varepsilon}{}^{\zeta \eta}  \nn\\&&+ c_{869} H_{\alpha \beta \gamma} H^{\alpha \beta \gamma} H_{\delta \epsilon}{}^{\mu} H^{\delta \epsilon \varepsilon} \nabla_{\theta}H_{\mu \zeta \eta} \nabla^{\theta}H_{\varepsilon}{}^{\zeta \eta} + c_{870} H_{\alpha}{}^{\delta \epsilon} H^{\alpha \beta \gamma} H_{\beta \delta}{}^{\varepsilon} H_{\gamma \epsilon \varepsilon} \nabla_{\theta}H_{\mu \zeta \eta} \nabla^{\theta}H^{\mu \zeta \eta}  \nn\\&&+ c_{871} H_{\alpha \beta}{}^{\delta} H^{\alpha \beta \gamma} H_{\gamma}{}^{\epsilon \varepsilon} H_{\delta \epsilon \varepsilon} \nabla_{\theta}H_{\mu \zeta \eta} \nabla^{\theta}H^{\mu \zeta \eta} + c_{872} H_{\alpha \beta \gamma} H^{\alpha \beta \gamma} H_{\delta \epsilon \varepsilon} H^{\delta \epsilon \varepsilon} \nabla_{\theta}H_{\mu \zeta \eta} \nabla^{\theta}H^{\mu \zeta \eta}\labell{T54}\nn
\eeqa
\beqa
{\cal L}_3^{H(\prt H)^3\prt\Phi}&\!\!\!\!\!=\!\!\!\!\! & c_{623} H_{\alpha}{}^{\beta \gamma} \nabla^{\alpha}\Phi \nabla^{\varepsilon}H_{\beta}{}^{\delta \epsilon} \nabla_{\zeta}H_{\delta \epsilon \mu} \nabla^{\zeta}H_{\gamma \varepsilon}{}^{\mu} + c_{624} H_{\alpha}{}^{\beta \gamma} \nabla^{\alpha}\Phi \nabla_{\gamma}H_{\varepsilon \mu \zeta} \nabla^{\varepsilon}H_{\beta}{}^{\delta \epsilon} \nabla^{\zeta}H_{\delta \epsilon}{}^{\mu}  \nn\\&&+ c_{625} H^{\beta \gamma \delta} \nabla_{\alpha}H_{\epsilon \mu \zeta} \nabla^{\alpha}\Phi \nabla^{\varepsilon}H_{\beta \gamma}{}^{\epsilon} \nabla^{\zeta}H_{\delta \varepsilon}{}^{\mu} + c_{626} H_{\alpha}{}^{\beta \gamma} \nabla^{\alpha}\Phi \nabla_{\gamma}H_{\epsilon \mu \zeta} \nabla^{\varepsilon}H_{\beta}{}^{\delta \epsilon} \nabla^{\zeta}H_{\delta \varepsilon}{}^{\mu} \nn\\&& + c_{627} H_{\alpha}{}^{\beta \gamma} \nabla^{\alpha}\Phi \nabla^{\epsilon}H_{\beta \gamma}{}^{\delta} \nabla_{\zeta}H_{\epsilon \varepsilon \mu} \nabla^{\zeta}H_{\delta}{}^{\varepsilon \mu} + c_{628} H^{\beta \gamma \delta} \nabla^{\alpha}\Phi \nabla_{\gamma}H_{\alpha \beta}{}^{\epsilon} \nabla_{\delta}H_{\varepsilon \mu \zeta} \nabla^{\zeta}H_{\epsilon}{}^{\varepsilon \mu} \nn\\&& + c_{629} H_{\alpha}{}^{\beta \gamma} \nabla^{\alpha}\Phi \nabla_{\delta}H_{\varepsilon \mu \zeta} \nabla^{\epsilon}H_{\beta \gamma}{}^{\delta} \nabla^{\zeta}H_{\epsilon}{}^{\varepsilon \mu}\labell{T55}
\eeqa
where $c_1,\cdots, c_{872}$ are 872 arbitrary parameters that can not be fixed by the gauge symmetry.

We have found the above minimal couplings in the theory which has no four-derivative and six-derivative couplings as in the superstring theory, \ie we have used the field redefinition \reef{eq.13}.  It has been shown in \cite{Garousi:2019cdn} that even if the theory has four-derivative and six-derivative couplings with fixed parameters, up to some total derivative terms, the field redefinition at order $\alpha'^3$ can be written as 
\beqa
\delta S_0+\delta S_1+\delta S_2&=&\frac{\delta S_0}{\delta g_{\alpha\beta}}\delta g^{(3)}_{\alpha\beta}+\frac{\delta S_0}{\delta B_{\alpha\beta}}\delta B^{(3)}_{\alpha\beta}+\frac{\delta S_0}{\delta \Phi}\delta \Phi^{(3)}
\eeqa
where the deformations $\delta {g}_{\mu\nu}^{(3)}, \delta {B}_{\mu\nu}^{(3)}, \delta {\Phi}^{(3)}$  are arbitrary functions of $R,H,\nabla\Phi$ and their derivatives at order $\alpha'^3$. 
Hence, the minimal gauge invariant  couplings that we have found in this paper are valid  couplings in any higher-derivative theory which has metric, dilaton and B-field. %  which has no four- and six-derivative couplings. 
 
 Even though the total number of minimal gauge invariant  couplings at order $\alpha'^3$ are fixed, \ie 872,  the number of couplings in each structure are not fixed. In different schemes, one may find  different structures and different number of couplings in each structure. The above structures and the number of terms in each structure are fixed in the scheme that we have chosen. Note, however, that the couplings with coefficients $c_1,c_2,c_5,c_7,c_8$ in $\cL_3^{H^8}$  are invariant under field redefinition, Bianchi identities and total derivative terms. They are scheme independent. All other couplings dependent on the scheme that one uses for the  couplings. The values of the 872 parameters are fixed in a specific theory by impose various techniques in the theory. 

In the superstring theory the above 872 parameters may be found by calculating various S-matrix elements in the effective field theory \reef{seff} and comparing them with the corresponding S-matrix elements in the string theory which has no arbitrary parameters. In this method one has to calculate in the  string theory various S-matrix elements which produces 872 independent contact terms. In the next section we illustrate this method for four-point functions to fix some of the parameters.

\section{Constraint from 4-point functions}\label{sec.3}

The S-matrix element of four NS-NS vertex operators in the superstring theory has been calculated in \cite{ Gross:1986iv,Schwarz:1982jn}. The low energy expansion of this S-matrix element produces the following eight-derivative couplings in the string frame \cite{Gross:1986mw,Myers:1987qx,Policastro:2008hg,Paban:1998ea,Garousi:2013zca}:
\beqa
 S_3&\supset& \frac{\gamma_3}{\kappa^2}\int d^{10}x\sqrt{-g} e^{-2\Phi} \cL(\cR) \labell{S2}
 \eeqa  
 where the normalization factor is $\gamma_3=\z(3)/2^5$, and the Lagrangian density has the following eight independent terms:
 \beqa
 \cL(\cR)&=&  \frac{1}{8}\cR_{\kappa \gamma}{}^{\delta \beta} \cR^{\kappa \gamma \tau \alpha}\cR_{\mu \nu \delta \tau} \cR^{\mu \nu}{}_{\alpha \beta} + \frac{1}{32}\cR_{\alpha \beta \kappa \gamma} \cR^{\alpha \beta \kappa \gamma}\cR_{\mu \nu \delta \tau} \cR^{\mu \nu \delta \tau}\nn\\&& + \frac{1}{16}\cR_{\alpha \beta}{}^{\delta \tau} \cR^{\alpha \beta \kappa \gamma}\cR_{\mu \nu \delta \tau} \cR^{\mu \nu}{}_{\kappa \gamma} -  \frac{1}{4}\cR_{\kappa \gamma}{}^{\alpha \beta} \cR^{\kappa \gamma \delta}{}_{\alpha}\cR_{\mu \nu \delta \tau} \cR^{\mu \nu \tau}{}_{\beta}\nn\\&& + \frac{1}{4} \cR^{\alpha}{}_{\beta \kappa \gamma}\cR_{\mu \nu \delta \tau} \cR^{\mu \beta \kappa \gamma} \cR^{\nu}{}_{\alpha}{}^{\delta \tau} + \frac{1}{8} \cR^{\alpha}{}_{\beta}{}^{\delta \tau}\cR_{\mu \nu \delta \tau} \cR^{\mu \beta \kappa \gamma} \cR^{\nu}{}_{\alpha \kappa \gamma}\nn\\&& + \frac{1}{2} \cR^{\alpha}{}_{\beta}{}^{\kappa}{}_{\gamma}\cR_{\mu \nu \delta \tau} \cR^{\mu \beta \delta \gamma} \cR^{\nu}{}_{\alpha}{}^{\tau}{}_{\kappa} + \cR^{\alpha}{}_{\beta}{}^{\delta}{}_{\gamma}\cR_{\mu \nu \delta \tau} \cR^{\mu \beta \kappa \gamma} \cR^{\nu}{}_{\alpha}{}^{\tau}{}_{\kappa}\labell{L}
 \eeqa
 where   $\cR_{\mu\nu\alpha\beta}$ is the linear part of the following tensor:
\beqa
\cR_{\mu\nu\alpha\beta}&=& R_{\mu\nu\alpha\beta}+H_{\mu\nu[\alpha;\beta]}
\eeqa 
Here $H_{\mu\nu[\alpha;\beta]}=\frac{1}{2}\nabla_{\beta}H_{\mu\nu\alpha}-\frac{1}{2}\nabla_{\alpha}H_{\mu\nu\beta}$. While the dilaton  appears non-trivially in the Einstein frame, it appears in the string frame only as the overall factor of $e^{-2\Phi}$ \cite{Garousi:2012jp,Liu:2013dna}. Note that if one ignores the B-field coupling, then the symmetries of  the Riemann curvature reduces the eight terms in \reef{L} to   six independent terms \cite{Gross:1986mw}, however, in the presence of B-field the string theory S-matrix element of four NS-NS vertex operators are reproduced by the above Lagrangian \cite{Garousi:2013zca}. The heterotic and bosonic string theories have the above couplings as well as some other couplings at this order.

Now, requiring  the Lagrangian \reef{T3}, to produce the  four graviton couplings in \reef{L} after using on-shell conditions $k_i\cdot k_i=0$ and $ k_i\cdot \z_i=0$ for $i=1,2,3,4$ where the graviton polarization is $\z_i\z_j$ and momentum of gravition is $k_i$, one finds the following relations for the parameters in \reef{T3}:
\beqa
&&c_{11}= 2 + 2 c_{10}, c_{12}= 1 + 2 c_{10}, c_{20}= 
 - c_{10},\nn\\&& c_{21}= 4 c_{10}, c_{22}= - c_{10}, c_{33}=  - c_{10}/4
 \eeqa
As can be seen not all parameters are fixed by the four-point function. The above result indicates that there is one combination of the seven couplings in \reef{T3} which produces zero effect on the four-point function. Consistency with five-point function should fix  the overall  parameter of this combination, \ie $c_{10}$. Therefore, imposing the four-point  function on the parameters in \reef{T3}, one finds the following couplings
\beqa
{\cal L}_3^{R^4}&\!\!\!\!\!=\!\!\!\!\! &2 R_{\alpha}{}^{\epsilon}{}_{\gamma}{}^{\varepsilon} R^{\alpha \beta \gamma \delta} R_{\beta}{}^{\mu}{}_{\epsilon}{}^{\zeta} R_{\delta \zeta \varepsilon \mu} + R_{\alpha \beta}{}^{\epsilon \varepsilon} R^{\alpha \beta \gamma \delta} R_{\gamma}{}^{\mu}{}_{\epsilon}{}^{\zeta} R_{\delta \zeta \varepsilon \mu}\labell{RRf}
\eeqa 
and some other couplings with coefficient  $c_{10}$. The above couplings are exactly the couplings that have been found from the sigma-model \cite{Grisaru:1986vi,Freeman:1986zh}. Hence, the five-point function must constrain $c_{10}=0$. It is important to note that if
one uses the  KLT constraint \cite{Kawai:1985xq}  to write the couplings in terms of $t_8t_8R^4$ and  $\eps_{10}\inn\eps_{10}R^4$, then the four-point function can fix only the coefficient of $t_8t_8R^4$ and the five-point function fixes the coefficient of  $\eps_{10}\inn\eps_{10}R^4$ to be  non-zero. In the scheme that we have used to write the couplings in the previous section, however, the structure  $\eps_{10}\inn\eps_{10}R^4$ which includes the Ricci and scalar curvatures, does not appear at all. 

Requiring  the Lagrangian \reef{T48} to produce the  four B-field couplings in \reef{L} after using on-shell conditions $k_i\cdot k_i=0$, $ k_i\cdot \z_i=0$ and $ k_i\cdot \z'_i=0$ for $i=1,2,3,4$ where the B-field polarization is $\z_i\z'_j-\z_i'\z_j$ and momentum of B-field is $k_i$, one finds all 9 parameters in \reef{T48} are fixed, \ie
\beqa
&& c_{410}= 1/8, c_{615}= -1/8, c_{616}= -1/4, c_{617}= 1/4,\nn\\&& c_{618}= -1/8, 
 c_{619}= -1/16, c_{620}= 0, c_{621}= -3/16, c_{622}= 1/144
\eeqa
Note that in the scheme that we have chosen in this paper, there are 9 independent couplings with structure $(\prt H)^4$ whose coefficients are fixed by the four-point function.  We could chosen a different scheme in which there would be less couplings with structure $(\prt H)^4$. In fact it has been shown in \cite{Garousi:2012yr} that the four-point function  can be reproduced by three couplings with structure $(\prt H)^4$. If one chooses a scheme in which there are three  couplings with structure $(\prt H)^4$, then the  extra six  independent couplings would appear in other structures such that the total number of independent couplings remains fix, \ie 872.

Similarly, requiring   the Lagrangians \reef{T53} and \reef{T27} to produce two gravitons and two B-field couplings in \reef{L}, one finds parameter in \reef{T53} to be zero and finds 17 relations between the 22 parameters in  \reef{T27}, \ie 
\beqa
&&c_{369}= -24 c_{282} - c_{283}, c_{407}= 1 - 24 c_{282}, c_{408}= 12 c_{282}, 
 c_{409}= 12 c_{282},\nn\\&& c_{416}= 2 - 24 c_{282} + 2 c_{284}, 
 c_{417}= -2 - 4 c_{283} + 2 c_{285}, c_{418}= 4 c_{283} + 2 c_{286}, \nn\\&&
 c_{475}= 2 + 2 c_{283}, c_{476}= -4 c_{283}, c_{477}= 1 + 2 c_{283}, 
 c_{478}= 2 c_{283},\nn\\&& c_{514}= 4 c_{284} + 2 c_{285} + 2 c_{286}, 
 c_{515}= 1 - c_{283} + c_{285} - c_{286},\nn\\&& c_{516}= -2 c_{283} + 2 c_{285} - 2 c_{286}, 
 c_{520}= 0, c_{697}= -2 c_{284} - c_{285} - c_{286},\nn\\&& c_{698}= 1 + c_{284} + c_{285}, 
 c_{699}= 1/2 + c_{284}/2 + c_{285}/2
\eeqa
Hence there are five  different combinations of the 22 couplings in \reef{T27} that produce zero effect on four-point function. They can be found by studying five-point functions of two B-field and three gravitons in which we are not interested in this paper. 

Requiring   the Lagrangian \reef{T38}  to produce no four-point function of two dilatons and two B-fields as in \reef{S2}, one finds the parameter $c_{413}$ does not appear in the on-shell amplitude, and the  following relations between the other parameters in \reef{T38}:
\beqa
&&c_{357}=-c_{353}, \, c_{426}=-3c_{353}+c_{415},\, c_{430}=2c_{353},\, c_{435}=-\frac{1}{6}c_{353}
\eeqa
This indicates there are two other combinations of terms in \reef{T38} that produce zero four-point functions.

Requiring   the Lagrangian \reef{T50}  to produce no four-point function of one dilaton, one graviton  and two B-fields, one finds
\beqa
c_{480}=0,\,c_{486}=0,\,c_{499}=0,\,c_{500}=0,\,c_{502}=0,\,c_{503}=0,\,c_{506}=0
\eeqa
There are also no four-point functions of four dilatons, three dilatons and one graviton, two dilatons and two gravitons, one dilaton and three gravitons in the string frame. Their corresponding parameters are all zero. This can also be seen  by the T-duality constraint when metric is diagonal that we are going to discuss in the next section. 
  
\section{Constraint from T-duality when $B=0$}\label{sec.4}

It is very hard to continue the above S-matrix method to find all 872 parameters in ${\cal L}_3$. In particular, to find the parameters in ${\cal L}_3^{H^8}$, one would need to calculate S-matrix element of eight NS-NS vertex operators which is tremendously difficult. Fortunately, there is a simple method to find all parameters by imposing the T-duality constraint. The calculations here also is very lengthy to perform, however, using the computer one can perform it, see \eg \cite{Garousi:2019mca}. 

The calculation in the absence of B-field has been already done in \cite{Razaghian:2018svg}. In that paper, it has been argued that in any higher-derivative   action  which contains   only metric and dilaton, the dilaton couplings can be set to zero by the T-duality constraint.  The argument in that paper, however, can not be extended to the case that B-field is non-zero. Explicit calculation at order $\alpha'^3$ in the superstring and in the heterotic string theory has been also done in \cite{Razaghian:2018svg}. That is, assuming the dilaton couplings are zero, it has been shown  in \cite{Razaghian:2018svg} that the gravity couplings \reef{T3} at order $\alpha'^3$ are consistent with the T-duality constraint when metric is diagonal and B-field is zero, provided that the parameters in \reef{T3} to be the same as those in \reef{RRf} in the superstring theory. These parameters in the heterotic theory, however, are  fixed up to two parameters by the T-duality which is consistent with the S-matrix calculations in the heterotic theory \cite{Gross:1986mw}. This conforms that even if one considers all independent  gravity and dilaton couplings that we have considered in \reef{T3}, \reef{T7}, \reef{T11}, \reef{T14}, \reef{T19}, \reef{T21}, \reef{T30}, \reef{T31}, \reef{T32}, \reef{T35}, \reef{T36}, \reef{T42}, \reef{T43}, \reef{T44}, then the T-duality satisfied when all the 36  dilaton couplings are zero, \ie     
\beqa
&&c_{94}=0\,,c_{95}=0\,,c_{148}=0\,,c_{149}=0\,,c_{150}=0\,,c_{178}=0\,,c_{289}=0\,,c_{181}=0\,,\nn\\&&c_{261}=0\,,c_{277}=0\,,c_{321}=0\,,c_{322}=0\,,c_{326}=0\,,c_{223}=0\,,c_{224}=0\,,c_{225}=0\,,\nn\\&& c_{254}=0\,,c_{269}=0\,,c_{341}=0\,,c_{342}=0\,,c_{346}=0\,,c_{293}=0\,,c_{294}=0\,,c_{295}=0\,,\nn\\&& c_{304}=0\,,c_{359}=0\,,c_{305}=0\,,c_{307}=0\,,c_{360}=0\,,c_{306}=0\,,c_{361}=0\,,c_{372}=0\,,\nn\\&& c_{377}=0\,,c_{381}=0\,,c_{374}=0\,,c_{379}=0
\eeqa
The dilaton couplings \reef{T19}, \reef{T21}, \reef{T36}, \reef{T44} with the above coefficients are consistent with the four-point functions in \reef{L}.

The minimal gauge  invariant couplings at order $\alpha'$ and $\alpha'^2$ in the bosonic theory are 8 and 60, respectively, however, in the absence of B-field, the dilaton couplings constraint to be zero by explicit T-duality calculations    \cite{Razaghian:2017okr}. Note however that the couplings involving dilaton and B-field may not  be zero by the T-duality constraint, see \eg \cite{Garousi:2019mca}.

When B-field is not zero, one may use the T-duality constraint to fix all 872 couplings. Similar calculations at orders $\alpha'$ and $\alpha'^2$ in the bosonic  string theory have been done in \cite{Garousi:2019mca} to fix all parameters of the minimal  couplings. We postpone this calculation at order $\alpha'^3$ for the  future works.

We have found that  the minimum number of  independent couplings at order $\alpha'^3$ is 872 in any  higher-derivative theory containing the NS-NS  fields. In string theory, using the T-duality constraint one may hopefully find all parameters. Using different schemes, the T-duality can fix the corresponding non-zero terms. A priori one can not argue which minimal scheme would produce minimum number of couplings in string theory after imposing the T-duality constraint. To find the minimum number of couplings in the string theory, one should first find the 872 parameters in a specific scheme, \eg the one we have used in section 2. Then using once again the field redefinitions, the  total derivative terms and using various Bianchi identities, one may rewrite   the couplings such that the number of couplings would be  minimum. 

{\bf Acknowledgments}:   This work is supported by Ferdowsi University of Mashhad.% under grant  1/50070(1398/04/10).
 
% \newpage

\end{document}